\documentclass[11pt]{article}

\usepackage{psfrag,amsfonts,varioref,epsfig,tabularx,color}

\definecolor{mygreen}{rgb}{0.08860759493670886, 0.8, 0.16877637130801687}
\definecolor{mycyan}{rgb}{0, 0.75, 1}
\definecolor{mygold}{rgb}{0.6, 0.6, 0}
\definecolor{myred}{rgb}{0.8, 0, 0}
\usepackage{hhline}
\usepackage{multirow}

\newcommand{\rmt}[1]{\textrm{\tiny{#1}}}
\newcommand{\MSbar}{\ensuremath{\overline{{\rm MS}}}}
\newcommand{\LMS}{\Lambda_{\rmt{\MSbar}}}
\newcommand{\pmpm}[2]{\scriptsize{$\begin{array}{l}\!\!\!#1\\\!\!\!#2\end{array}$}}
\newcommand{\pmpmc}[2]{\scriptsize{$\begin{array}{c}\!\!\!#1\\\!\!\!#2\end{array}$}}
\def\nn{\nonumber}
\def\qq {\qquad}
\def\MI{{\cal I}}
\def\MC{{\cal C}}
\def\MO{{\cal O}}
\def\MM{{\cal M}}

\newcommand{\Rt}{\tilde{R}}
\newcommand{\Rh}{\hat{R}}

\newcommand{\lb}{\left}
\newcommand{\rb}{\right}
\newcommand{\rt}{\tilde{r}}
\newcommand{\pt}{\tilde{p}}
\newcommand{\tF}{\tilde{F}}
\newcommand{\tL}{\tilde{\Lambda}}
\newcommand{\nf}{\ensuremath{N_{f}}}
\newcommand{\tx}{\tilde{x}}
\newcommand{\tX}{\tilde{X}}
\newcommand{\Xh}{\hat{X}}
\newcommand{\rmGeV}{{\rm GeV}}
\newcommand{\GeVV}{{\rm GeV}^{2}}
\newcommand{\rmss}[1]{\textrm{\scriptsize{#1}}}
\newcommand{\D}{\Delta}
\newcommand{\hA}{\hat{A}}

\begin{document}

\thispagestyle{empty}

\title{Improved analysis of moments of $F_3$ in neutrino-nucleon scattering using the Bernstein polynomial method}
\author{\textbf{Paul M. Brooks\footnote{e-mail:\texttt{brooks.pm@googlemail.com}}\,\, and C.J.
Maxwell\footnote{e-mail:\texttt{\texttt{c.j.maxwell@durham.ac.uk}}}}}
\date{\vskip10mm\emph{Institute for Particle Physics Phenomenology, University of Durham,\\South Road, Durham, DH1 3LE,
UK.}}
\vspace{5cm}
\maketitle 
\large \vspace{-8.45cm}\hfill\vbox{\hbox{IPPP/06/64}
            \hbox{DCPT/06/128}
             \hbox{October 2006}}
\normalsize \vspace{8.45cm}
\begin{abstract}
\setlength{\baselineskip}{13pt}

We use recently calculated next-to-next-to-leading order (NNLO)
anomalous dimension coefficients for the moments of the $x{F}_{3}$
structure function in ${\nu}N$ scattering, together with the
corresponding three-loop Wilson coefficients, to obtain improved QCD
predictions for both odd and even moments of $F_{3}$. To investigate
the issue of renormalization scheme dependence, the Complete
Renormalization Group Improvement (CORGI) approach is used, in which
all dependence on renormalization and factorization scales is
avoided by a complete resummation of RG-predictable scale
logarithms. We also consider predictions using the method of
effective charges, and compare with the standard `physical scale'
choice. The Bernstein polynomial method is used to construct
experimental moments (from the $x{F}_{3}$ data of the CCFR
collaboration) that are insensitive to the value of $xF_{3}$ in the
region of $x$ which is inaccessible experimentally. Direct fits for
$\LMS^{(5)}$ ($\alpha_s(M_Z)$) are then performed. The CORGI fits
including target mass corrections give a value
$\alpha_s(M_Z)={0.1189}^{+0.0019}_{-0.0019}$, consistent with the
world average. The effective charge and physical scale fits give
slightly smaller values, which are still consistent within the
errors.

\end{abstract}
\newpage

\renewcommand{\thepage}{\arabic{page}}
\pagestyle{headings} \setlength{\baselineskip}{16pt}
\setcounter{page}{2} \setlength{\parskip}{8pt}

\pagestyle{plain}
\section{Introduction}\vspace{-\parskip}
The measurements of the CCFR collaboration provide a precise
determination of the non-singlet deep inelastic scattering (DIS)
structure functions of neutrinos and anti-neutrinos on nucleons,
$xF_{3}(x,Q^2)$ \cite{r1}. Recently the NuTev collaboration have
also published measurements of $F_2(x,Q^2)$ and $F_3(x,Q^2)$
\cite{r1a}. The $Q^2$-dependence of moments of structure functions
can be predicted in perturbative QCD, and fits to the data can be
used to infer $\LMS$ (or equivalently $\alpha_s(M_Z)$). In doing
this, the principal difficulty is that there are upper and lower
limits on the experimentally accessible range of $x$ at low and high
$Q^2$, respectively. The moments are potentially sensitive to this
missing information, and this propagates into an additional level of
uncertainty in the resultant prediction of $\LMS$.

An approach which has been applied in the past is to use Bernstein
polynomials, which are peaked in a rather limited $x$-range, to
construct linear combinations of moments which are insensitive to
the missing $x$ regions \cite{r2,r3}. The analysis of Ref.~\cite{r2}
chose, as is customary, to work in the $\MSbar$ scheme and set both
renormalization and factorization scales to $Q$ (physical scale (PS)
choice). This was extended in Ref.~\cite{r4} to consider predictions
for $F_3$ obtained in the `complete renormalization group improved'
(CORGI) approach \cite{r5} in which all dependence on the
renormalization scale $\mu$ and the factorization scale $M$ is
eliminated by an all-orders resummation of RG-predictable scale
logarithms.  

The analyses of Refs.~\cite{r2,r3} and \cite{r4} used the then
state-of-the-art three-loop (NNLO) results for the anomalous
dimension and coefficient function, which were restricted to a
subset of odd moments $n=1,3,5,\ldots,13$ \cite{r6}. Recent progress
has yielded NNLO results for these quantities for {\it any} value of
$n$ \cite{r7,r8}. Consequently the set of Bernstein moments used in
the fits can now be greatly extended. The Bernstein polynomials
defined in Refs.~\cite{r2,r3} were linear combinations of {\it odd}
moments, but the new results of Refs.~\cite{r7,r8} mean that, by a
slight redefinition of the polynomials, even moments can now also be
studied.

In this paper we intend to perform such an extended analysis. We
shall fit the CCFR data \cite{r1} to PS and CORGI NNLO QCD
predictions, and will in addition compare with the predictions in
the closely-related method of effective charges approach (EC)
\cite{r9}. Target mass corrections and higher twist effects will
also be considered. We shall also compare our results to those
obtained using a fitting technique based on Jacobi Polynomials
\cite{r9A,r9B,r9C}.

The plan of the paper is to give a brief review of the factorization
and renormalization scheme dependence of structure function moments
in Section 2. In Section 3 we discuss the  CORGI and effective
charge approaches for leptoproduction moments. We take this
opportunity to correct an error in the expression for the NNLO CORGI
result for the scheme invariant $X_2$ derived in Ref.~\cite{r4}.
Section 4 will contain a  description of the Bernstein polynomial
averages to be employed in the fits. We shall show how to modify the
definition of the polynomials to accommodate both odd and even
moments. We then constrain the set of acceptable Bernstein moments
to be used in the fits by comparing how four different methods of
extrapolation (to obtain $xF_3$ on the full $x$-range) differ; this
enables us to define a `modelling error' to be combined with the
other sources of error in our analysis. In section 5 we give details
of the fitting procedure and in section 6 we present the results of
the fits to the PS, CORGI and EC predictions for the moments, and
consider how the fits change if target mass corrections and higher
twist corrections are included. Section 7 contains a discussion and
conclusions.

\section{Factorization and renormalization scheme dependence of the moments}\vspace{-\parskip}

\label{s:FRS}

The moments we are concerned with in this paper are those derived from
$F_{3}$ in (anti)neutrino-nucleon scattering. They are defined as:
\begin{eqnarray}
\MM_{3}^{\nu N}(n;Q^{2})&=&\int_{0}^{1}dx\;x^{n-1}F_{3}^{\nu
N}(x,Q^{2}). \label{eq:6moms}
\end{eqnarray}
These moments can be factorized in the following form,
\begin{eqnarray}
\MM_{3}^{\nu N}(n;Q^{2})&=&\langle
N|\MO_{n,\;\rmt{NS}}(M)|N\rangle\,\MC^{(3)}_{n}(Q,M,\mu,a(\mu)),
\label{eq:momfac}
\end{eqnarray}
where $\langle N|\MO_{n,NS}(M)|N\rangle$ is the non-singlet (NS)
operator matrix element of nucleon states and
$\MC^{(3)}_{n}(Q,M,\mu,a(\mu))$ is the coefficient function. Here
$a\equiv{\alpha}_{s}/\pi$ is the RG-improved coupling. The operator
matrix element is factorized at the scale $M$ into a
non-perturbative component and a perturbative expression, written in
terms of the coupling evaluated at the factorization scale,
$a=a(M)$. The factorization scale dependence is governed by the
anomalous dimension equation,
\begin{eqnarray}
M\frac{\partial}{\partial M}\langle
N|\MO_{n,\;\rmt{NS}}(M)|N\rangle&=&\langle
N|\MO_{n,\;\rmt{NS}}(M)|N\rangle\gamma_{n,\;\rmt{NS}}(a)
\label{eq:ad6a}.
\end{eqnarray}
Here ${\gamma}_{n,\;\rmt{NS}}(a)$ is the anomalous dimension of the
moment. It has the following perturbative expansion,
\begin{eqnarray}
{\gamma}_{n,\;\rmt{NS}}(a)&=&-d(n)a-d_{1}(n)a^{2}-d_{2}(n)a^{3}-d_{3}(n)a^{4}-\ldots,
\label{eq:ad6b}
\end{eqnarray}
where $d(n)$ is factorization scheme invariant, and the higher
coefficients serve to label the factorization scheme dependence. The
$M$-dependence of the coupling is governed by the beta-function
equation,
\begin{eqnarray}
M\frac{\partial a}{\partial M}&=&{\beta}(a)\;\equiv\;
-ba^{2}(1+ca+c_{2}a^{2}+c_{3}a^{3}+\ldots). \label{eq:6bfe}
\end{eqnarray}
Here $b=(33-2{N}_{f})/6$ and $c=(153-19N_f)/12b$ are renormalization
scheme (RS) invariant. The higher coefficients serve to label the RS
dependence. Together, these two equations determine the perturbative
behaviour of the operator matrix element. For the remainder of this
paper we simplify our notation by dropping the sub- and superscripts
`$\nu N$', `$n$', `$(3)$' and `NS', from the quantities in
Eqs.~(\ref{eq:momfac}) and (\ref{eq:ad6a}). Also, although the
coefficients $d_{i}(n)$ in Eq.~(\ref{eq:ad6b}) are $n$-dependent, we
also suppress this.

A solution to Eq.~(\ref{eq:ad6a}) can be obtained in the form,
\begin{eqnarray}
\langle\mathcal{O}(M)\rangle&=&A_{n}\exp\lb\{\int_{0}^{a(M)}\frac{\gamma(x)}{\beta(x)}dx-\int_{0}^{\infty}\frac{\gamma^{(1)}(x)}{\beta^{(2)}(x)}dx\rb\},
\label{eq:6operatorme}
 \end{eqnarray}
where $\gamma^{(i)}$ and $\beta^{(i)}$ denote the anomalous
dimension and beta-function equations truncated after $i$ terms.
There is a distinct parallel between the above equation and the
solution to the beta function equation. The second integral in
Eq.~(\ref{eq:6operatorme}) is an infinite constant. We are free to
choose any form we wish for this term, subject to the constraint
that it must have the same singularity structure as the first
integral. However, a particular choice for this constant corresponds
to a particular definition of $A_{n}$. Consequently, $A_{n}$ can be
likened to the dimensional transmutation parameter, $\Lambda$, in
that it defines the missing boundary condition in
Eq.~(\ref{eq:ad6a}). $A_{n}$  is actually a (set of)
non-perturbative constant(s), generated by the factorization
process, and they are factorization and renormalization scheme (FRS)
invariant. Their precise values cannot be calculated within
perturbation theory, and hence must be obtained by comparison with
experimental data.

The coefficient function $\MC(Q,M,\mu,a(\mu))$, depends on both the
renormalization {\it and} factorization scheme adopted, and it takes
the form of an expansion in powers of the coupling evaluated at the
renormalization scale,
\begin{eqnarray}
  \MC(Q,M,\mu,{\tilde{a}}(\mu))&=&1+r_{1}\tilde{a}+r_{2}\tilde{a}^{2}
+r_{3}\tilde{a}^{3}+\ldots,
\end{eqnarray}
where $\tilde{a}=a(M=\mu)$. Using the above equation together with
Eqs.~(\ref{eq:momfac}) and (\ref{eq:6operatorme}), the moments can
be written as \cite{r4,r10,r11},
\begin{eqnarray}
\MM(n;Q^{2})&=&A_{n}\lb(\frac{ca}{1+ca}\rb)^{d/b}\exp(\mathcal{I}(a))(1+r_{1}\tilde{a}+r_{2}\tilde{a}^{2}+r_{3}\tilde{a}^{3}+\cdots),
\label{eq:6protomoments}
\end{eqnarray}
where,
\begin{eqnarray}
\MI(a)&=&\int_{0}^{a}dx\frac{d_{1}+(d_{1}c+d_{2}-dc_{2})x+(d_{3}+cd_{2}-c_{3}d)x^{2}+\cdots)}{b(1+cx)(1+cx+c_{2}x^{2}+c_{3}x^{3}+\cdots)}.
\label{MI}
\end{eqnarray}
The explicit $M$ dependence of the coupling can be obtained by solving the
following transcendental equation \cite{r9a},
\begin{eqnarray}
\frac{1}{a}+c\ln\frac{ca}{1+ca}&=&b\ln\frac{M}{\tilde{\Lambda}}-b\int_{0}^{a}\left[\frac{1}{\beta(x)}-\frac{1}{\beta^{(2)}(x)}\right].
\label{eq:6trans}
\end{eqnarray}
Equation (\ref{eq:6protomoments}) serves as a prototypical
expression for the moments, from which CORGI, EC and PS predictions
can be derived.

The self-consistency of  perturbation theory means that the
perturbative coefficient $r_1$ has a dependence on $M$ and $d_1$.
The higher coefficients also have a dependence on the parameters
specifying the FRS, ${r_k} ({\mu},M,{c_2},c_3,{\ldots},{c_k}
;{d_1},{d_2},{\ldots},{d_k})$. The explicit form of this
FRS-dependence can be determined by demanding that on calculating
the moments up to O($a^k$) the partial derivative with respect to
each FRS parameter is O($a^{k+1}$) \cite{r10,r11} . The complete set
of partial derivatives required to derive the FRS-dependence of
$r_1$, $r_2$ and $r_3$ is, for the $\mu$-dependence
\begin{eqnarray}
\mu\frac{\partial r_{1}}{\partial\mu}\;=\;0, \qq \mu\frac{\partial
r_{2}}{\partial\mu}\;=\;r_{1}b, \qq\mu\frac{\partial
r_{3}}{\partial\mu}\;=\;2r_{2}b+br_{1}c.\label{eq:r3mu}
\end{eqnarray}
For the $M$-dependence we have
\begin{eqnarray}
&&M\frac{\partial r_{1}}{\partial M}\;=\;d,
\qq M\frac{\partial r_{2}}{\partial M}\;=\;dr_{1}-dL+d_{1}, \nn
\\[10pt]
&&M\frac{\partial r_{3}}{\partial
M}\;=\;d_{2}+d_{1}r_{1}+dr_{2}-dr_{1}L-2d_{1}L+dL^{2}-dcL,\label{eq:r3M}
\end{eqnarray}
where we have defined $L\equiv b{\ln}(M/\mu)$. For the
$c_2$-dependence we have
\begin{eqnarray}
&&\frac{\partial r_{1}}{\partial c_{2}}\;=\;0, \label{eq:r1c2}
\qq\frac{\partial r_{2}}{\partial c_{2}}\;=\;-\frac{d}{2b},
\label{eq:r2c2}
\nn \\[10pt]
&&\frac{\partial r_{3}}{\partial
c_{2}}\;=\;-\frac{r_{1}d}{2b}+\frac{Ld}{b}+\frac{cd}{3b}-\frac{2d_{1}}{3b}-r_{1}.\label{eq:r3c2}
\end{eqnarray}
For the $c_3$-dependence
\begin{eqnarray}
\frac{\partial r_{1}}{\partial c_{3}}\;=\;0, \qq \frac{\partial
r_{2}}{\partial c_{3}}\;=\;0, \qq\frac{\partial r_{3}}{\partial
c_{3}}\;=\;-\frac{d}{6b}.\label{eq:r3c3}
\end{eqnarray}
For the $d_1$-dependence
\begin{eqnarray}
&&\frac{\partial r_{1}}{\partial d_{1}}\;=\;-\frac{1}{b}, \qq
\frac{\partial r_{2}}{\partial
d_{1}}\;=\;\frac{c}{2b}-\frac{r_{1}}{b}+\frac{L}{b},\nn\\[10pt]
&&\frac{\partial r_{3}}{\partial
d_{1}}\;=\;\frac{cr_{1}}{2b}-\frac{c^{2}}{3b}-\frac{r_{2}}{b}+\frac{c_{2}}{3b}-\frac{L^{2}}{b}+\frac{Lr_{1}}{b}.\label{eq:r3d1}
\end{eqnarray}
For the $d_2$-dependence we have
\begin{eqnarray}
\frac{\partial r_{1}}{\partial d_{2}}\;=\;0,\qq \frac{\partial
r_{2}}{\partial d_{2}}\;=\;-\frac{1}{2b},\qq \frac{\partial
r_{3}}{\partial
d_{2}}\;=\;\frac{c}{3b}+\frac{1}{2b}(2L-r_{1}).\label{eq:r3d2}
\end{eqnarray}
Finally, for the $d_3$-dependence we have
\begin{eqnarray}
\frac{\partial{r}_{1}}{\partial{d_3}}\;=\;0, \qq
\frac{\partial{r_2}}{\partial{d_3}}\;=\;0, \qq \frac{\partial
r_{3}}{\partial d_{3}}\;=\;-\frac{1}{3b}.\label{17}
\end{eqnarray}
These results may now be integrated to obtain $r_1$, $r_2$ and $r_3$. For $r_1$ we obtain
\begin{eqnarray}
r_{1}&=&\frac{d}{b}\tau_{M}-\frac{d_{1}}{b}-X_{0}(Q),
\label{eq:corgi1}
\end{eqnarray}
where $\tau_{M}=b\ln\lb(M/\tL\rb)$. $X_{0}(Q)$ is an FRS invariant
quantity, generated as a constant of integration. One can define an
FRS invariant, non-universal scale parameter, $\Lambda_{\MM}$, via
the FRS invariant $X_{0}(Q)$. Thus,
\begin{eqnarray}
\frac{d}{b}\tau_{M}-\frac{d_{1}}{b}-r_{1}&=&X_{0}(Q)\;\equiv\;
d\;\ln\lb(\frac{Q}{\Lambda_{\MM}}\rb). \label{eq:X0Lm}
\end{eqnarray}
For $r_2$ we obtain
\begin{eqnarray}
r_{2}&=&\Bigg{(}\frac{1}{2}-\frac{b}{2d}\Bigg{)}r_{1}^{2}+\frac{b}{d}r_{1}\tilde{r}_{1}+\frac{d_{1}}{d}r_{1}-\frac{dc_{2}}{2b}+\frac{d_{1}^{2}}{2bd}
+\frac{cd_{1}}{2b}-\frac{d_{2}}{2b}+X_{2}, \label{eq:corgi2}
\end{eqnarray}
where we have defined,
\begin{eqnarray}
\tilde{r}_{1}&\equiv&r_{1}(M=\mu) \nn\\
&=&\frac{d}{b}\tau_{\mu}-\frac{d_{1}}{b}-X_{0}(Q).
\label{eq:rtau}
\end{eqnarray}
Here $X_2$ is another FRS-invariant constant of integration.
Crucially $X_2$ and higher invariants are independent of $Q$. Hence,
the complete $Q$-dependence of the observable is generated by
$X_0(Q)$. Similarly, for $r_3$ we obtain
\begin{eqnarray}
r_{3}&=&\frac{c_{2}dc}{3b}
-\frac{c_{3}d}{6b}
+\frac{cd_{2}}{3b}
-\frac{d_3}{3b}
-\frac{2d_{1}^{3}}{3bd^{2}}
-\frac{cd_{1}^{2}}{2bd}
-\frac{c^{2}d_{1}}{3b}
+\frac{d_{1}d_{2}}{bd}
+\frac{c_{2}d_{1}}{3b}\nn\\
&-&\frac{2d_{1}^{2}r_{1}}{d^{2}}
-\frac{d_{1}r_{1}^{2}}{d}
+\frac{b^{2}r_{1}^{3}}{3d^{2}}-
\frac{r_{1}^{3}}{3}
-\frac{bcr_{1}^{2}}{2d}
+\frac{d_{2}r_{1}}{d}\nn\\
&-&\frac{2bd_{1}r_{1}\rt_{1}}{d^{2}}
-\frac{b^{2}r_{1}\rt_{1}^{2}}{d^{2}}
 - \frac{br_{1}^{2}\rt_{1}}{d}
 +\frac{bcr_{1}\rt_{1}}{d}\nn\\
&+&\frac{2b\rt_{1}r_{2}}{d}
+\frac{2d_{1}r_{2}}{d}
+ r_{1}r_{2}
+X_{3}.
\label{eq:r3FRS}
\end{eqnarray}
Again $X_3$ is a $Q$-independent FRS-invariant constant of
integration. Using Eq.~(\ref{eq:corgi2}) $X_3$ can be written in
terms of $r_1$, ${\tilde{r}}_{1}$ and the other FRS parameters. This
also holds for the higher invariants. The results of
Eqs.~(\ref{eq:r3mu}) - (\ref{17}) and of Eqs.~(\ref{eq:corgi2}) and
(\ref{eq:r3FRS}) replace, respectively, Eqs.~(15) and (18) of
Ref.~\cite{r4} which contain several errors. The invariant $X_2$ can
be obtained from NNLO results for the anomalous dimension
coefficients and coefficient function in {\it any} FRS. For instance
if we make the customary choice of $\MSbar$ with $M=\mu=Q$ then
$r_1={\tilde{r_1}}$ and we obtain
\begin{eqnarray}
X_{2}&=&\lb.r_{2}-\Bigg{(}\frac{1}{2}+\frac{b}{2d}\Bigg{)}r_{1}^{2}-\frac{d_{1}}{d}r_{1}+\frac{dc_{2}}{2b}-\frac{d_{1}^{2}}{2bd}
-\frac{cd_{1}}{2b}+\frac{d_{2}}{2b}\rb|_{\MSbar}. \label{eq:6:X2MS}
\end{eqnarray}

In summary, through Eqs.~(\ref{eq:corgi1}), (\ref{eq:corgi2}) and
(\ref{eq:r3FRS}), we have determined the explicit FRS dependence of
the coefficients $r_{i}$. In doing so, we have generated a set of
FRS invariant quantities $X_{i}$, the importance of which will
become clear when we come to consider the CORGI form of the moments
in the following section.

\section{PS, CORGI and EC predictions}\vspace{-\parskip}

The standard physical scale approach is to set $M=\mu=Q$ and adopt
$\MSbar$ subtraction. Setting $M=\mu$ implies that $a=\tilde{a}$,
and hence the moments have the form,
\begin{eqnarray}
\MM(n;Q^{2})&=&A_{n}\lb(\frac{ca}{1+ca}\rb)^{d/b}\lb(1+R_{1}a+R_{2}a^{2}+\ldots\rb).
\label{eq:PSM}
\end{eqnarray}
The coefficients $R_{i}$ can be determined by expanding
Eq.~(\ref{eq:6protomoments}) in powers of $a$,
\begin{eqnarray}
R_{1}&=&r_{1}+\frac{d_{1}}{b}\label{eq:6:R1}\\
R_{2}&=&r_{2}+\frac{d_{1}^{2}}{2b^{2}}-\frac{cd_{1}}{2b}+\frac{r_{1}d_{1}}{b}-\frac{dc_{2}}{2b}+\frac{d_{2}}{2b},
\label{eq:6:R2}
\end{eqnarray}
and the coupling in this expression is the three-loop $\MSbar$
coupling with $\mu=Q$.

The CORGI idea (see Ref.~\cite{r5} for a detailed discussion) is
that all RG-predictable information about higher perturbative
coefficients, available at a given fixed-order of calculation should
be resummed to all-orders. Given an NLO calculation for instance one
knows $X_0(Q)$ but not $X_2,\;X_3$ or higher FRS-invariants. One
should therefore resum to all-orders all the terms {\it not}
involving these unknown invariants. As discussed in Section 2 these
terms are multinomials in
$r_1,{\tilde{r}_{1}},c_2,\ldots,c_i,d_1,d_2,\ldots,d_i,\ldots$.

Crucially this all-orders sum must be FRS-invariant, as separately
must be the subset of terms involving $X_2$, $X_3\ldots$. One may
exploit this invariance and choose to use the FRS where all the FRS
parameters are zero, ${r_1}={\tilde{r}_{1}}=
c_2=\ldots={c}_{i}=\ldots=d_1=d_2=\ldots={d}_{i}=\ldots=0$.

Setting $r_1={\tilde{r}_{1}}=0$ means that $\mu=M$, setting  $d_1=0$
then implies that (from Eq.~(\ref{eq:X0Lm})) ${\tau}_{M}=
b{\ln}(Q/{\Lambda}_{\MM})$. Also, with $c_{i}={d}_{i}=0$ the
integral $\MI(a)$ of Eq.~(\ref{MI}) vanishes, and one finally
obtains the CORGI form of the moments,
\begin{eqnarray}
\MM(n;Q^{2})&=&A_{n}\lb(\frac{ca_0}{1+ca_0}\rb)^{d/b}\lb(1+X_{2}a_{0}^{2}+X_{3}a_{0}^{3}+\ldots\rb).
\label{eq:CORGImoms}
\end{eqnarray}
Here the CORGI coupling $a_0$ is the coupling in a 't Hooft scheme
\cite{r11a}, in which $c_i=0$ $(i>1)$. This can be written in terms
of the Lambert $W$ function defined implicitly by $W(z){e}^{W(z)}=z$
\cite{r12,r13},
\begin{eqnarray}
a_{0}(Q)&=&\frac{-1}{c\lb[1+W_{-1}(z(Q))\rb]}, \label{eq:6tHooft}
\end{eqnarray}
with,
\begin{eqnarray}
z(Q)&=&-\frac{1}{{\rm e}}\lb(\frac{Q}{{\Lambda}_{\MM}}\rb)^{-b/c}.
\end{eqnarray}
$W_{-1}$ refers to the branch of the Lambert $W$ function required
for asymptotic freedom, the nomenclature being that of
Ref.~\cite{r14}. $\Lambda_{\MM}$ is the invariant scale connected
with the $X_0(Q)$ FRS-invariant, defined in Eq.~(\ref{eq:X0Lm}).
Since it is an FRS-invariant it can be evaluated in any FRS.
Choosing the $\MSbar$ scheme with $M=\mu=Q$ one finds
\begin{eqnarray}
\Lambda_{\MM}&=&\LMS{\left(\frac{2c}{b}\right)}^{-c/b}\exp\lb\{\frac{d_{1}}{db}+\frac{r_{1}}{d}\rb\},
\label{eq:6:LMM}
\end{eqnarray}
with $r_{1}$ and $d_{1}$ calculated in \MSbar~with $M=\mu=Q$. The
factor of ${\left(\frac{2c}{b}\right)}^{-c/b}$ converts to the
standard convention for integrating the beta-function equation and
defining $\LMS$ (see Ref.~\cite{r14a} for further details). The
second factor on the r.h.s.~of Eq.~(\ref{eq:CORGImoms}) resums to
all-orders the RG-predictable terms not involving $X_2,X_3,\ldots$.
The $a_0^2$ term sums to all-orders the RG-predictable terms
involving $X_2$, but not $X_3,X_4,\ldots$, etc.

The CORGI result corresponds to  an $\MSbar$ scale choice
$M=\mu=xQ$, with
\begin{eqnarray}
x&=&x_{\rmt{CORGI}}\;\equiv\;\exp\lb\{-\frac{d_{1}}{db}-\frac{r_{1}}{d}\rb\}.
\end{eqnarray}
To illustrate how the CORGI scale differs from the PS choice ($x=1$)
we plot in table 1 the $x_{\rmt{CORGI}}$, for the first $20$ moments
$n=1,2,\ldots,20$. We also tabulate the corresponding $X_2(n)$ NNLO
CORGI invariants obtained from Eq.~(\ref{eq:6:X2MS}). The anomalous
dimension coefficients up to NNLO are taken from Refs.~\cite{r7,r8},
and the coefficient function from Ref.~\cite{r6}. We assume $N_f=5$
active quark flavours.
\begin{table}[t]
\begin{center}
\begin{tabular}{|r|rr|r|}
\hline $n$&$x_{\rmt{CORGI}}$&&$X_2(n)$
\\\hline
1&0.4688  &&-1\\
2&0.7156&&-3.01\\
3&0.5074&&-3.28\\
4&0.42966&&-3.485\\
5&0.3838&&-3.627\\
6&0.3530&&-3.713\\
7&0.3300&&-3.766\\
8&0.312 &&-3.792\\
9&0.2974&&-3.8\\
10&0.2853&&-3.793\\
\hline
\end{tabular}
\qq\qq\qq\begin{tabular}{|r|rr|r|} \hline
$n$&$x_{\rmt{CORGI}}$&&$X_2(n)$
\\\hline
11&0.2749&&-3.776\\
12&0.2659&&-3.749\\
13&0.2580&&-3.716\\
14&0.2510&&-3.677\\
15&0.2447&&-3.633\\
16&0.2390&&-3.586\\
17&0.2338&&-3.536\\
18&0.2291&&-3.483\\
19&0.2248&&-3.428\\
20&0.2207&&-3.372\\
\hline
\end{tabular}
\end{center}
\caption{The numerical values of $x_{\rmt{CORGI}}$ and the NNLO
CORGI invariants ${X}_{2}(n)$ for the $n=1-20$ moments of $F_3$.}
\label{xcorgi}
\end{table}
We see from table \ref{xcorgi} that as $n$ increases the CORGI scale
decreases, becoming significantly less than $x=1$ (PS). The $X_2(n)$
invariants are seen to be moderate in size.

Finally, we discuss the third variant of perturbative QCD which we
shall consider. By setting $M=\mu$ and rearranging, we can recast
the perturbation series for $\MM(n;Q^{2})$ of
Eq.~(\ref{eq:6protomoments}) in the form
\begin{eqnarray}
\MM(n;Q^{2})&=&A_{n}\lb(c\Rt(a)\rb)^{d/b}, \label{eq:ECmoms}
\end{eqnarray}
where
\begin{eqnarray}
\Rt(a)&=&a+\Rt_{1}a^{2}+\Rt_{2}a^{3},
\end{eqnarray}
is an effective charge \cite{r9},
and the coefficients $\Rt_{i}$ have the form,
\begin{eqnarray}
\Rt_{1}&=&\frac{bcR_{1}}{d}-c^{2}\nn\\
&=&\frac{bcr_{1}}{d}+\frac{d_{1}c}{d}-c^{2},\\
\Rt_{2}&=&\frac{bcR_{2}}{d}+\frac{b^{2}R^{2}_{1}c}{2d^{2}}-\frac{bR_{1}^{2}c}{2d}-\frac{bR_{1}c^{2}}{d}+c^{3}\nn\\
&=&\frac{bcr_{2}}{d}+\frac{bd_{1}r_{1}c}{d^{2}}+\frac{d_{2}c}{2d}-\frac{c_{2}c}{2}\nn\\
&+&\frac{b^{2}r^{2}_{1}c}{2d^{2}}
-\frac{br_{1}^{2}c}{2d}
+\frac{d_{1}^{2}c}{2d^{2}}
-\frac{br_{1}c^{2}}{d}-\frac{3d_{1}c^{2}}{2d}+c^{3}.
\end{eqnarray}
Here $R_1$ and $R_2$ are the coefficients defined in
Eqs.~(\ref{eq:6:R1}) and (\ref{eq:6:R2}). Rather than integrating
the effective charge beta-function we shall instead apply CORGI to
the effective charge, avoiding the need to numerically solve a
transcendental equation which would make the fitting to data
considerably more complicated. At NLO the CORGI and EC results agree
exactly. We have the CORGI result
\begin{eqnarray}
\MM(n;Q^{2})&=&A_{n}c^{d/b}\lb(a_{0}+\tX_{2}a_{0}^{3}+\tX_{3}a_{0}^{4}+\ldots\rb)^{d/b}.
\label{eq:6ECH}
\end{eqnarray}
In this case, the $\tX_{i}$ coefficients are the CORGI invariants
corresponding to single scale RS-dependence \cite{r5}. They have the
form,
\begin{eqnarray}
\tX_{2}&=&\Rt_{2}-\Rt_{1}^{2}-c\Rt_{1}+c_{2},
\label{eq:x2Rt}\\
\tX_{3}&=&\Rt_{3}-3\Rt_{1}\Rt_{2}+2\Rt_{1}^{3}+\frac{c\Rt_{1}^{2}}{2}-\Rt_{1}c_{2}+\frac{1}{2}c_{3}.\label{eq:x3Rt}
\end{eqnarray}
The CORGI coupling $a_0$ is that of Eq.~(\ref{eq:6tHooft}) but with
the scale $\Lambda_{\MM}$ now defined by,
\begin{eqnarray}
\Lambda_{\MM}^{\rmt{EC}}&=&\lb(\frac{2c}{b}\rb)^{-c/b}\exp\lb(\frac{\Rt_{1}}{b}\rb)\LMS.
\label{eq:6echL}
\end{eqnarray}
We shall refer to this variant of perturbation theory as `EC' for
simplicity, even though as noted above it is really CORGI applied to
a single-scale effective charge.

We note that we can streamline the calculation of the FRS-invariants
$X_i$ by using the single-scale effective charge. If we set $M=\mu$
in Eq.~(\ref{eq:6protomoments}), then the moments reduce to a
single-scale problem \cite{r5}. We can then rearrange the resultant
expression in terms of an effective charge, $\Rh(a)$,
\begin{eqnarray}
\MM(n;Q^{2})&=&A_{n}\lb(\frac{c\Rh(a)}{1+c\Rh(a)}\rb)^{d/b}.\label{Malt}
\end{eqnarray}
$\Rh(a)$ has the form,
\begin{eqnarray}
\Rh(a)&=&a+\Rh_{1}a^{2}+\Rh_{2}a^{3}+\Rh_{3}a^{4}+\ldots.
\end{eqnarray}
The coefficients $\Rh_{i}$ can be determined by expanding
Eqs.~(\ref{eq:PSM}) and (\ref{Malt}) in powers of $a$ and then
equating coefficients. They are found to be,
\begin{eqnarray}
\Rh_{1}&=&\frac{b}{d}R_{1},\label{eq:6:Rh1}\\
\Rh_{2}&=&\frac{b}{d}\lb(R_{2}+cR_{1}-\frac{R_{1}^{2}}{2}+\frac{bR_{1}^{2}}{2d}\rb),
\label{eq:6:Rh2}
\end{eqnarray}
where $R_{1}$ and $R_{2}$ are given by Eqs.~(\ref{eq:6:R1}) and
(\ref{eq:6:R2}). If we then CORGI-ize this effective charge, we have
a new set of FRS
invariants \cite{r5},
\begin{eqnarray}
\Xh_{0}&=&b\ln\frac{M}{\tL}-\Rh_{1},
\label{eq:6:xh0}\\
\Xh_{2}&=&\Rh_{2}-\Rh_{1}^{2}-c\Rh_{1}+c_{2},
\label{eq:6:xh2}
\end{eqnarray}
and the moments become,
\begin{eqnarray}
\MM(n;Q^{2})&=&A_{n}\lb(\frac{c\lb(a_{0}+\Xh_{2}
a_{0}^{3}\rb)}{1+c\lb(a_{0}+\Xh_{2} a_{0}^{3}\rb)}\rb)^{d/b}.
\end{eqnarray}
Expanding this into a form which we can compare with Eq.~(\ref{eq:CORGImoms}), gives,
\begin{eqnarray}
\MM(n;Q^{2})&=&A_{n}\lb(\frac{ca_{0}}{1+ca_{0}}\rb)^{d/b}\lb(1+\frac{d}{b}\Xh_{2}\,a_{0}^{2}+\ldots\rb).
\label{eq:6:momsX2alt}
\end{eqnarray}
Isolating the $\MO(a_{0}^{2})$ in the RHS bracket of the above
equation, and then using Eqs.~(\ref{eq:6:xh2}), (\ref{eq:6:Rh1}),
(\ref{eq:6:Rh2}), (\ref{eq:6:R1}) and (\ref{eq:6:R2}), gives,
\begin{eqnarray}
\frac{d}{b}\Xh_{2}&=& \frac{d}{b}\lb(\Rh_{2}-\Rh_{1}^{2}-c\Rh_{1}+c_{2}\rb)
\label{eq:6:X2altRh}\\
&=&r_{2}-\Bigg{(}\frac{1}{2}+\frac{b}{2d}\Bigg{)}r_{1}^{2}-\frac{d_{1}}{d}r_{1}+\frac{dc_{2}}{2b}-\frac{d_{1}^{2}}{2bd}
-\frac{cd_{1}}{2b}+\frac{d_{2}}{2b}.
\end{eqnarray}
So we see that the coefficient of the $\MO(a^{2})$ term in
Eq.~(\ref{eq:6:momsX2alt}) is the FRS invariant $X_{2}$, of
Eq.~(\ref{eq:6:X2MS}) with $\mu=M$ ($r_{1}=\rt_{1}$). Isolating the
$a^{3}$ term will yield $X_{3}$, and so on for higher $X_{i}$.

The coupling in Eq.~(\ref{eq:6:momsX2alt}) is the 't Hooft coupling
of Eq.~(\ref{eq:6tHooft}), but with the scale parameter determined
by Eq.~(\ref{eq:6:xh0}). Evaluating Eq.~(\ref{eq:6:xh0}) in
\MSbar~with $M=\mu=Q$ and using the standard definition of the
single-scale RS invariant $\Xh_{0}$ \cite{r9,r9a} gives,
\begin{eqnarray}
\Xh_{0}(Q)&\equiv&b\ln\frac{Q}{{\Lambda}_{\MM}}\\
&=&b\ln\frac{Q}{\tL}-\Rh_{1}.
\end{eqnarray}
Comparison with Eq.~(\ref{eq:X0Lm}) then reveals that
$\Xh_{0}(Q)=(b/d)X_0(Q)$. Using the same procedure we can obtain
expressions for $X_3$ and higher CORGI invariants.

\subsubsection*{Non-perturbative effects}\vspace{-\parskip}

The three variants of NNLO perturbative QCD, PS, CORGI, and EC, can
all be computed given $\MSbar$ anomalous dimension coefficients up
to NNLO \cite{r7,r8}, and the coefficient function \cite{r6}.
However, these perturbative predictions will be subject to
non-perturbative corrections in the form of $\MO\lb(1/Q^{2}\rb)$
terms. The two principal sources of these terms are: higher twist
terms and effects due to the mass of the target hadron.

The perturbative form of the moments is derived under the assumption
that the mass of the target hadron is zero (in the limit
$Q^{2}\rightarrow\infty$). At intermediate and low $Q^{2}$ this
assumption will begin to break down and the moments will be subject
to potentially significant power corrections, of order
$\MO\lb(m_{N}^{2}/Q^{2}\rb)$, where $m_N$ is the mass of the
nucleon. These are known as target mass corrections (TMCs) and when
included, the $F_{3}$ moments have the form \cite{r15,r15a},
\begin{eqnarray}
\MM^{\rmt{TMC}}(n;Q^{2})&=&\MM(n;Q^{2})+\frac{n(n+1)}{n+2}\frac{m_{N}^{2}}{Q^{2}}\MM(n+2;Q^{2})+\MO\lb(\frac{m_{N}^{4}}{Q^{4}}\rb).\label{eq:6:TMC}
\end{eqnarray}

The moments will also be subject to corrections from sub-leading
twist contributions to the OPE. These effects are poorly understood
and hence we only estimate them; this is done by means of an unknown
parameter, $A_{\rmt{HT}}$. The estimate has the form \cite{r2},
\begin{eqnarray}
\MM^{\rmt{HT}}(n;Q^{2})&=&n\lb(A^{\rmt{HT}}\frac{\LMS^{2}}{Q^{2}}\rb)\MM(n;Q^{2}),
\end{eqnarray}
and the value of $A^{\rmt{HT}}$ is obtained by fitting to data. Due
to the poorly understood nature of these effects, we do not include
the above term in the full analysis. Rather, we perform the analysis
with and without this term included, and take the difference in the
results as an estimate of the error associated with our ignorance of
the true nature of these effects.

The bottom quark mass threshold is within the range of $Q^{2}$
spanned by the available data for $F_{3}$. It is therefore necessary
to evolve the expressions for the moments over this threshold, and
in order to do this we use the formalism of Ref.~\cite{r16a}. We use
massless QCD with 4 quarks for $Q^{2}\leq m_{b}^{2}$ and massless
QCD with 5 quarks for $Q^{2}>m_{b}^{2}$. Here $m_b$ is the pole mass
of the $b$-quark with ${m}_{b}=4.85\pm{0.15}$ MeV \cite{r16}. From
the decoupling theorem, one finds the following relation between the
coupling above and below a quark threshold (denoted by
$a_{f+1}(Q^{2})$ and $a_{f}(Q^{2})$ respectively) \cite{r16a}
\begin{eqnarray}
a_{f}(m_{b}^{2})&=&a_{f+1}(m_{b}^{2})+\frac{11}{72}\lb(a_{f+1}(m_{b}^{2})\rb)^{3}.
\label{eq:5:decoup}
\end{eqnarray}
In practice, this matching is implemented by adopting different
values of the scale parameter in different $N_{f}$ regions. This is
governed by the following equations \cite{r16a},
\begin{eqnarray}
{\Lambda}_{N_{f}+1}^{2}&=&{\Lambda}_{N_{f}}^{2}\left(\frac{m_{N_{f}+1}^{2}}{{\Lambda}_{N_{f}}^{2}}\right)^{1-\frac{b^{N_{f}}}{b^{N_{f}+1}}}\times\exp\left(\frac{\delta_{\rmt{NLO}}+\delta_{\rmt{NNLO}}}{2b^{N_{f}+1}}\right),
\label{eq:matching}
\end{eqnarray}
where $\delta_{\rmt{NLO}}$ and $\delta_{\rmt{NNLO}}$ are given by
\begin{eqnarray}
 \delta_{\rmt{NLO}}&=&4(c^{N_{f}+1}-c^{N_{f}})\ln
L_{m}-4c^{N_{f}+1}\ln\frac{b^{N_{f}+1}}{b^{N_{f}}},\\
\delta_{\rmt{NNLO}}&=&\frac{8}{b^{N_{f}}L_{m}}\left(\left(c^{N_{f}+1}-c^{n_f}\right)c^{N_{f}}\ln
  L_{m}+\left(c^{N_{f}+1}\right)^{2}-\left(c^{N_{f}}\right)^{2}\nn\right.\\
&+&\left.c_{2}^{N_{f}}-c_{2}^{N_{f}+1}+\frac{7}{384}\right).\label{eq:5:mat3}
\end{eqnarray}
Here, ${\Lambda}_{N_{f}}$ is the scale parameter in the region where
$N_{f}$ quarks are active, $m_{N_{f}}$ is the pole mass of the $f$
quark, $b^{N_{f}}$, $c^{N_{f}}$ and $c_{2}^{N_{f}}$ are simply $b$,
$c$ and $c_{2}$ evaluated for $N_{f}$ quark flavours and we have
defined $L_{m}\equiv\ln\lb(m_{\nf+1}^{2}/{\Lambda}_{\nf}^{2}\rb)$.
Furthermore, we also demand continuity of the moments at the
threshold i.e.
\begin{eqnarray}
\lb.\MM(n;m_{b}^{2})\rb|_{\nf=4}&=&\lb.\MM(n;m_{b}^{2})\rb|_{\nf=5}.
\end{eqnarray}
As a consequence of this, the parameters $A_{n}$ also have different values in the
 $\nf=4$ and $\nf=5$ regions and their values are related by,
\begin{eqnarray}
A_{n}^{(5)}&=&\lb(\lb.\frac{A_{n}}{\MM(n;m_{b}^{2})}\rb|_{\nf=5}\,\rb)\lb(\lb.\frac{\MM(n;m_{b}^{2})}{A_{n}}\rb|_{\nf=4}\,\rb)A_{n}^{\nf=4}.
\end{eqnarray}

\section{The method of Bernstein averages}\vspace{-\parskip}
\begin{figure}
\begin{center}
\begin{tabular}{c c c}
\hspace{-.4cm}\includegraphics[angle=270,width=0.36\textwidth]{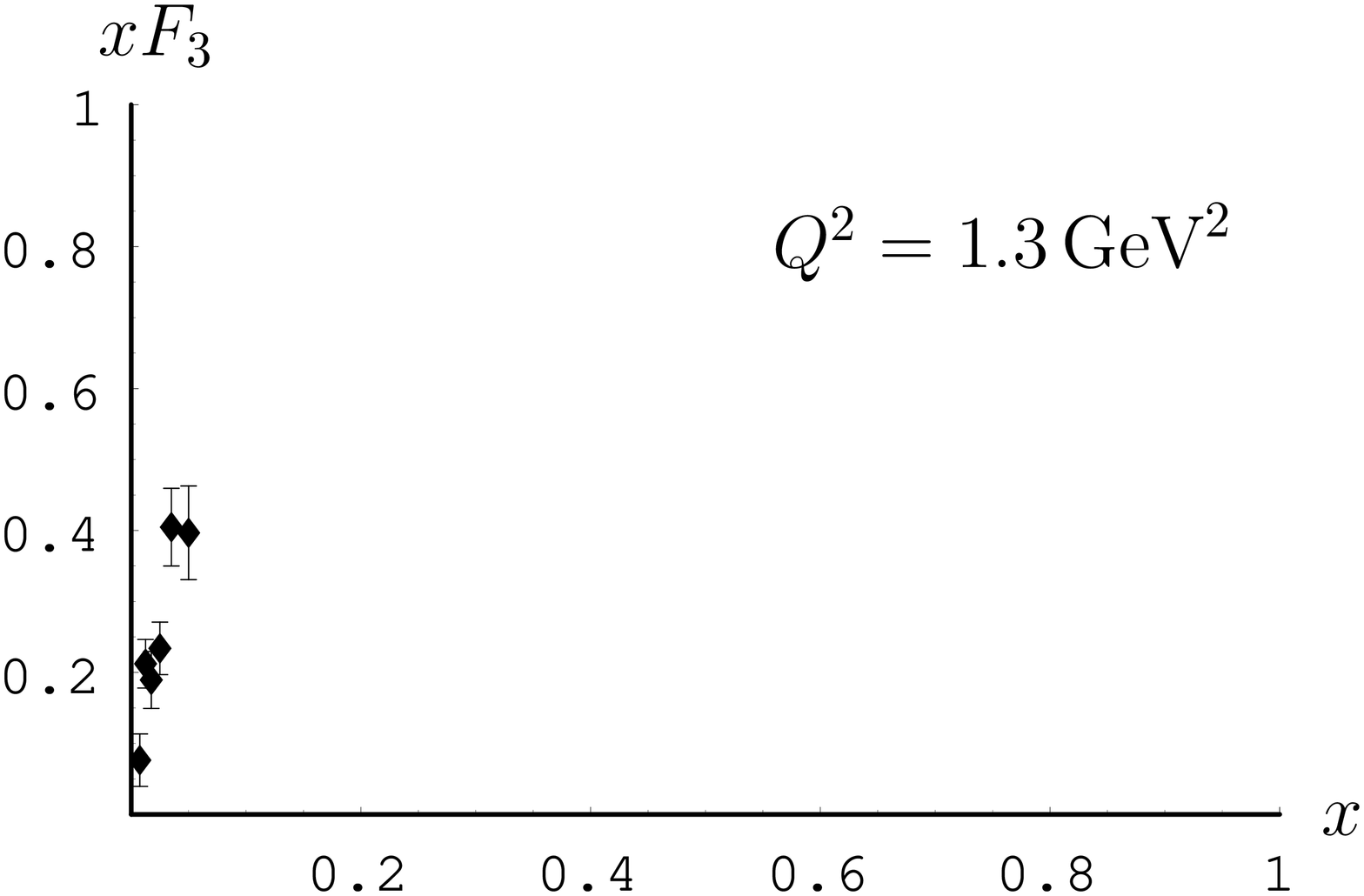}&\hspace{-1cm}
\includegraphics[angle=270,width=0.36\textwidth]{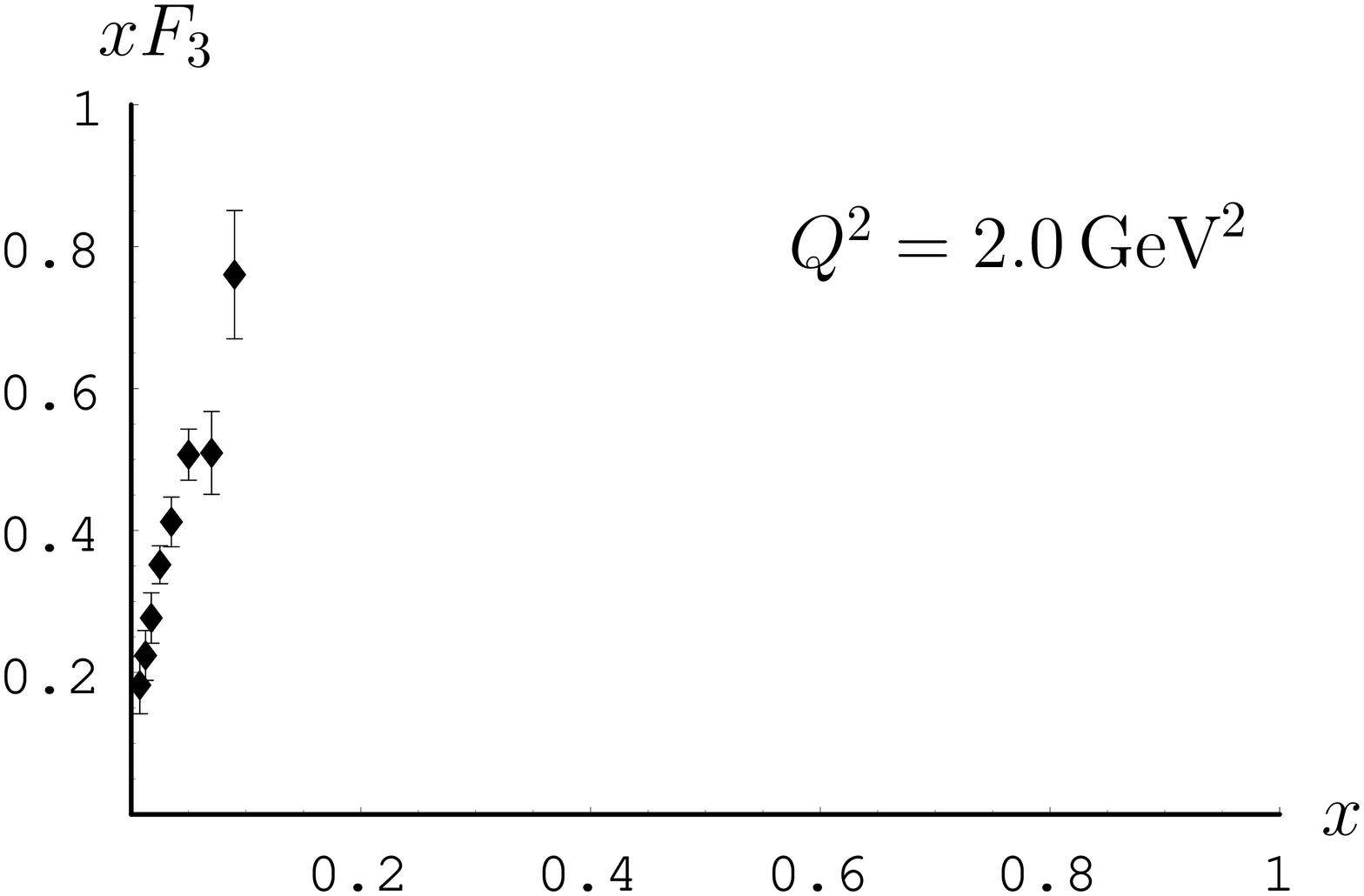}&\hspace{-1cm}
\includegraphics[angle=270,width=0.36\textwidth]{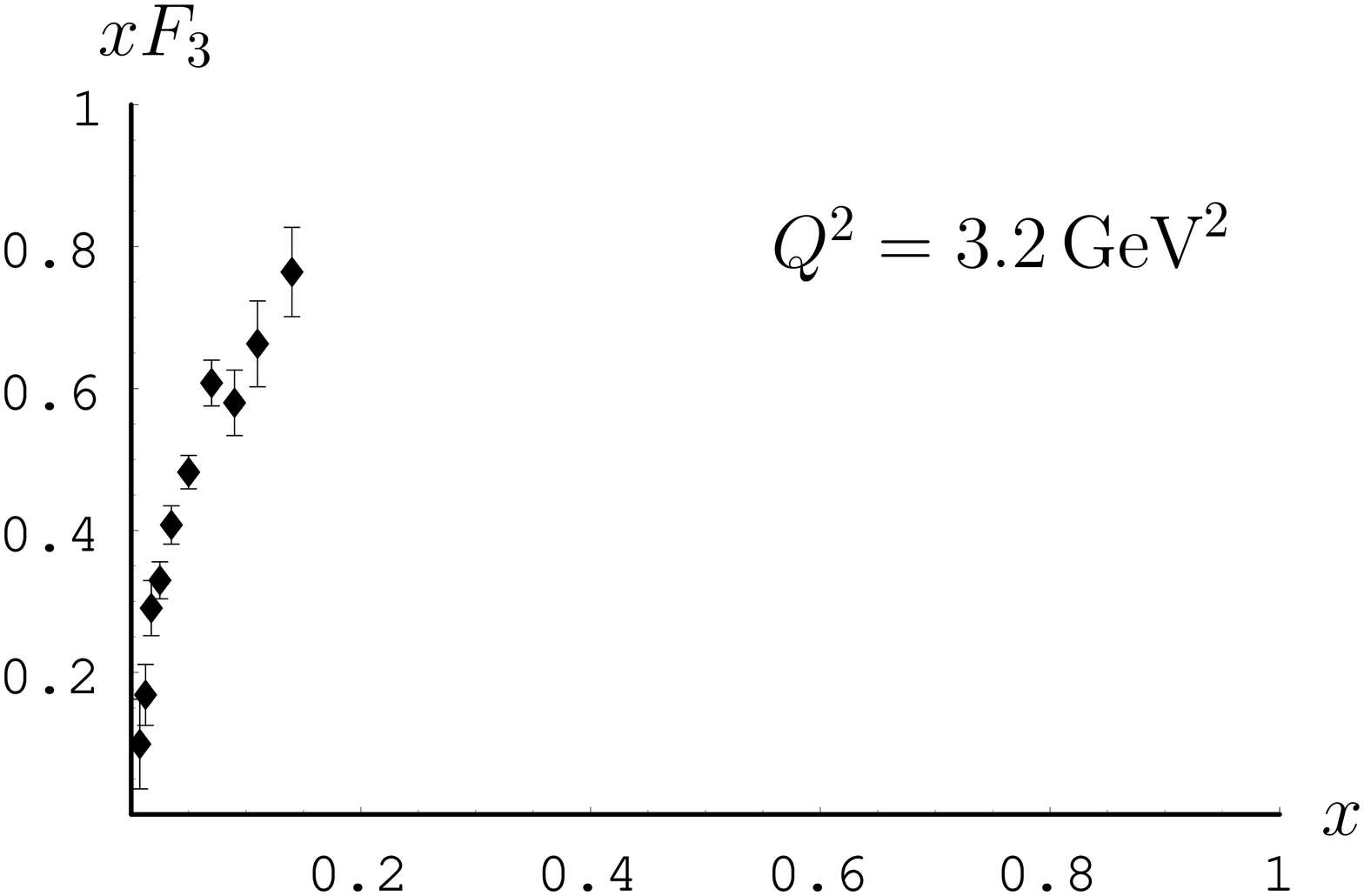}\\
\hspace{-.4cm}\includegraphics[angle=270,width=0.36\textwidth]{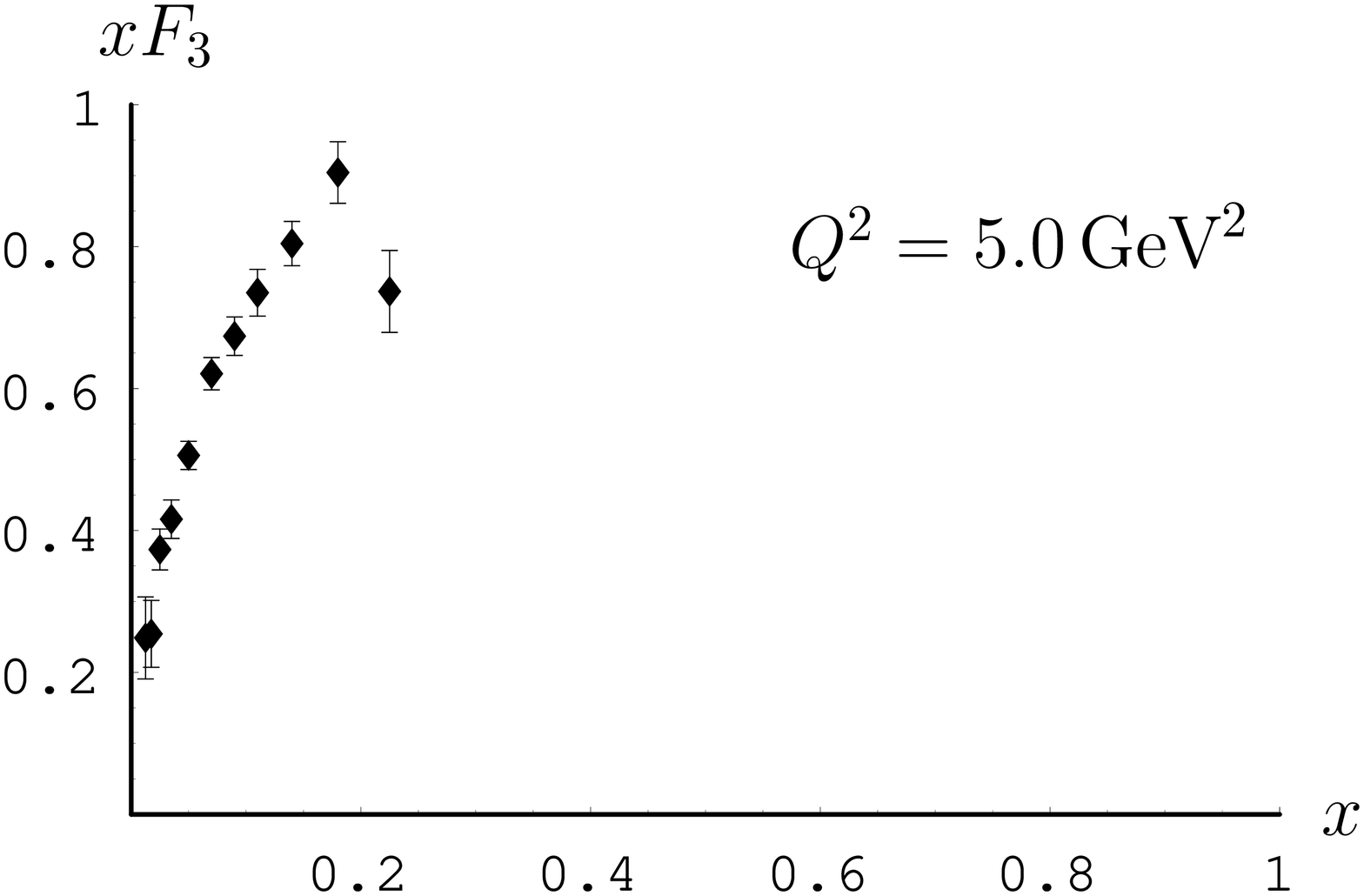}&\hspace{-1cm}
\includegraphics[angle=270,width=0.36\textwidth]{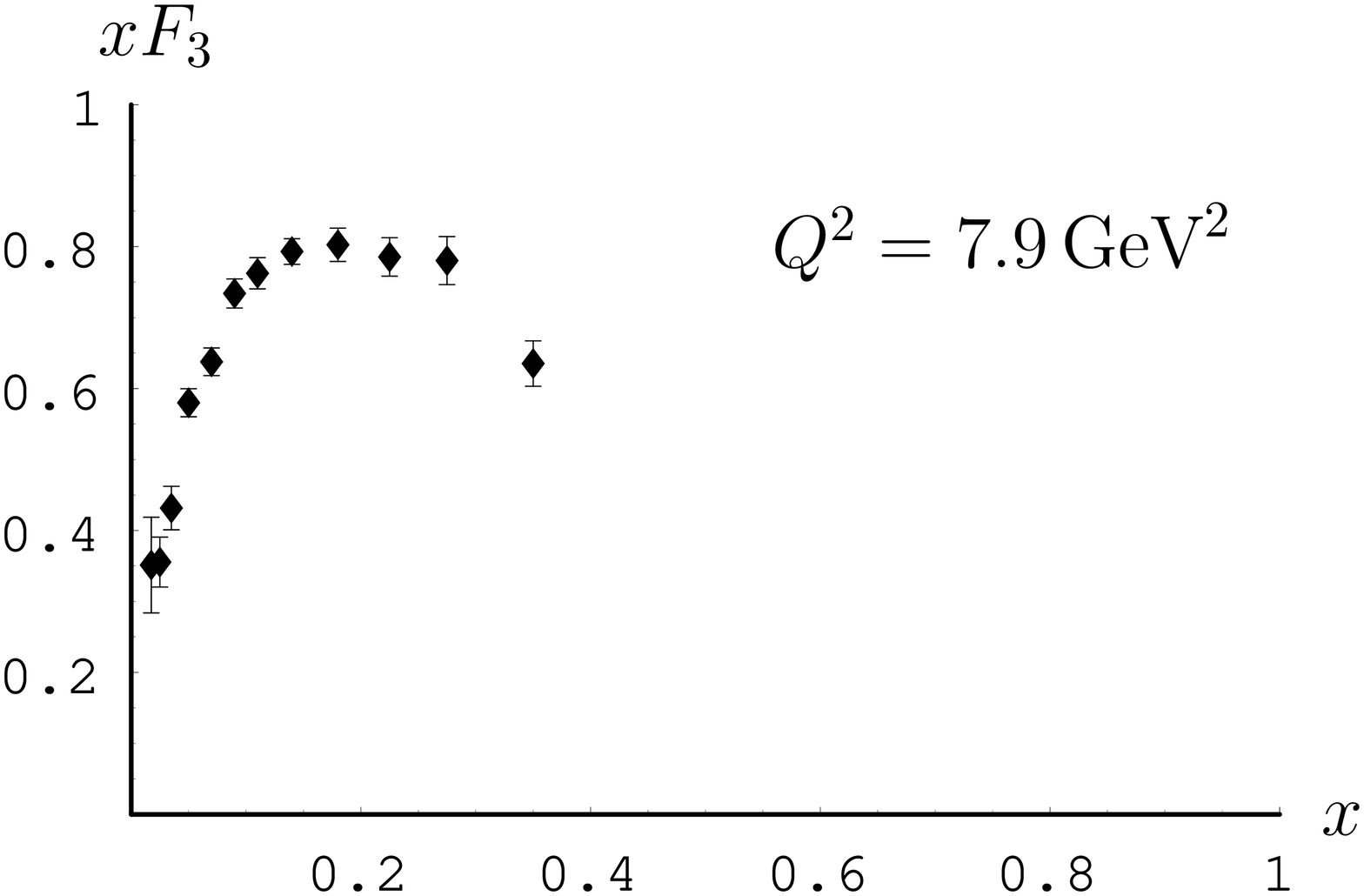}&\hspace{-1cm}
\includegraphics[angle=270,width=0.36\textwidth]{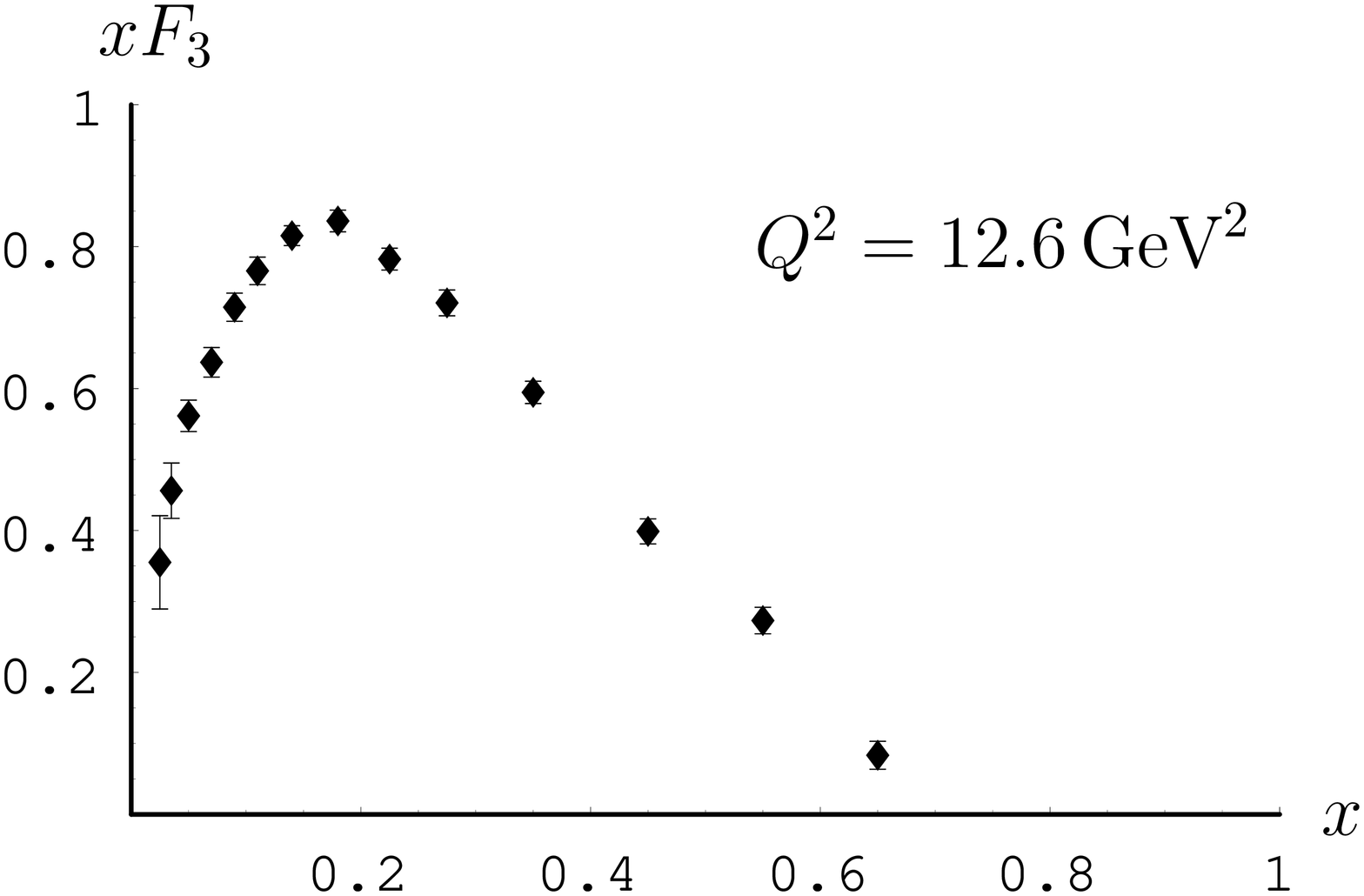}\\
\hspace{-.4cm}\includegraphics[angle=270,width=0.36\textwidth]{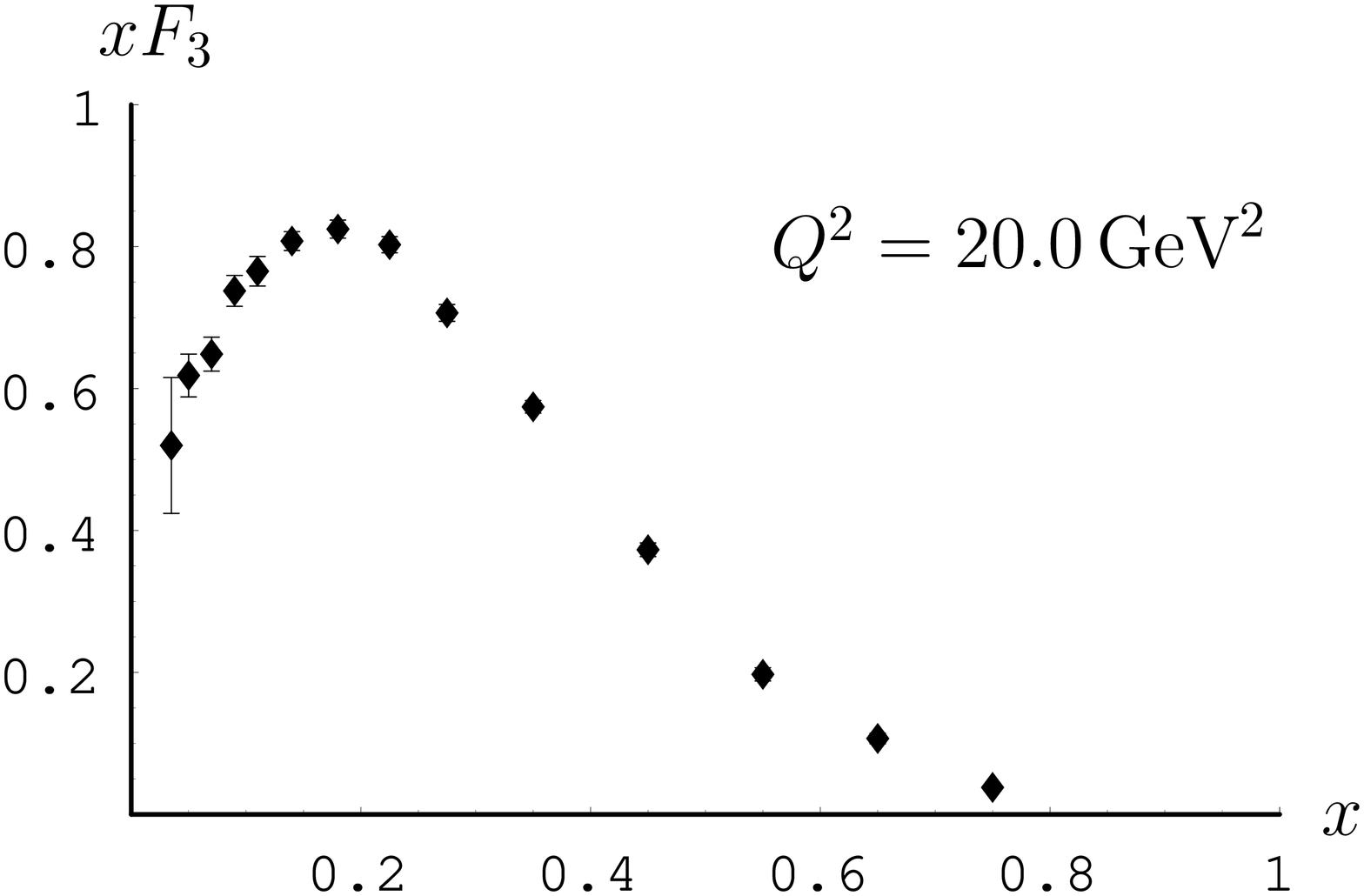}&\hspace{-1cm}
\includegraphics[angle=270,width=0.36\textwidth]{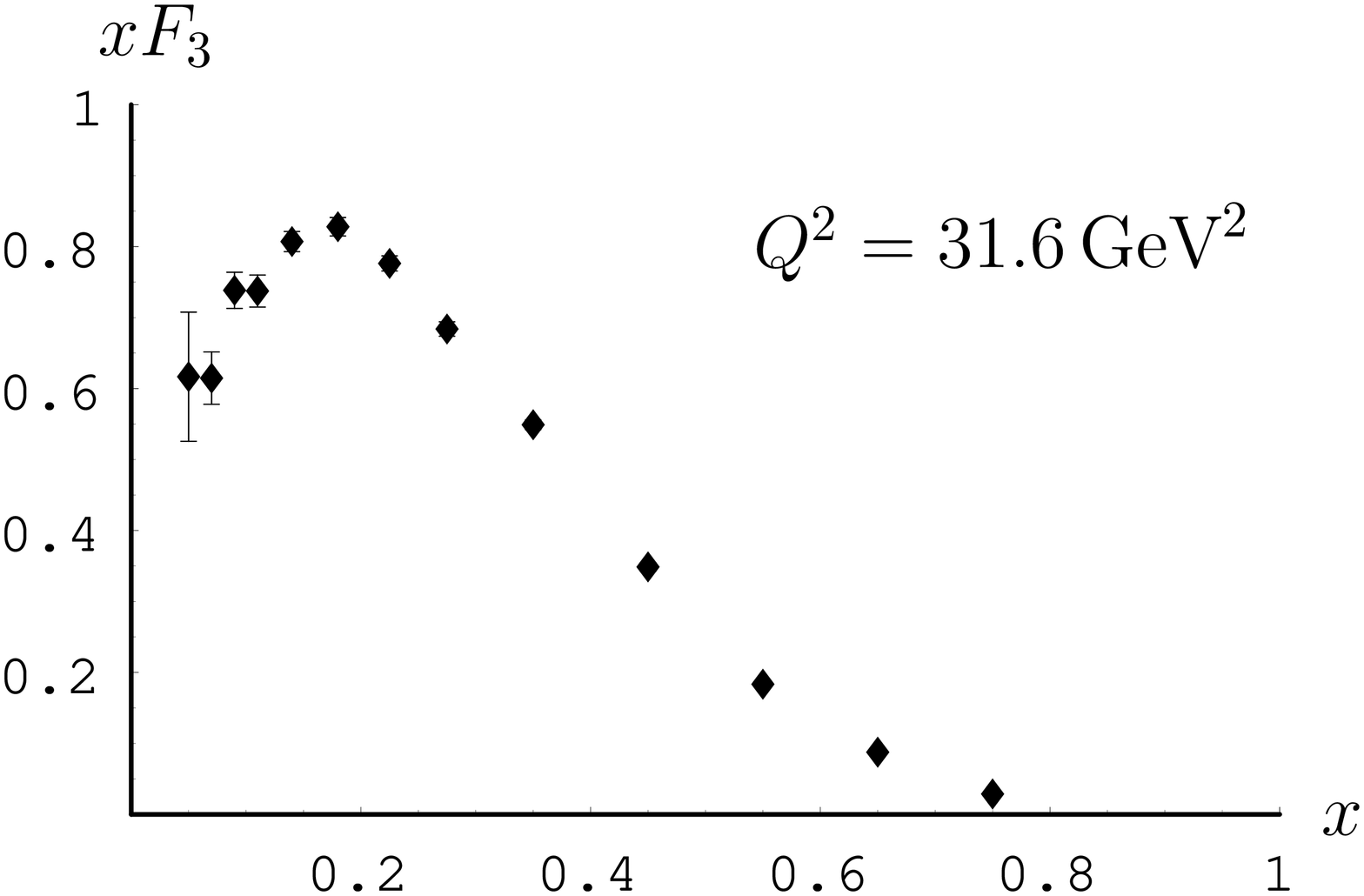}&\hspace{-1cm}
\includegraphics[angle=270,width=0.36\textwidth]{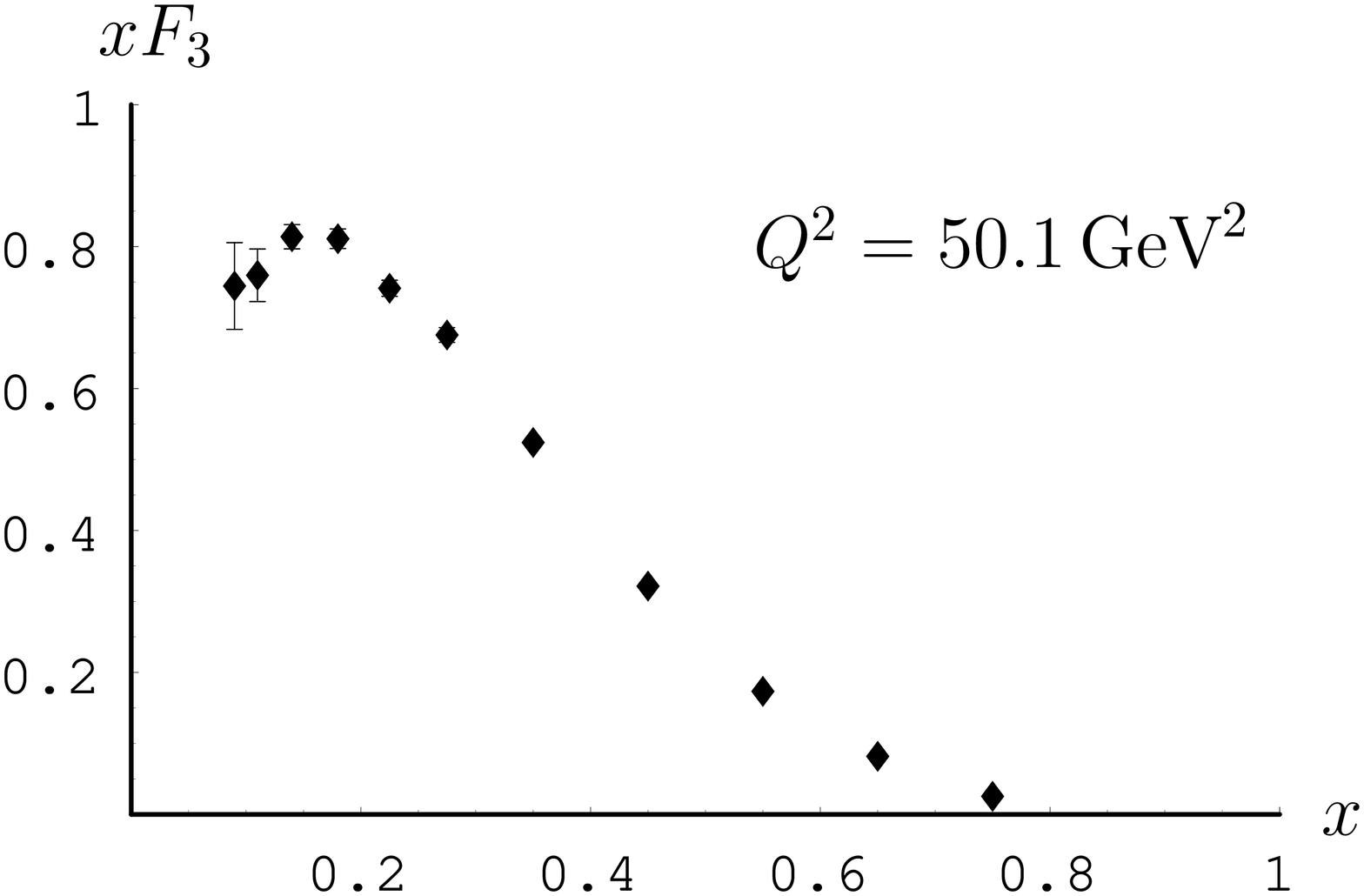}\\
\hspace{-.4cm}\includegraphics[angle=270,width=0.36\textwidth]{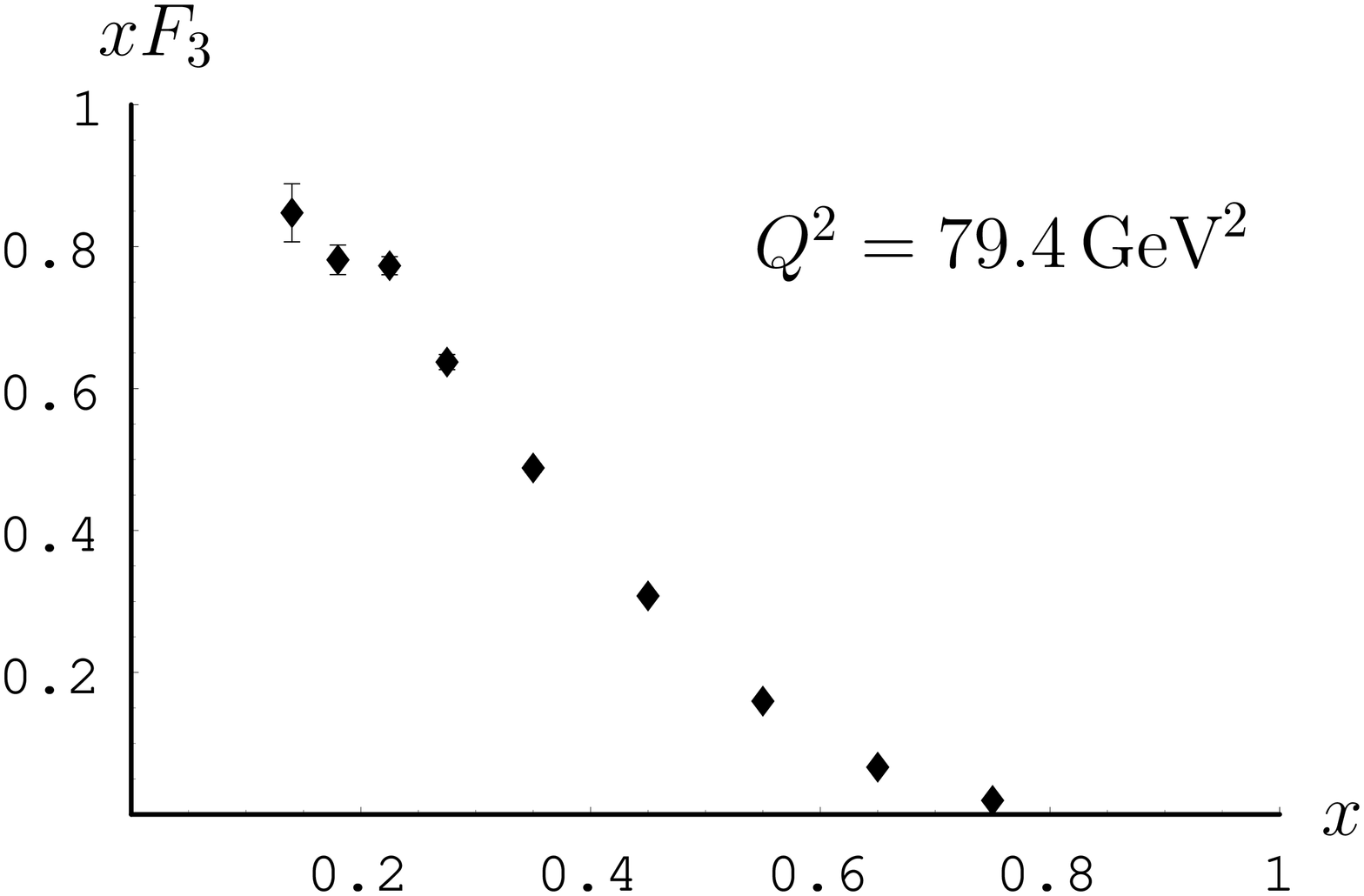}&\hspace{-1cm}
\includegraphics[angle=270,width=0.36\textwidth]{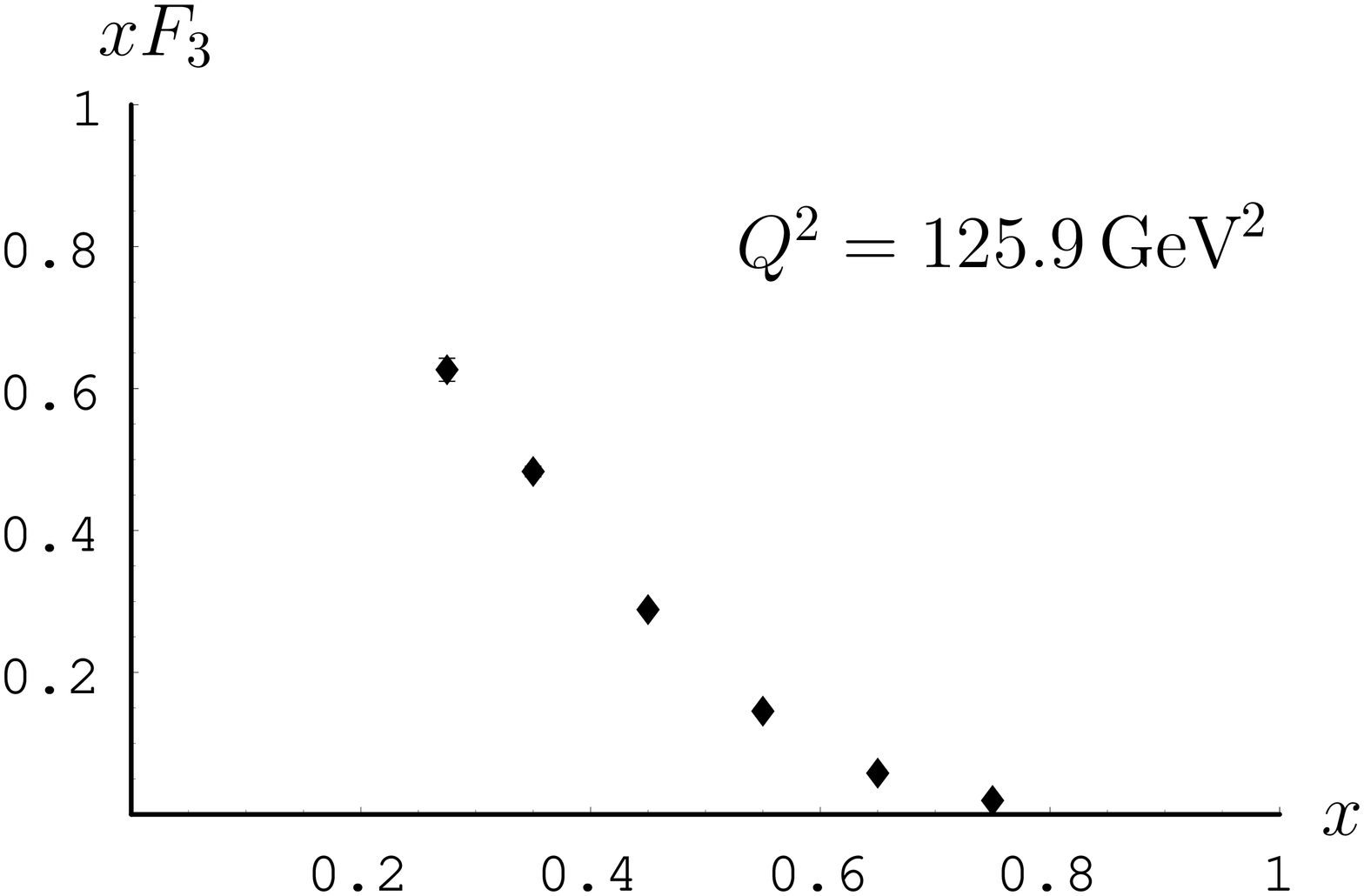}&\hspace{-1cm}
\includegraphics[angle=270,width=0.36\textwidth]{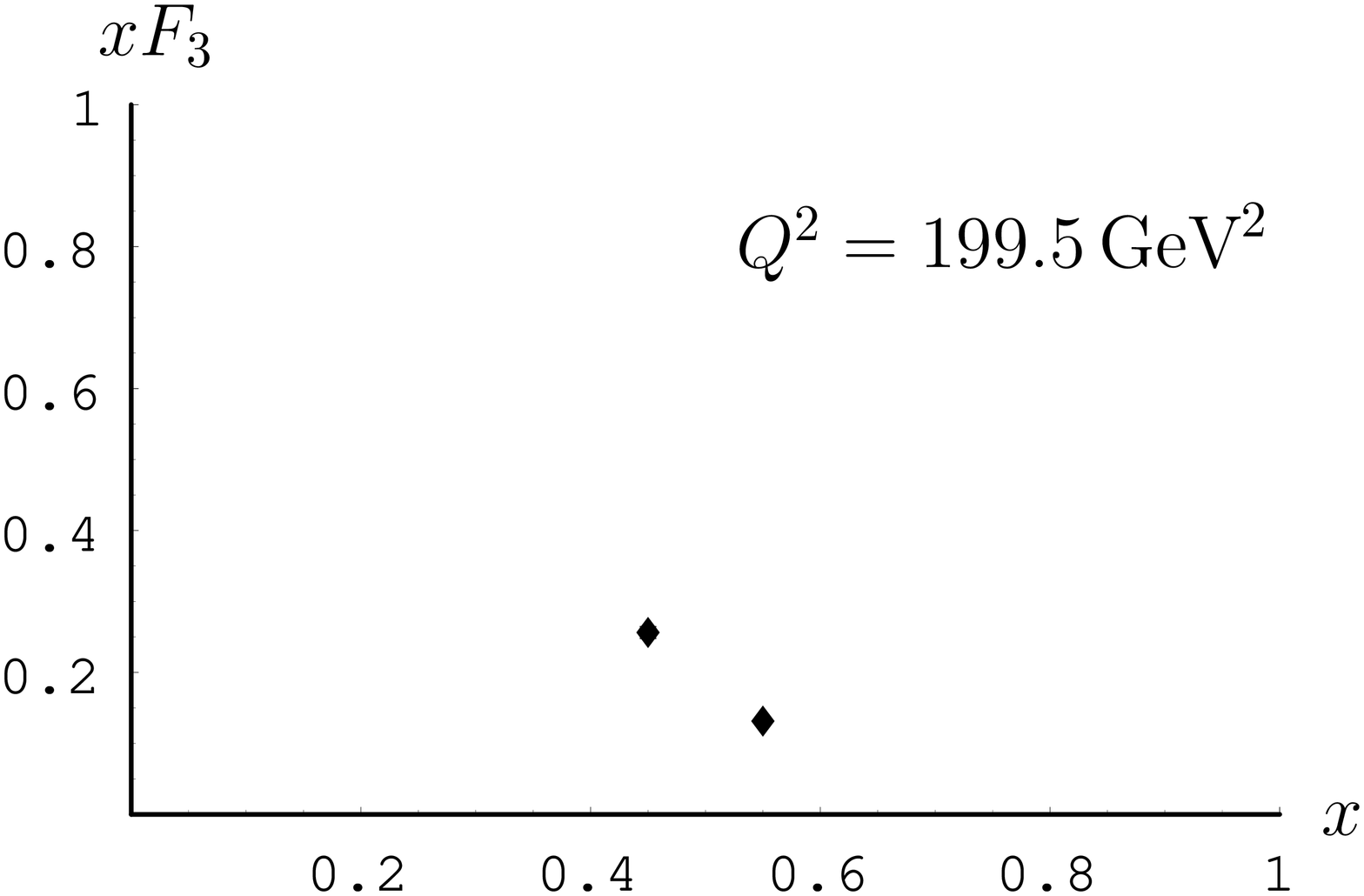}
\end{tabular}
\end{center}
\caption{Data for $xF_{3}$ plotted against $x$ for the 12 different $Q^{2}$
  bins of the CCFR data.}
\label{f:CCFR}
\end{figure}
When comparing theoretical predictions for moments of structure
functions with experimental data, we are faced with the
long-standing issue of missing data regions at high and low $x$ for
low and high $Q^{2}$ respectively. This is demonstrated in
Fig.~\ref{f:CCFR} in which we plot the CCFR data \cite{r1} for 12
different values of $Q^{2}$. We can see that at the lower range of
$Q^{2}$ we are limited to low-$x$ data, and that at high $Q^{2}$ we
are limited to the high $x$ range.

In order to reliably evaluate a moment at a particular $Q^{2}$, we
require data for the whole range of $x$. This being unavailable, we
are forced to make some guess about how the structure function
behaves in the missing data region. That is to say, we have to
choose some method of modelling (extrapolating and interpolating)
the data to cover the full range of $x$. However we wish to make the
evaluation of the experimental moments as free from QCD input as
possible, thus making the comparison between theory and experiment
as direct as possible. To this end, we shall adopt the approach
involving Bernstein averages \cite{r2,r3}; objects which, though
related to the moments, have negligible dependence on the modelling
method adopted (and hence on the behaviour of the structure function
in the missing data regions).

We define the Bernstein
polynomials as follows,
\begin{eqnarray}
\qq
p_{nk}(x^{2})&=&2\frac{\Gamma\lb(n+\frac{3}{2}\rb)}{\Gamma\lb(k+\frac{1}{2}\rb)\Gamma\lb(n-k+1\rb)}x^{2k}(1-x^{2})^{n-k},\qq
n,k\in\mathbb{I}. \label{eq:BP}
\end{eqnarray}
These functions are constructed such that they are zero at the endpoints
$x=0$ and $x=1$, and they are also normalized such that
$\int_{0}^{1}p_{nk}(x)dx=1$. Furthermore, if we constrain $n$ and $k$ such that $n\geq
k\geq0$, then $p_{nk}(x)$ are peaked sharply in some region between the
two endpoints.

The Bernstein polynomials can be treated as a distribution, with a
mean,
\begin{eqnarray}
\overline{x}_{nk}&=&\int_{0}^{1}x\;p_{nk}(x)\;dx\\
&=&\frac{\Gamma(k+1)\Gamma(n+\frac{3}{2})}{\Gamma(k+\frac{1}{2})\Gamma(n+2)},
\label{eq:BPmean}
\end{eqnarray}
and variance,
\begin{eqnarray}
\Delta x_{nk}&=&\int_{0}^{1}\lb(x-\overline{x}_{nk}\rb)^{2}\;p_{nk}(x)\;dx\\
&=&\frac{k+\frac{1}{2}}{n+\frac{3}{2}}-\lb(\frac{\Gamma(k+1)\Gamma(n+\frac{3}{2})}{\Gamma(k+\frac{1}{2})\Gamma(n+2)}\rb)^{2}.
\label{eq:BPvar}
\end{eqnarray}
The Bernstein averages of $F_{3}$ are then defined by,
\begin{eqnarray}
F_{nk}(Q^{2})&=&\int_{0}^{1}p_{nk}(x^{2})F_{3}(x,Q^{2})dx.
\label{eq:BAs}
\end{eqnarray}
Thus $F_{nk}(Q^{2})$ is the average of the structure function
weighted such that the region around $\overline{x}_{nk}$ is
emphasized. By picking the values of $n$ and $k$ wisely, we can
construct a set of averages which enhance the region for which we
have data for $F_{3}$ and de-emphasize the regions where there are
gaps. Therefore, in the resultant averages, the dependence on the
missing data regions will be heavily suppressed.

Defining this more carefully,
for a given value of $Q^{2}$, we only consider averages for which the range,
\begin{eqnarray}
\overline{x}_{nk}-\sqrt{\Delta x_{nk}}\;\le\;
x\;\le\;\overline{x}_{nk}+\sqrt{\Delta x_{nk}}, \label{eq:interval}
\end{eqnarray}
lies entirely within the region for which we have data. The only exception to this is
that if the
highest-$x$ data point lies within this range, then we {\it do} accept this
average, but only if the data suggests that $xF_{3}$ vanishes rapidly beyond this point.

\begin{figure}
\begin{center}
\begin{tabular}{c c c}
\hspace{-.018\textwidth}\includegraphics[angle=270,width=0.33\textwidth]{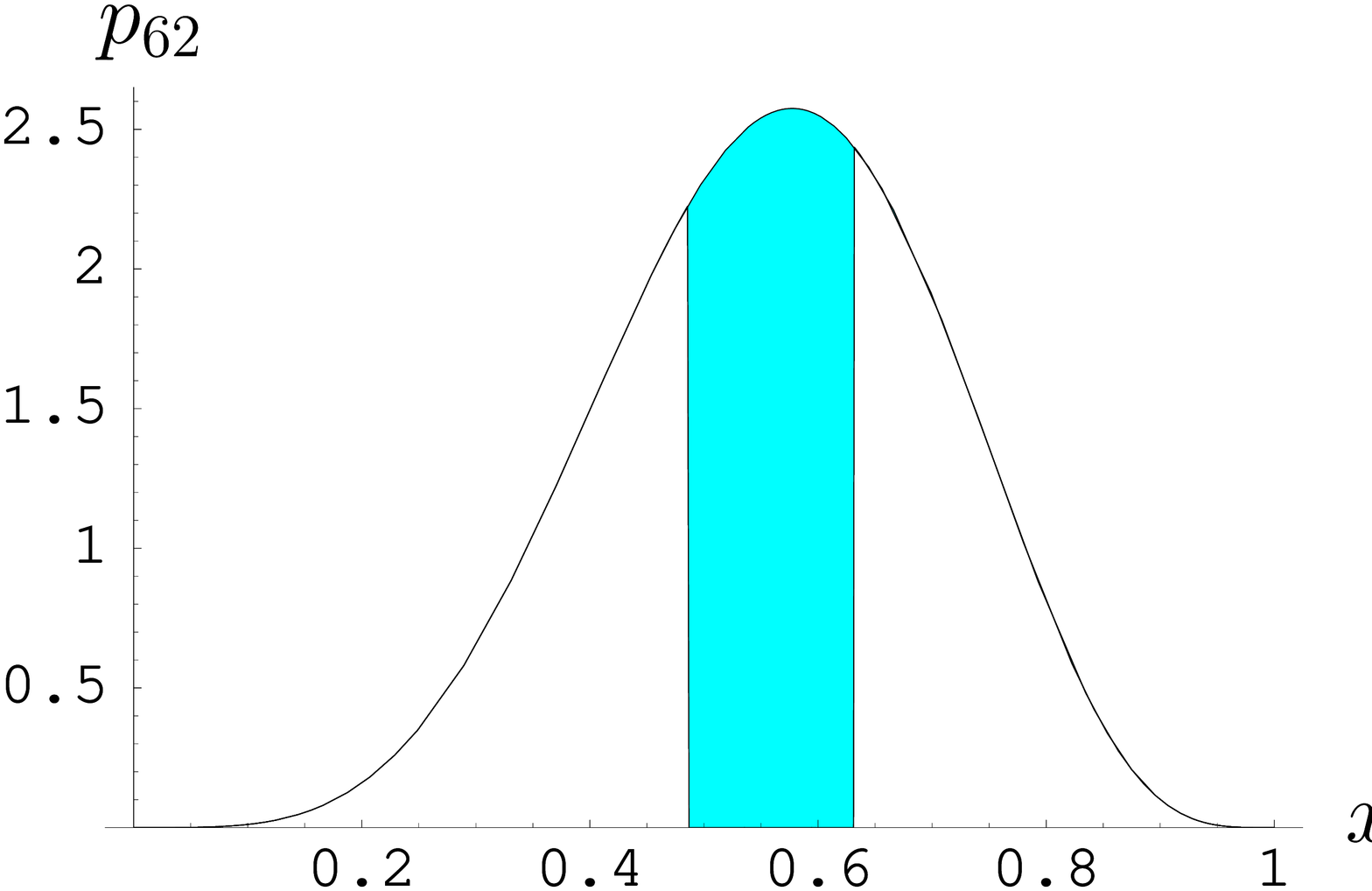}&\hspace{-.036\textwidth}
\includegraphics[angle=270,width=0.33\textwidth]{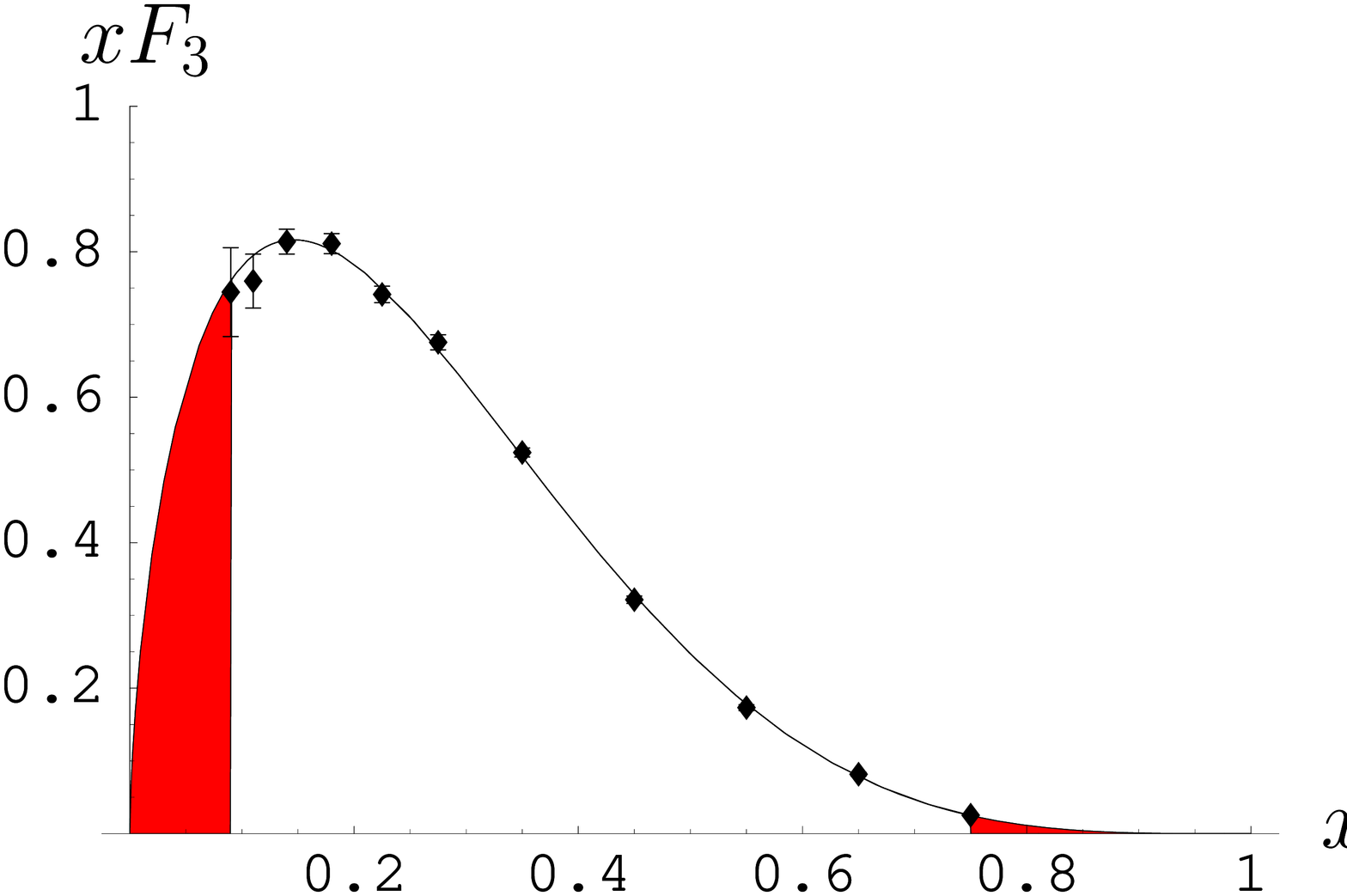}&\hspace{-.036\textwidth}
\includegraphics[angle=270,width=0.33\textwidth]{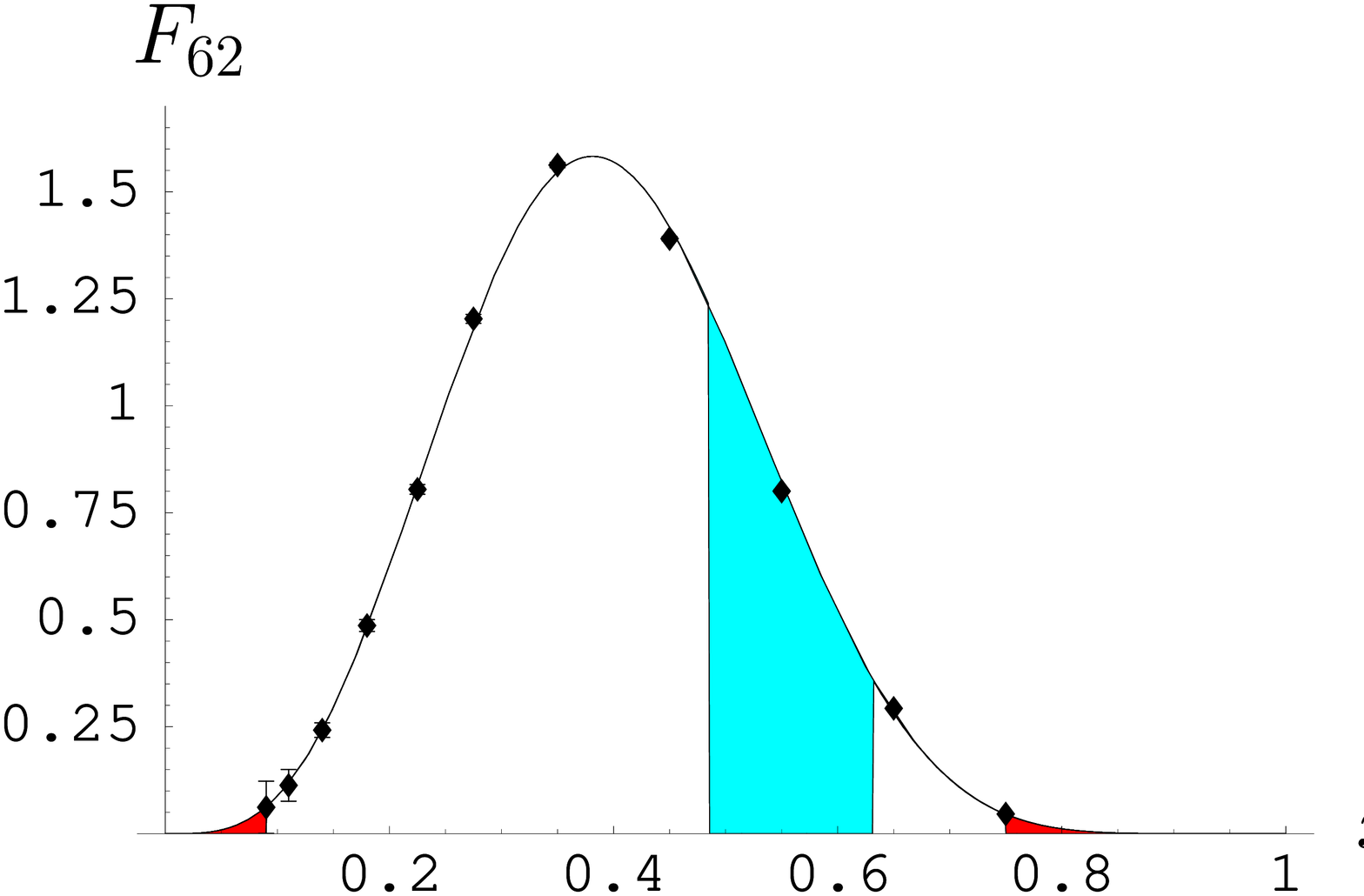}
\end{tabular}
\end{center}
\vspace*{-0.15\textheight}
\begin{eqnarray*}
\qquad\!\!\!\!\times\frac{1}{x}\qquad
\qquad\qquad\qquad\quad\qquad\;=\qquad
\end{eqnarray*}
\vspace*{0.8truecm}
\caption{Constructing the Bernstein average, $F_{62}(Q^{2}=50.1\;{\rm{GeV}}^{2})$. The
  light grey region represents the interval in Eq.~(\ref{eq:interval}) and the dark grey
  areas represent the missing data regions. The small size of the
  dark grey
  region in the right hand plot demonstrates that this average will have
  negligible dependence on the missing data regions. Note that the right
  hand plot actually shows the {\it integrand} of the Bernstein average. The average
  itself will be this function integrated over $[0,1]$.}
\label{f:berndemo}
\end{figure}

The construction of an acceptable average, and the resultant suppression of
the missing data region is demonstrated in Fig.~\ref{f:berndemo}. We see that
the shaded (dark grey) missing data regions almost disappear in the right
hand plot.

By expanding the integrand of Eq.~(\ref{eq:BAs}) in powers of $x$, and using
Eq.~(\ref{eq:6moms}), we can relate the averages directly
to the moments,
\begin{eqnarray}
F_{nk}(Q^{2})&=&\frac{2\Gamma(n+\frac{3}{2})}{\Gamma(k+\frac{1}{2})}\sum_{l=0}^{n-k}\frac{(-1)^{l}}{l!(n-k-l)!}\MM(2(k+l)+1;Q^{2}),
\label{eq:BAsMoms}
\end{eqnarray}
and so theoretical predictions for the averages can be obtained by
substitution of Eqs.~(\ref{eq:CORGImoms}), (\ref{eq:PSM}) and
(\ref{eq:6ECH}) into the above expression. The Bernstein average is
seen to be a linear combination of {\it odd} moments. Due to the
unavailability of results for $d_{2}$ for even $n$, previous NNLO
analyses of this kind have been limited to the inclusion of only odd
$F_{3}$ moments. However, now that the NNLO calculation of the NS
anomalous dimension is complete \cite{r7,r8}, we are no longer
constrained in such a way. In light of this, we define a new set of
{\it modified} Bernstein polynomials,
\begin{eqnarray}
\qq
\pt_{nk}(x^{2})&=&2\frac{\Gamma\lb(n+2\rb)}{\Gamma\lb(k+1\rb)\Gamma\lb(n-k+1\rb)}x^{2k+1}(1-x^{2})^{n-k},\qq
n,k\in\mathbb{I}, \label{eq:mBP}
\end{eqnarray}
which include only odd powers of $x$ and hence whose averages are
related to even moments. These modified Bernstein polynomials are
simply the original polynomials of Eq.~(\ref{eq:BP}), multiplied be
$x$, and then `re-normalized' such that they still satisfy
$\int_{0}^{1}\pt_{nk}(x)dx=1$. We can calculate the mean and
variance of $\pt_{nk}(x)$,
\begin{eqnarray}
\overline{\tx}&=&
\frac{\Gamma(k +\frac{3}{2})\Gamma(n+2)}{\Gamma(k+1)\Gamma(n+\frac{5}{2})},\label{eq:mBPmean}\\
\Delta \tx&=&
\frac{k+1}{n+2}
-\lb[\frac{\Gamma(k+\frac{3}{2})\Gamma(n+2)}{\Gamma(k+1)\Gamma(n+\frac{5}{2})}\rb]^{2},
\label{eq:mBPvar}
\end{eqnarray}
and in analogy with Eq.~(\ref{eq:BAs}), we define the modified
Bernstein averages,
\begin{eqnarray}
\tF_{nk}(Q^{2})&=&\int_{0}^{1}\pt_{nk}(x^{2})F_{3}(x,Q^{2})dx.
\label{eq:mBAs}
\end{eqnarray}
Again, we only accept experimental modified averages for which the range,
\begin{eqnarray}
\overline{\tx}_{nk}-\sqrt{\Delta \tx_{nk}}\;\le\; x\;\le\;\overline{\tx}_{nk}+\sqrt{\Delta \tx_{nk}},
\label{eq:minterval}
\end{eqnarray}
lies within the region for which we have data. We obtain theoretical
predictions for the modified averages using the equation,
\begin{eqnarray}
\tF_{nk}(Q^{2})&=&\frac{2\Gamma(n+2)}{\Gamma(k+1)}\sum_{l=0}^{n-k}\frac{(-1)^{l}}{l!(n-k-l)!}\MM(2(k+l)+2;Q^{2}),
\label{eq:mBAsMoms}
\end{eqnarray}
which is seen to be a linear combination of even moments.

In order to calculate averages from data for $F_{3}$, we need an
expression for $xF_{3}$ covering the entire range of $x$, for each
value of $Q^{2}$. As mentioned above, the values of the moments
calculated in this way will depend on how we model the structure
functions in the missing data regions but for the averages, this
dependence is suppressed. However, we would like to test this
assertion, and so we use four different methods of modelling $F_{3}$
and perform our analysis separately for each method. Significant
differences between the results would signify a failure of the
Bernstein average method, and in instances where this is the case,
we reject that particular average at that particular $Q^{2}$.
Moderate deviation however, is acceptable, provided that we use the
magnitude of the deviation as an estimate of the error associated
with the missing data region. This error is then included as a
`modelling error' in the final result. In this way, we can almost
completely remove any dependence on missing data regions, and also
quantify the error associated with any residual dependence.

The four extrapolation methods we use are described below:
\renewcommand{\labelenumi}{{\bf \Roman{enumi}}}
\begin{enumerate}
\item In the first method, we fit the function,
\begin{eqnarray}
xF_{3}(x)&=&{\cal{A}}x^{\cal{B}}(1-x)^{\cal{C}}, \label{eq:ABC}
\end{eqnarray}
to the data for each fixed value of $Q^{2}$. The parameters ${\cal{A}}$, ${\cal{B}}$ and ${\cal{C}}$
are obtained by performing $\chi^{2}$ fitting of Eq.~(\ref{eq:ABC}) to data
for $F_{3}$. They are $Q^{2}$-dependent
quantities, and errors on their values are obtained by performing
the fitting with the data for $F_{3}$ shifted to the two extremes of the
error bars.

A justification for the particular form of fitting function in
Eq.~(\ref{eq:ABC}) can be found in Ref.~\cite{r17}. However, the
simple fact that this function fits the data well is justification
enough, since the Bernstein averages are independent of the
extrapolation method.

\item The second method we use is linear interpolation between successive data
points. We also extrapolate beyond the data range, to the endpoints $xF_{3}(x)|_{x=0}=xF_{3}
(x)|_{x=1}=0$, in order to be consistent with method {\bf I}.

\item The third method consists of using the fitting function of Eq.~(\ref{eq:ABC}), but
setting $xF_{3}(x)=0$ everywhere outside the region for which we have data.

\item In analogy with {\bf III}, in this method we use the linear interpolation of
method {\bf II} but setting $xF_{3}(x)=0$ everywhere outside the data region.\label{en:mod4}

\end{enumerate}
\renewcommand{\labelenumi}{\arabic{enumi}}

The deviation between the results obtained from the above methods
(in particular, the difference between the first two and the last
two) will be a good measure of the effectiveness of the Bernstein
average method.

\begin{figure}
\begin{center}
  \includegraphics[angle=270,width=0.5\textwidth]{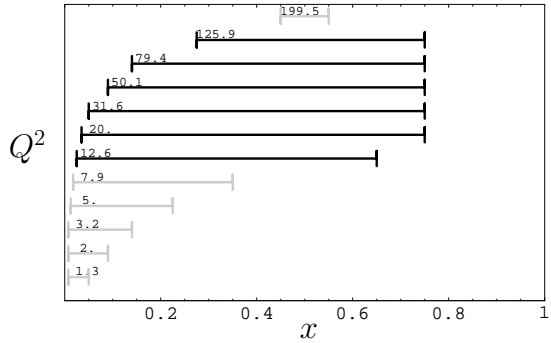}
\caption{Diagram depicting the $x$-ranges covered by the CCFR data,
at
  different $Q^{2}$.}\label{f:x-ranges}
\end{center}
\end{figure}

Data for $xF_{3}$ in neutrino-nucleon scattering is available from
the CCFR collaboration \cite{r1}. The data was obtained from the
scattering of neutrinos off iron nuclei and the measurements span
the  ranges $1.26\leq Q^{2}\leq199.5\;{\rm{GeV}}^2$ and $0.015\leq
x\leq0.75$. The $x$-ranges covered at each $Q^{2}$ are depicted in
Fig.~\ref{f:x-ranges}.

\begin{figure}[h]
\begin{center}
\begin{tabular}{c c}
\hspace{-.45cm}\includegraphics[angle=270,width=0.5\textwidth]{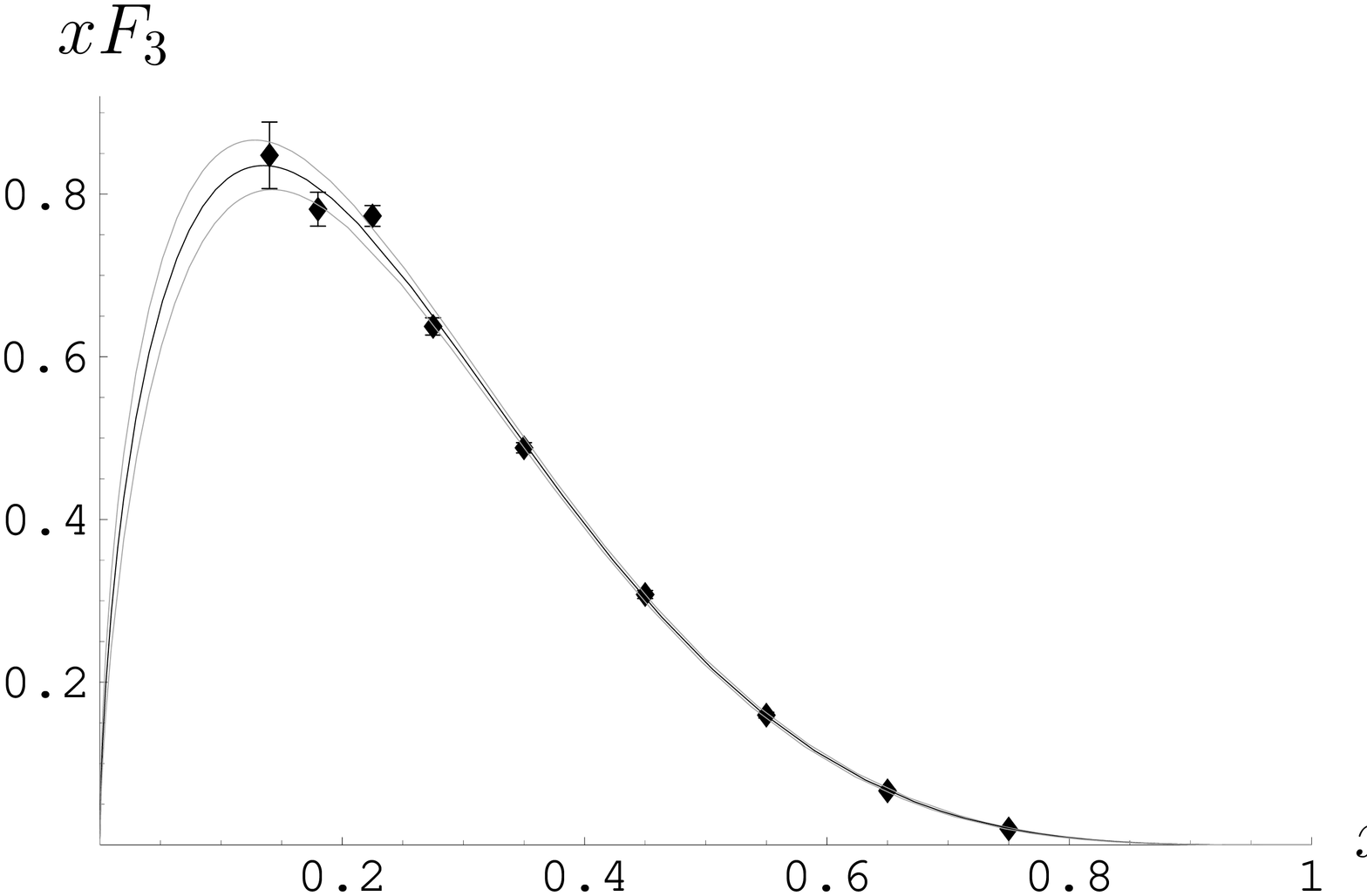}&
\includegraphics[angle=270,width=0.5\textwidth]{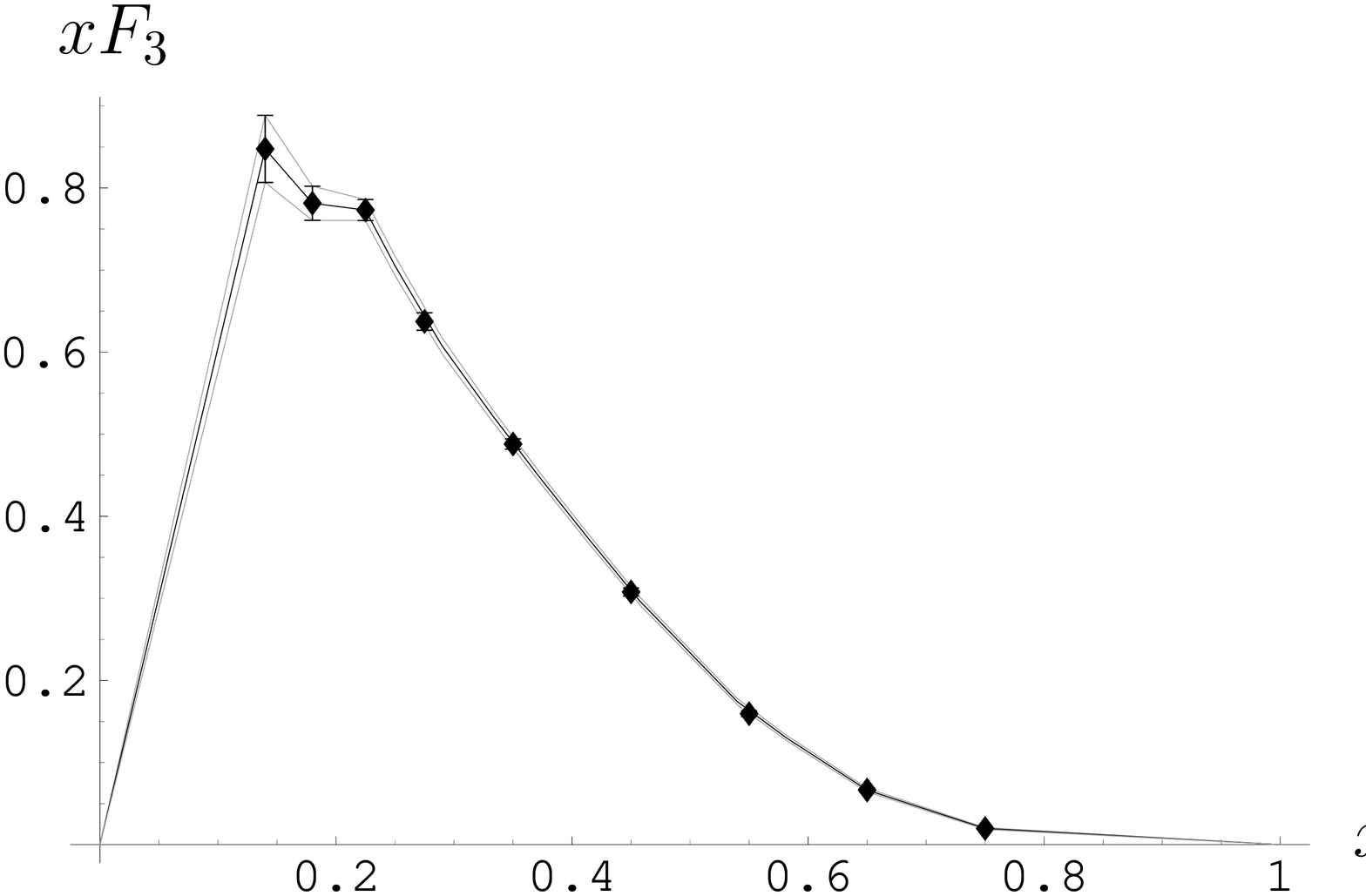}\\[-20pt]
\vspace{-0.6cm}\hspace{-.45cm}\includegraphics[angle=270,width=0.5\textwidth]{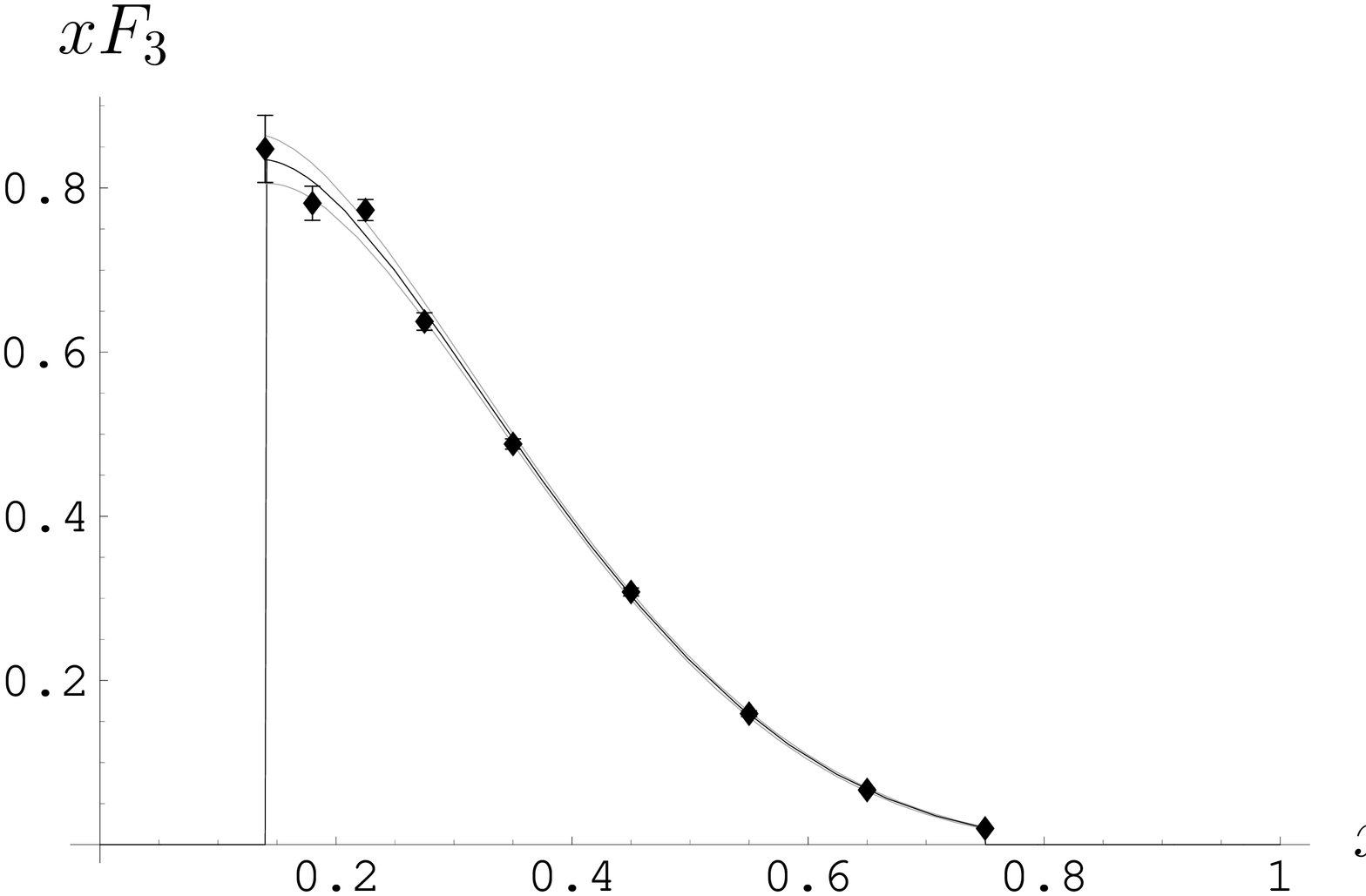}&
\includegraphics[angle=270,width=0.5\textwidth]{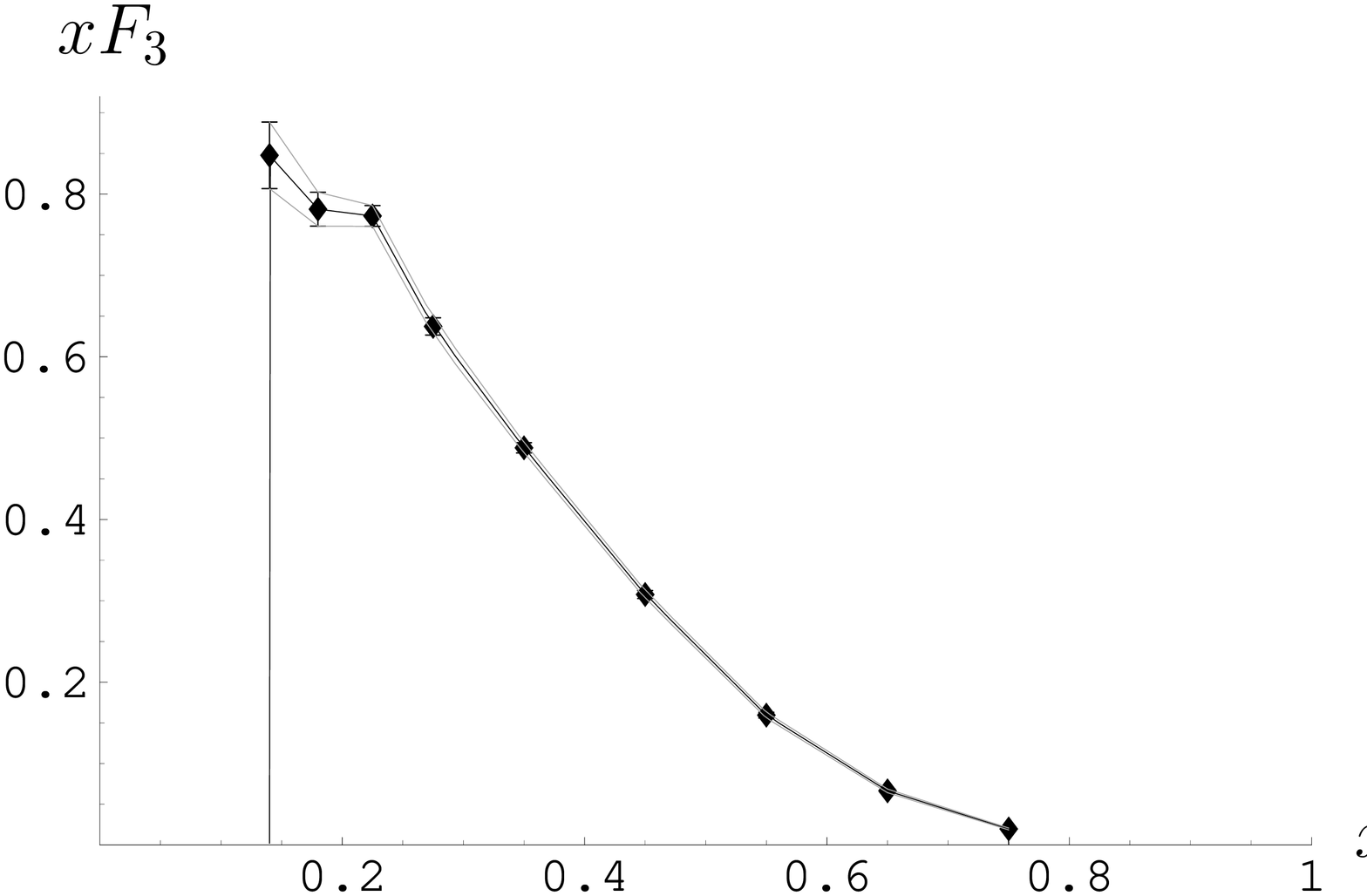}
\end{tabular}
\caption{The four methods used for fitting the structure functions.
Here we show the
  measured values of $xF_{3}$ at $Q^{2}=79.4\;\GeVV$. The errors are
  determined by re-performing the fitting for the data shifted to the extremes
  of the error bars, and this is denoted by grey lines.}\label{f:4fits}
\end{center}
\end{figure}

In Fig.~\ref{f:4fits} we show each of the four modelling methods
applied to $F_{3}$ measured at $Q^{2}=79.4\;{\rm{GeV}}^2$. Also
shown on these figures (in grey) are the fits which are used to
determine the errors on the modelling, which propagate through to
errors on the averages.

From the CCFR (and using the methods {\bf I} - {\bf IV} outlined
above) we can obtain expressions describing the behaviour of the
structure function over the full range of $x$, for each value of
$Q^{2}$. It is then possible to extract experimental values of the
averages, using the methods outlined below:

In the case of {\bf I}, obtaining the averages is particularly simple. Substituting Eq.~(\ref{eq:ABC}) into
Eqs.~(\ref{eq:BAs}) and (\ref{eq:mBAs}) gives,
\begin{eqnarray}
F^{(\rmss{exp})}_{nk}&=&\mathcal{A}\frac{2\Gamma(n+\frac{3}{2})}{\Gamma(k+\frac{1}{2})}\sum_{l=0}^{n-k}\frac{(-1)^{l}}{l!(n-k-l)!}{\rm
  B}(2(k+l)+\mathcal{B},\mathcal{C}+1),
\label{eq:BAexp}
\end{eqnarray}
for the Bernstein averages and,
\begin{eqnarray}
\tF^{(\rmss{exp})}_{nk}&=&\mathcal{A}\frac{2\Gamma(n+2)}{\Gamma(k+1)}\sum_{l=0}^{n-k}\frac{(-1)^{l}}{l!(n-k-l)!}{\rm
  B}(2(k+l)+\mathcal{B}+1,\mathcal{C}+1),
\label{eq:mBAexp}
\end{eqnarray}
for the modified Bernstein averages. Here,
B$(x,y)\equiv\Gamma(x)\Gamma(y)/\Gamma(x+y)$ is the Beta function.
Once values for ${\cal{A}}$, ${\cal{B}}$, and ${\cal{C}}$, have been
obtained, substitution into the above expressions leads directly to
the averages.

In the case of {\bf II}, each of the averages is split into $j+1$ sections (where $j$
is the number of data points) and each section is an integral over a polynomial of
order $2n+1$. It is then reasonably simple to evaluate the averages by
computing this set of integrals. This approach also applies to method {\bf IV}, but
in this case there are only $j-1$ integrals.

For method {\bf III} we simply integrate the fitting function,
multiplied by the Bernstein polynomials, with the integration limits
being the values of $x$ at the first and last data point.

Having outlined the method for obtaining the experimental averages
we now turn our attention to which averages are acceptable at which
energies. The highest moment we use is the 18th moment and the
lowest the 1st. Inclusion of higher moments than this leads to {\it
no} significant increase in the number of acceptable Bernstein
averages. The upper limit of $n=18$ implies that the highest
Bernstein averages included are $F_{8k}$ and $\tF_{8k}$ and that the
lowest used are $F_{10}$ and $\tilde{F}_{10}$. We exclude the
averages for which $n=k$ as they simply correspond to individual
moments themselves. This leaves us with a total of 72 (36 +
$\tilde{36}$) potential averages at our disposal for each value of
$Q^{2}$. This number will be reduced when we come to exclude
averages on the basis of the acceptance criteria. After applying the
acceptance criteria, we are left with 132 data points for the
standard Bernstein averages and 141 for the modified Bernstein
averages. Exactly which averages we use at a particular $Q^{2}$, can
be determined by inspecting the plots in the results section.

\begin{figure}[h!]
\begin{center}
\hspace*{-6pt}\begin{tabular}{c c} \psfrag{x}{$x$}
\psfrag{Qt}{{\scriptsize$\!\!\!\!Q^{2}/\GeVV$}}
\psfrag{F10}{\textcolor{white}{\scriptsize{$\!\!\!\!F_{10}$}}}
\psfrag{F20}{\textcolor{white}{\scriptsize{$\!\!\!\!F_{20}$}}}
\psfrag{F30}{\textcolor{white}{\scriptsize{$\!\!\!\!F_{30}$}}}
\psfrag{F40}{\textcolor{white}{\scriptsize{$\!\!\!\!F_{40}$}}}
\psfrag{F50}{\textcolor{white}{\scriptsize{$\!\!\!\!F_{50}$}}}
\psfrag{F60}{\textcolor{white}{\scriptsize{$\!\!\!\!F_{60}$}}}
\psfrag{F70}{\textcolor{white}{\scriptsize{$\!\!\!\!F_{70}$}}}
\psfrag{F80}{\textcolor{white}{\scriptsize{$\!\!\!\!F_{80}$}}}
\psfrag{F87}{\textcolor{black}{\scriptsize{$\!\!\!\!F_{87}$}}}
\includegraphics[width=0.49\textwidth]{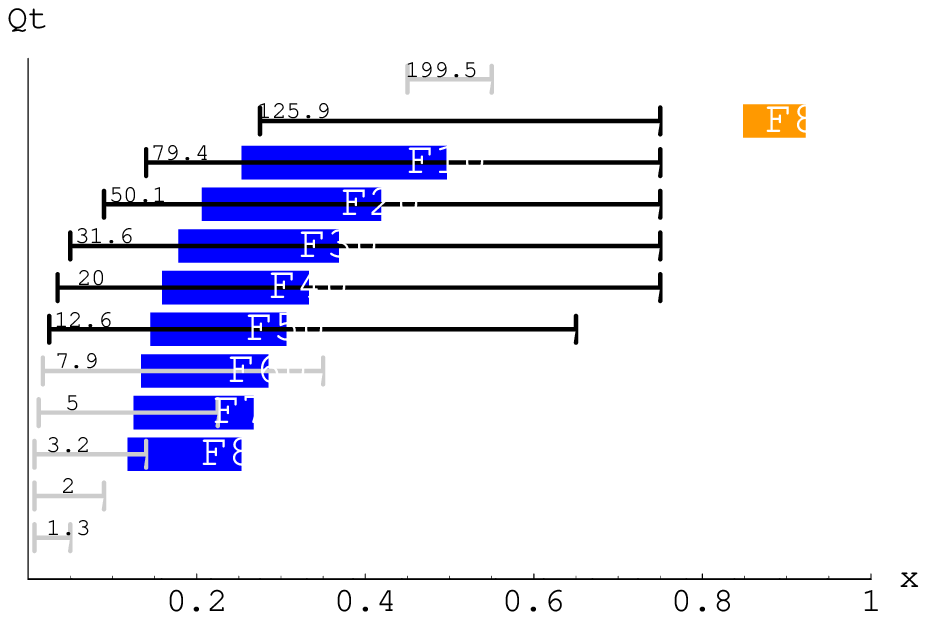}&
\psfrag{x}{$x$} \psfrag{Qt}{{\scriptsize$\!\!\!\!Q^{2}/\GeVV$}}
\psfrag{F21}{\textcolor{white}{\scriptsize{$\!\!\!\!F_{21}$}}}
\psfrag{F31}{\textcolor{white}{\scriptsize{$\!\!\!\!F_{31}$}}}
\psfrag{F41}{\textcolor{white}{\scriptsize{$\!\!\!\!F_{41}$}}}
\psfrag{F51}{\textcolor{white}{\scriptsize{$\!\!\!\!F_{51}$}}}
\psfrag{F61}{\textcolor{white}{\scriptsize{$\!\!\!\!F_{61}$}}}
\psfrag{F71}{\textcolor{white}{\scriptsize{$\!\!\!\!F_{71}$}}}
\psfrag{F81}{\textcolor{white}{\scriptsize{$\!\!\!\!F_{81}$}}}
\psfrag{F76}{\textcolor{black}{\scriptsize{$\!\!\!\!F_{76}$}}}
\psfrag{F86}{\textcolor{black}{\scriptsize{$\!\!\!\!F_{86}$}}}
\includegraphics[width=0.49\textwidth]{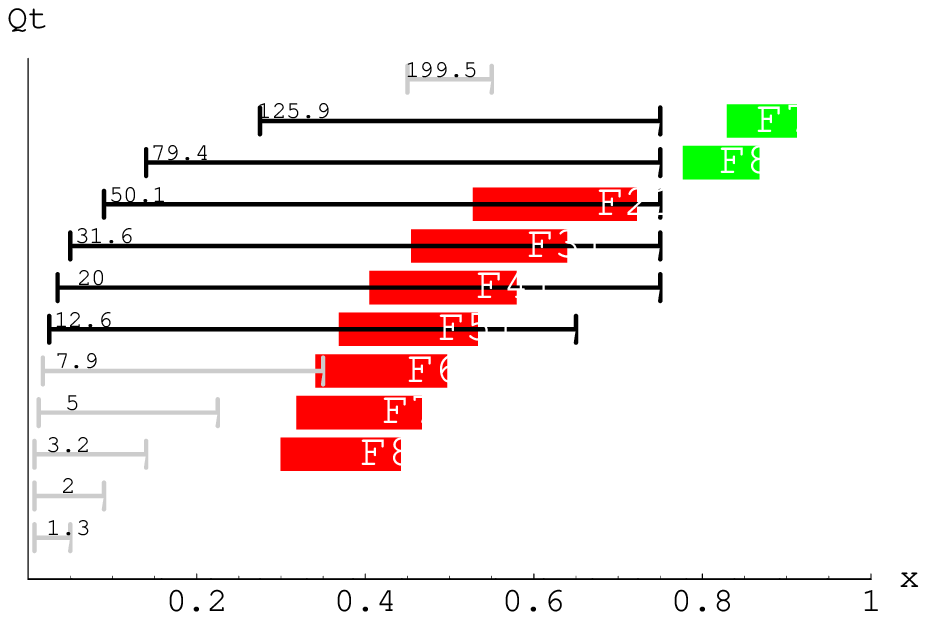}\\
\psfrag{x}{$x$}
\psfrag{Qt}{{\scriptsize$\!\!\!\!Q^{2}/\GeVV$}}
\psfrag{F32}{\textcolor{white}{\scriptsize{$\!\!\!\!F_{32}$}}}
\psfrag{F42}{\textcolor{white}{\scriptsize{$\!\!\!\!F_{42}$}}}
\psfrag{F52}{\textcolor{white}{\scriptsize{$\!\!\!\!F_{52}$}}}
\psfrag{F62}{\textcolor{white}{\scriptsize{$\!\!\!\!F_{62}$}}}
\psfrag{F72}{\textcolor{white}{\scriptsize{$\!\!\!\!F_{72}$}}}
\psfrag{F82}{\textcolor{white}{\scriptsize{$\!\!\!\!F_{82}$}}}
\psfrag{F65}{\textcolor{black}{\scriptsize{$\!\!\!\!F_{65}$}}}
\psfrag{F75}{\textcolor{black}{\scriptsize{$\!\!\!\!F_{75}$}}}
\psfrag{F85}{\textcolor{black}{\scriptsize{$\!\!\!\!F_{85}$}}}
\includegraphics[width=0.49\textwidth]{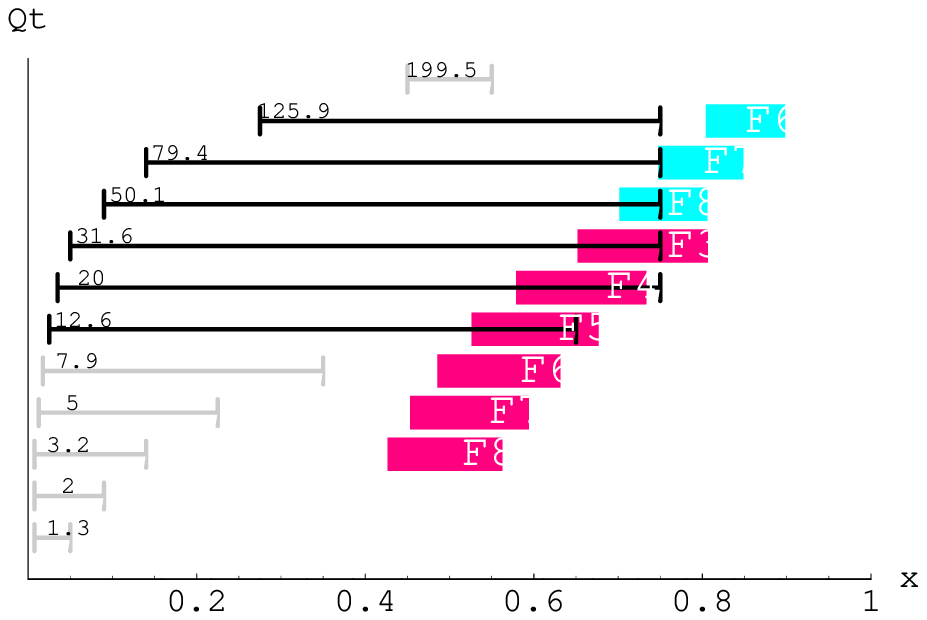}&
\psfrag{x}{$x$}
\psfrag{Qt}{{\scriptsize$\!\!\!\!Q^{2}/\GeVV$}}
\psfrag{F43}{\textcolor{white}{\scriptsize{$\!\!\!\!F_{43}$}}}
\psfrag{F53}{\textcolor{white}{\scriptsize{$\!\!\!\!F_{53}$}}}
\psfrag{F63}{\textcolor{white}{\scriptsize{$\!\!\!\!F_{63}$}}}
\psfrag{F73}{\textcolor{white}{\scriptsize{$\!\!\!\!F_{73}$}}}
\psfrag{F83}{\textcolor{white}{\scriptsize{$\!\!\!\!F_{83}$}}}
\psfrag{F54}{\textcolor{black}{\scriptsize{$\!\!\!\!F_{54}$}}}
\psfrag{F64}{\textcolor{black}{\scriptsize{$\!\!\!\!F_{64}$}}}
\psfrag{F74}{\textcolor{black}{\scriptsize{$\!\!\!\!F_{74}$}}}
\psfrag{F84}{\textcolor{black}{\scriptsize{$\!\!\!\!F_{84}$}}}
\includegraphics[width=0.49\textwidth]{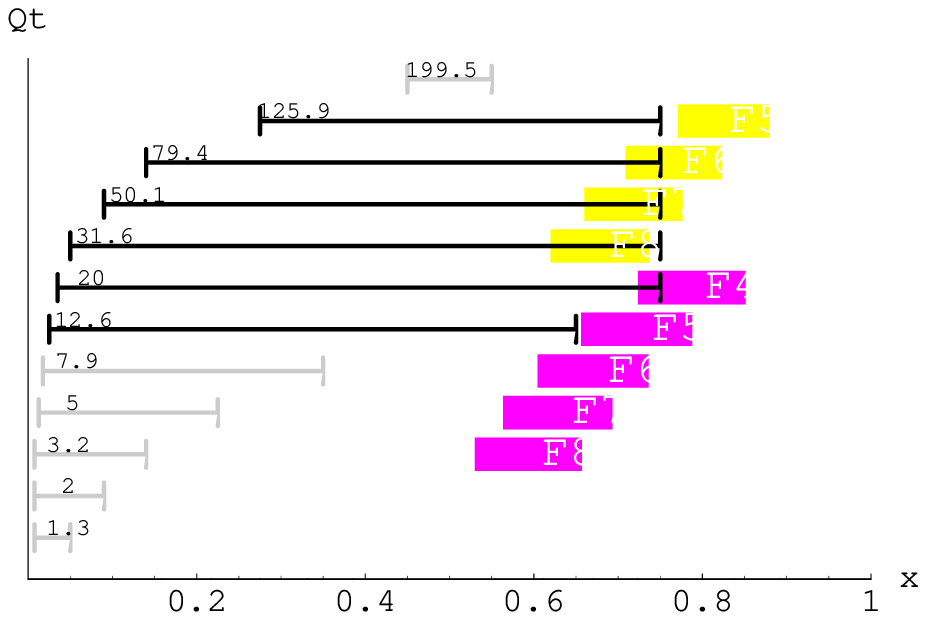}
\end{tabular}
\end{center}

\caption{The black and light grey bars ($\vdash\!\dashv$) show  $x$ ranges covered by the CCFR data at
  different energies. Superimposed onto these, in various colours,
  are the peaked regions of the individual Bernstein polynomials, defined by the interval in
 Eq.~(\ref{eq:interval}).}
\label{f:ranges}
\end{figure}

\begin{figure}[h!]
\begin{center}
\hspace*{-6pt}\begin{tabular}{c c}
&\\
\psfrag{x}{$x$}
\psfrag{Qt}{{\scriptsize$\!\!\!\!Q^{2}/\GeVV$}}
\psfrag{F10}{\textcolor{white}{\tiny{$\!\!\!\!\tF_{10}$}}}
\psfrag{F20}{\textcolor{white}{\tiny{$\!\!\!\!\tF_{20}$}}}
\psfrag{F30}{\textcolor{white}{\tiny{$\!\!\!\!\tF_{30}$}}}
\psfrag{F40}{\textcolor{white}{\tiny{$\!\!\!\!\tF_{40}$}}}
\psfrag{F50}{\textcolor{white}{\tiny{$\!\!\!\!\tF_{50}$}}}
\psfrag{F60}{\textcolor{white}{\tiny{$\!\!\!\!\tF_{60}$}}}
\psfrag{F70}{\textcolor{white}{\tiny{$\!\!\!\!\tF_{70}$}}}
\psfrag{F80}{\textcolor{white}{\tiny{$\!\!\!\!\tF_{80}$}}}
\psfrag{F87}{\textcolor{black}{\tiny{$\!\!\!\!\tF_{87}$}}}
\includegraphics[width=0.49\textwidth]{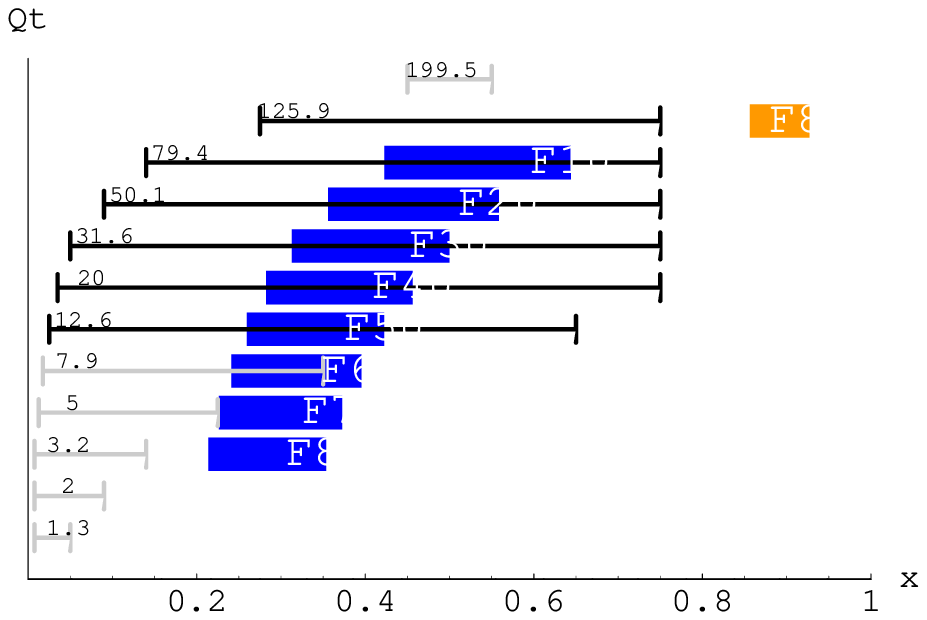}&
\psfrag{x}{$x$}
\psfrag{Qt}{{\scriptsize$\!\!\!\!Q^{2}/\GeVV$}}
\psfrag{F21}{\textcolor{white}{\tiny{$\!\!\!\!\tF_{21}$}}}
\psfrag{F31}{\textcolor{white}{\tiny{$\!\!\!\!\tF_{31}$}}}
\psfrag{F41}{\textcolor{white}{\tiny{$\!\!\!\!\tF_{41}$}}}
\psfrag{F51}{\textcolor{white}{\tiny{$\!\!\!\!\tF_{51}$}}}
\psfrag{F61}{\textcolor{white}{\tiny{$\!\!\!\!\tF_{61}$}}}
\psfrag{F71}{\textcolor{white}{\tiny{$\!\!\!\!\tF_{71}$}}}
\psfrag{F81}{\textcolor{white}{\tiny{$\!\!\!\!\tF_{81}$}}}
\psfrag{F76}{\textcolor{black}{\tiny{$\!\!\!\!\tF_{76}$}}}
\psfrag{F86}{\textcolor{black}{\tiny{$\!\!\!\!\tF_{86}$}}}
\includegraphics[width=0.49\textwidth]{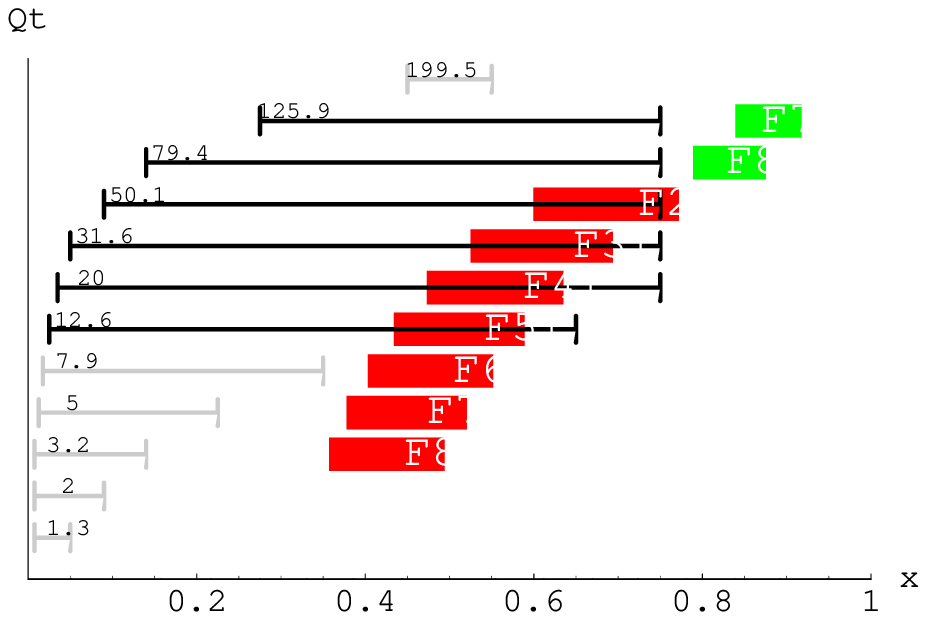}\\
\psfrag{x}{$x$}
\psfrag{Qt}{{\scriptsize$\!\!\!\!Q^{2}/\GeVV$}}
\psfrag{F32}{\textcolor{white}{\tiny{$\!\!\!\!\tF_{32}$}}}
\psfrag{F42}{\textcolor{white}{\tiny{$\!\!\!\!\tF_{42}$}}}
\psfrag{F52}{\textcolor{white}{\tiny{$\!\!\!\!\tF_{52}$}}}
\psfrag{F62}{\textcolor{white}{\tiny{$\!\!\!\!\tF_{62}$}}}
\psfrag{F72}{\textcolor{white}{\tiny{$\!\!\!\!\tF_{72}$}}}
\psfrag{F82}{\textcolor{white}{\tiny{$\!\!\!\!\tF_{82}$}}}
\psfrag{F65}{\textcolor{black}{\tiny{$\!\!\!\!\tF_{65}$}}}
\psfrag{F75}{\textcolor{black}{\tiny{$\!\!\!\!\tF_{75}$}}}
\psfrag{F85}{\textcolor{black}{\tiny{$\!\!\!\!\tF_{85}$}}}
\includegraphics[width=0.49\textwidth]{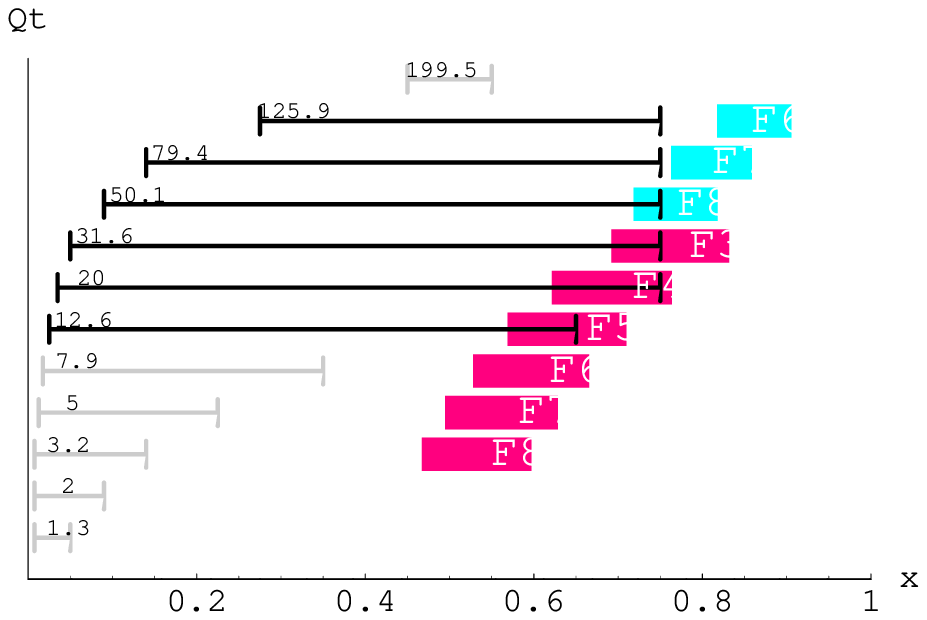}&
\psfrag{x}{$x$}
\psfrag{Qt}{{\scriptsize$\!\!\!\!Q^{2}/\GeVV$}}
\psfrag{F43}{\textcolor{white}{\tiny{$\!\!\!\!\tF_{43}$}}}
\psfrag{F53}{\textcolor{white}{\tiny{$\!\!\!\!\tF_{53}$}}}
\psfrag{F63}{\textcolor{white}{\tiny{$\!\!\!\!\tF_{63}$}}}
\psfrag{F73}{\textcolor{white}{\tiny{$\!\!\!\!\tF_{73}$}}}
\psfrag{F83}{\textcolor{white}{\tiny{$\!\!\!\!\tF_{83}$}}}
\psfrag{F54}{\textcolor{black}{\tiny{$\!\!\!\!\tF_{54}$}}}
\psfrag{F64}{\textcolor{black}{\tiny{$\!\!\!\!\tF_{64}$}}}
\psfrag{F74}{\textcolor{black}{\tiny{$\!\!\!\!\tF_{74}$}}}
\psfrag{F84}{\textcolor{black}{\tiny{$\!\!\!\!\tF_{84}$}}}
\includegraphics[width=0.49\textwidth]{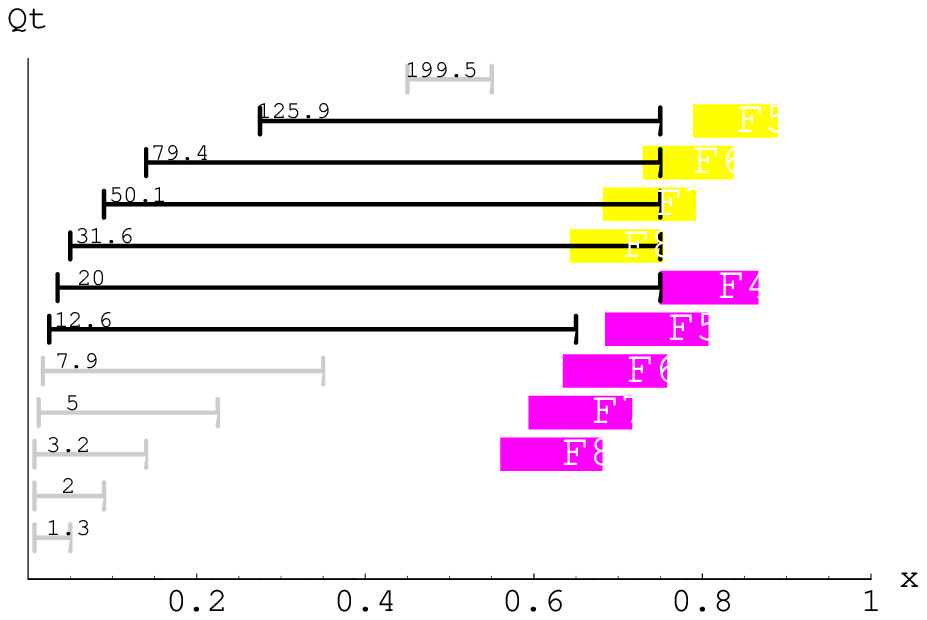}
\end{tabular}
\end{center}
\caption{The black and light grey bars ($\vdash\!\dashv$) show  $x$ ranges covered by the CCFR data at
  different energies. Superimposed onto these, in various colours,
  are the peaked regions of the individual modified Bernstein polynomials, defined by the interval in
Eq.~(\ref{eq:minterval}).}
\label{f:ranges2}
\end{figure}

In Fig.~\ref{f:ranges} we plot the dominant regions of the Bernstein
polynomials (given by Eq.~(\ref{eq:interval})) for each of the used
averages. This is superimposed onto the data-range diagram of
Fig.~\ref{f:x-ranges}. Figure \ref{f:ranges2} shows equivalent plots
for the modified Bernstein averages. These plots can be used to
identify which averages are acceptable for a particular value of
$Q^{2}$.

\section{Fitting procedure}\vspace{-\parskip} We use $\chi^{2}$
minimization to optimize the fits of the theoretical predictions to
the data. The highest moment included in the experimental averages
is the 18th, and so when TMCs are included, we will require
predictions for the first 20 moments. Therefore, the set of fitting
parameters comprises of $\{A_{1}\ldots A_{20}\}$ plus the QCD scale
parameter $\LMS$\@. When we include higher twist corrections this
set is expanded to include $A_{\rmt{HT}}$\@. To check consistency
between the odd and even moments we perform the analysis for each of
these sets of moments separately and then finally together, and
compare the results.

Although we stated previously that 132 standard Bernstein averages
are available to us, in the CORGI case this is reduced to 130 for
the following reason: When fitting predictions to the data, we scan
values of $\LMS^{(4)}$ between 0 and 590 MeV for a minimum in
$\chi^{2}$. Unfortunately, for $n=17$ and 19, the values of
$\Lambda_{\MM}/\LMS$ (see table \ref{xcorgi}) are such that
$Q^{2}=7.9\;\GeVV$ is below the Landau pole in
Eq.~(\ref{eq:6tHooft}). Consequently we cannot obtain CORGI
predictions for Bernstein averages which include these moments. From
Eqs.~(\ref{eq:6:TMC}) and (\ref{eq:BAsMoms}) we can determine that
this excludes $F_{80}(Q^{2}=7.9 \;{\rm{GeV}}^{2})$ and $F_{70}(Q^{2}=7.9\;{\rm{GeV}}^{2})$ from the
fit. In the case of the modified Bernstein averages,
$\tF_{80}(Q^{2}=7.9\;{\rm{GeV}}^{2})$ and $\tF_{70}(Q^{2}=7.9\;{\rm{GeV}}^{2})$ are already excluded
due to their failure to meet the acceptance criteria.

The CCFR data includes statistical errors and 18 different sources
of systematic error. These errors cannot be added in quadrature, and
so we perform the analysis for each of these 19 sources of error
separately and then add the variation in the results in quadrature
to obtain the final total error. We also include, as additional
sources of error the deviation in results associated with using the
four different modelling methods (this forms the `modelling error'
in our final result), and the deviation in the results obtained from
performing the analysis with and without HT corrections included
(forming the `HT error').

\subsubsection*{Correlation of errors}\vspace{-\parskip}
When fitting theoretical predictions to experimental data using
$\chi^{2}$ minimization, care must be taken in order to take into
account fully the correlation between data points.

To construct $\chi^{2}$ from a set of $N$ uncorrelated data points
$\{f_{i}^{\rmt{exp.}}\}$ ($i=1,\ldots N$), with errors
$\{\sigma_{f_{i}}\}$ and corresponding theoretical predictions
$\{f_{i}^{\rmt{theo.}}\}$, we have,
\begin{eqnarray}
\chi^{2}&=&\sum_{i=1}^{N}\lb(\frac{f_{i}^{\rmt{exp.}}-f_{i}^{\rmt{theo.}}}{\sigma_{f_{i}}}\rb)^{2}.\label{eq:6:chinaive}
\label{eq:6:chi2}
\end{eqnarray}
The raw data for $xF_{3}$ are uncorrelated. However, we are not
comparing predictions for the structure functions themselves with
data directly; rather we are doing so indirectly via the Bernstein
averages. For a given value of $Q^{2}$, the full set of Bernstein
averages (and modified Bernstein averages) we obtain {\it will} be
correlated, due to their being derived from the same set of
($xF_{3}$) data points (see Fig.~\ref{f:berndemo}).

In the case where $\{f_{i}^{\rmt{exp.}}\}$ are correlated the
$\chi^{2}$ function becomes \cite{r16b},
\begin{eqnarray}
\chi^{2}&=&\sum_{i=1}^{N}\sum_{j=1}^{N}\lb(f_{i}^{\rmt{exp.}}-f_{i}^{\rmt{theo.}}\rb)V^{-1}_{ij}\lb(f_{j}^{\rmt{exp.}}-f_{j}^{\rmt{theo.}}\rb)\nn\\[10pt]
&=&\lb(\textbf{f}^{\;\rmt{exp.}}-\textbf{f}^{\;\rmt{theo.}}\rb)^{
T}\textbf{V}^{-1}\lb(\textbf{f}^{\;\rmt{exp.}}-\textbf{f}^{\;\rmt{theo.}}\rb).\label{eq:6:chicovar}
\end{eqnarray}
In the second line of the above equation we have constructed vectors
from the data points and their predictions. \textbf{V} is known as
the covariance matrix. It encodes the correlation between each of
the data points; its elements are obtained as follows,
\begin{eqnarray}
V_{ij}&=&{\rm cov}\lb(f_{i}, f_{j}\rb)\nn\\
&=&\langle f_{i} f_{j}\rangle-\langle f_{i}\rangle \langle
f_{j}\rangle.
\end{eqnarray}
If the $f_{i}$ are themselves functions of $M$ variables $x_{k}$
(representing the `raw' data), then we have,
\begin{eqnarray}
{\rm cov}\lb(f_{k}, f_{l}\rb)&=&
\sum_{i=1}^{M}\sum_{j=1}^{M}\lb(\frac{\partial f_{k}}{\partial
x_{i}}\rb) \lb(\frac{\partial f_{l}}{\partial x_{j}}\rb) {\rm
cov}\lb(x_{i},x_{j}\rb).
\end{eqnarray}
If the data for $x_{i}$ are uncorrelated, this reduces to,
\begin{eqnarray}
{\rm cov}\lb(f_{k}, f_{l}\rb)&=& \sum_{i=1}^{M}\lb(\frac{\partial
f_{k}}{\partial x_{i}}\rb) \lb(\frac{\partial f_{l}}{\partial
x_{i}}\rb)\sigma_{x_{i}}^{2}.\label{eq:5:cov}
\end{eqnarray}
To obtain the covariance of the Bernstein averages we simply
substitute $F_{nk}(Q^{2})$ for $f_{i}$ and $F_{3}(x)$ for $x_{i}$ in
Eq.~(\ref{eq:5:cov}). We assume no correlation between different
values of $Q^{2}$ and so Eq.~(\ref{eq:6:chicovar}) decouples into 7
different matrix equations (one for each of the values of $Q^{2}$
between 7.9 and 125.9 $\GeVV$). By using the trapezium rule to
approximate the Bernstein average integrals, we obtain the following
expression for the covariance matrices:
\begin{eqnarray}
V_{lm}(Q_{i}^{2})&=&\sum_{j=1}^{N_{i}}\frac{1}{4}\lb(p_{nk}(x_{j})\rb)_{l}\lb(p_{nk}(x_{j})\rb)_{m}\lb(x_{j+1}-x_{j-1}\rb)^{2}\sigma_{F_{3}(x_{j})}^{2}.
\end{eqnarray}
Here, the index $j$ runs over the number of data points we have for
$xF_{3}$ at a given $Q_{i}^{2}$. The $j=0$ and $j=N_{i}+1$ terms are
simply the $x=0$ and $x=1$ endpoints (see parts {\bf I} and {\bf II}
of Fig.~\ref{f:4fits}). From this equation we can calculate elements
of the covariance matrices for each value of $Q^{2}$. All that
remains is for us to invert them. However, upon attempting to do so,
we find that these matrices are ill-conditioned, with some of their
eigenvalues being close to zero. Hence their inverses are
intractable. As a result of this, it is impossible to perform a
reliable $\chi^{2}$ analysis of the averages with their correlation
taken into account.

We believe that this is principally due to the fact that the
correlation between averages is significant in some cases, and this
in turn is an artefact of the fact that the selection criteria
systematically select Bernstein polynomials which are peaked in the
same region and hence are of fairly similar shape. This situation
arises because the intent behind the inclusion of more averages in
the analysis is not to increase the amount of `data', rather it is
to further ensure that the missing data regions are suppressed.

In light of this, we settle for the method adopted in
Refs.~\cite{r2,r3}, in which the na\"{i}ve $\chi^{2}$ function of
Eq.~(\ref{eq:6:chi2}) is used, but the error bars on the averages
are modified in order to account for the `over-counting of degrees
of freedom'. For example, for the standard averages, at each value
of $Q^{2}$ we have theoretical information on 9 moments, but the
number of experimental Bernstein averages we use is often more than
this; e.g.~for $Q^{2}=20\,\GeVV$ we use 27 standard Bernstein
averages. To remedy this, we adopt the following approach. For each
value of $Q^{2}$ we count the number of averages above 9 as
duplicate information. The number of duplicates we have altogether
is 73 and so for the values of $Q^{2}$ for which we have more
Bernstein averages than moments, we rescale the error on these
averages by $\sqrt{130/(130-73)}=1.510$. Correspondingly, for the
modified Bernstein averages we rescale the errors by a factor of
$\sqrt{141/(141-87)}=1.616$. This rescaling has the effect of
suppressing the contribution of the duplicate data points to
$\chi^{2}$, relative to those values of $Q^{2}$ for which we have
fewer Bernstein averages than moments.

\subsubsection*{Positivity constraints}\vspace{-\parskip}

The fact that $xF_{3}$ is a positive definite function, and that the
moments are simply integrals over these functions multiplied by a
single power of $x$, means that we can impose certain positivity
constraints on the parameters $A_{n}$, as follows.

We construct the following matrices from the moments,
\begin{eqnarray}
\hat{\MM}\;=\;\lb(\begin{array}{cccc}
\MM_{1}&\MM_{2}&\cdots\qq&\MM_{9}\\[5pt]
\MM_{2}&\MM_{3}& &\\
\MM_{3}& &\ddots\qq&\\
\vdots& & &\\
\MM_{9}& & &\MM_{17}\\
\end{array}\rb),\label{eq:6:det1}
\end{eqnarray}
and
\begin{eqnarray}
\D\hat{\MM}\;=\;\lb(\begin{array}{cccc}
\D\MM_{1}&\D\MM_{2}&\cdots\qq&\D\MM_{9}\\[5pt]
\D\MM_{2}&\D\MM_{3}& &\\
\D\MM_{3}& &\ddots\qq&\\
\vdots& & &\\
\D\MM_{9}& & &\D\MM_{17}\\
\end{array}\rb),\label{eq:6:det2}
\end{eqnarray}
where $\MM_{n}=\MM(n;Q^{2})$ and $\D\MM_{n}=\MM_{n}-\MM_{n+1}$.

In order for $\MM(n;Q^{2})$ to be moments of positive definite
functions (as the structure functions must be), the determinants of
the above matrices, and of all their minors, must be positive, for
all values of $Q^{2}$ \cite{r3}. Evaluating these determinants at
fixed $Q^{2}$ will translate to conditions on the parameters
$A_{n}$. We do not implement these constraints as part of the
fitting procedure. Rather, we perform checks on the values of the
fitting parameters resulting from the $\chi^{2}$ minimization to
ensure that they obey the above constraints.

However, we do impose positivity constraints on the moments
themselves. As a result of the determinantal constraints described
above, and from the general form of the moments given in
Eq.~(\ref{eq:6moms}), we can infer that the following inequalities
must be satisfied,
\begin{eqnarray}
\MM(n;Q^{2})&>&0,\label{eq:6:con1}\\
\MM(n;Q^{2})&>&\MM(n+1;Q^{2}),\label{eq:6:con2}
\end{eqnarray}
for fixed $Q^{2}$. Furthermore, we can implement these constraints
by defining our fitting parameters $A_{n}$ in terms of a new set of
parameters and then minimizing $\chi^{2}$ with respect to these new
parameters.

We begin by picking some value of $Q_{0}^{2}$ at which to implement
the conditions. We then take the last moment used in the analysis
($n=20$) and rewrite the constraint in Eq.~(\ref{eq:6:con1}) as,
\begin{eqnarray}
\MM(20;Q_{0}^{2})&=&\lb(\hA_{20}\rb)^{2},\label{eq:6:con1a}
\end{eqnarray}
where $\hA_{20}$ is a real number. The constraints in Eq.~(\ref{eq:6:con2})
can also be rewritten as,
\begin{eqnarray}
  \MM(n;Q_{0}^{2})&=&\MM(n+1;Q_{0}^{2})+\lb(\hA_{n}\rb)^{2},\label{eq:6:con2a}
\end{eqnarray}
for $1\leq n<20$, where $\hA_{n}$ are all real numbers. The LHSs of
Eqs.~(\ref{eq:6:con1a}) and (\ref{eq:6:con2a}) are simply a fitting
parameter times a number. For example, in the case of $n=2$ and
$Q^{2}_{0}=12.6\;\GeVV$ we have,
\begin{eqnarray}
  \MM(2;12.6\;\GeVV)&=&0.3932\,A_{2}.
\label{eq:6:momnum}
\end{eqnarray}
From this, and equivalent expressions for the rest of $A_{n}$, we
can obtain an expression for each $A_{n}$ in terms of the parameters
$\hA_{1}$ - $\hA_{20}$. This means that we can replace the
parameters $A_{1}$ - $A_{20}$ with $\hA_{1}$ - $\hA_{20}$ in the
$\chi^{2}$ function. By doing this and then minimizing with respect
to the $\hA_{n}$ parameters, we can find a minimum in $\chi^{2}$ for
which the constraints in Eqs.~(\ref{eq:6:con1}) and
(\ref{eq:6:con2}) are automatically satisfied. In effect, the
reparameterization embedded in Eqs.~(\ref{eq:6:con1a}) and
(\ref{eq:6:con2a}) restricts the parameter space to exclude
solutions for which the constraints are not satisfied.

To implement this reparameterization we must choose a value of
$Q_{0}^{2}$ at which to impose the constraints, whereas in reality
they must be satisfied for all $Q^{2}$. Because of this, we perform
the analysis for several different values of $Q_{0}^{2}$ and check
that the results remain stable.

\section{Results of fitting to the data}\vspace{-\parskip}
\label{s:Res}

We focus principally on the results from the CORGI analysis in which
both odd and even moments are included and in which we include
target mass corrections. This analysis results in a prediction for
the  QCD scale parameter of,
\begin{eqnarray}
\LMS^{(5)}&=&219.11_{-23.6}^{+22.1}\;{\rm MeV}.
\end{eqnarray}
The errors on this value can be broken down into four different
sources,
\begin{eqnarray}
\LMS^{(5)}&=&219.11\;_{-16.57}^{+18.36}\;({\rm
stat.})\quad_{-8.17}^{+8.36}\;({\rm
sys.})\quad_{-13.74}^{+14.47}\;({\rm mod.})\quad\pm8.97\;({\rm
HT})\;{\rm MeV}. \label{eq:6:errbrea}
\end{eqnarray}
We have used method {\bf I} to obtain experimental values of the
averages. The deviation between the results obtained using methods
{\bf I} and {\bf IV} is used to evaluate the modelling error since
these are the two methods which exhibit the largest deviation.

This result for $\LMS$ corresponds to a value of the strong coupling
constant (evaluated at the mass of the $Z$ particle) of,
\begin{eqnarray}
  \alpha_{s}(M_{Z})&=&0.1189_{-0.0019}^{+0.0019}
\end{eqnarray}
These values are in excellent agreement with the current global
averages of $\LMS^{(5)}=207.2\;\rmGeV$ and $\alpha_{s}=0.118\pm.002$
\cite{r16}. There is also good agreement with the result obtained
from fits using the Jacobi polynomial method  \cite{r9C} which yield
${\alpha}_{s}(M_Z)={0.119}^{+0.004}_{-0.004}$. This result is based
on fits using odd moments only and includes a contribution to the
error from scale dependence. Whilst close to the global average, the value of $\alpha_s(M_Z)$ we obtain
is significantly larger than the values found in a recent analysis
of the $F_2(x,Q^2)$ structure function \cite{r16c}, or from fits of
parton distibution functions where DIS and Drell-Yan data are
combined \cite{r16d}, it is also larger than the value which
minimizes the $\chi^2$ in global parton distribution function fits
such as Ref.~\cite{r16e}.

The $\chi^{2}/$d.o.f.~for our CORGI result is as follows,
\begin{eqnarray}
\frac{\chi^{2}}{\rm d.o.f.}&=&\frac{20.37}{271-(20+1)}\nn\\
&=&0.0815.\label{eq:6:chi}
\end{eqnarray}
Here `$271$' refers to the number of experimental Bernstein average
points used in the fits. Although this value is an order of
magnitude larger than the $\chi^{2}/$d.o.f. obtained in
Ref.~\cite{r3}, it is still significantly smaller than one would
expect, suggesting the errors on the Bernstein averages have been
over estimated. However, as discussed previously, the $\chi^{2}$
function we are using does not take into account the correlation
between data points. Indeed, if correlation was taken into account,
one might expect that a more reasonable value of $\chi^{2}$ would be
obtained.

\begin{figure}
\begin{center}
\hspace*{4pt}\begin{tabular}{c c} \psfrag{Fb}{$F_{nk}$}
\psfrag{F70}{\textcolor{magenta}{\small{$F_{70}$}}}
\psfrag{F50}{\textcolor{blue}{\small{$F_{50}$}}}
\psfrag{F30}{\textcolor{mygreen}{\small{$F_{30}$}}}
\psfrag{F10}{\textcolor{red}{\small{$F_{10}$}}}
\hspace{-.04\textwidth}
\includegraphics[width=0.5\textwidth]{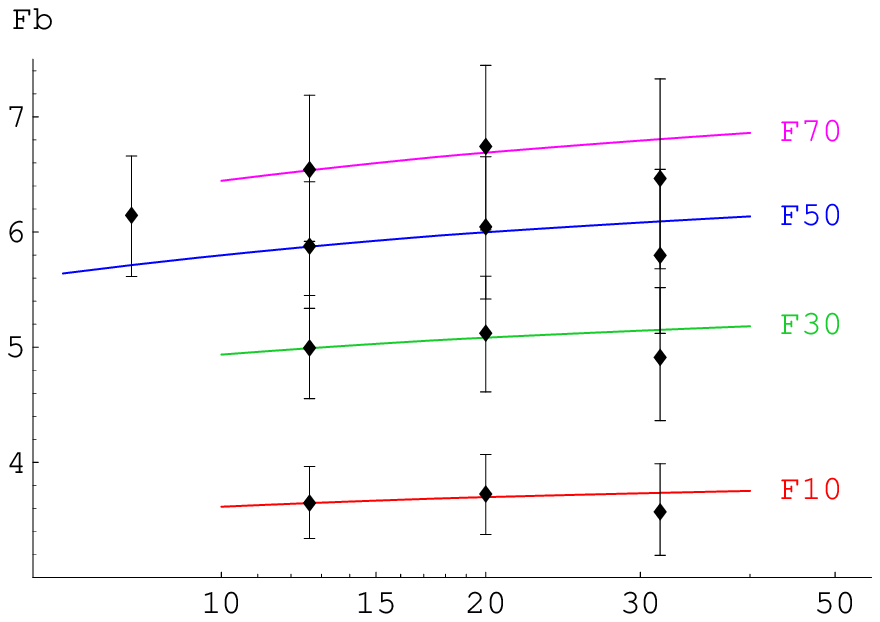}& \psfrag{Fb}{$F_{nk}$}
\psfrag{F80}{\textcolor{myred}{\small{$F_{80}$}}}
\psfrag{F60}{\textcolor{mygold}{\small{$F_{60}$}}}
\psfrag{F40}{\textcolor{mycyan}{\small{$F_{40}$}}}
\psfrag{F20}{\textcolor{magenta}{\small{$F_{20}$}}}
\includegraphics[width=0.5\textwidth]{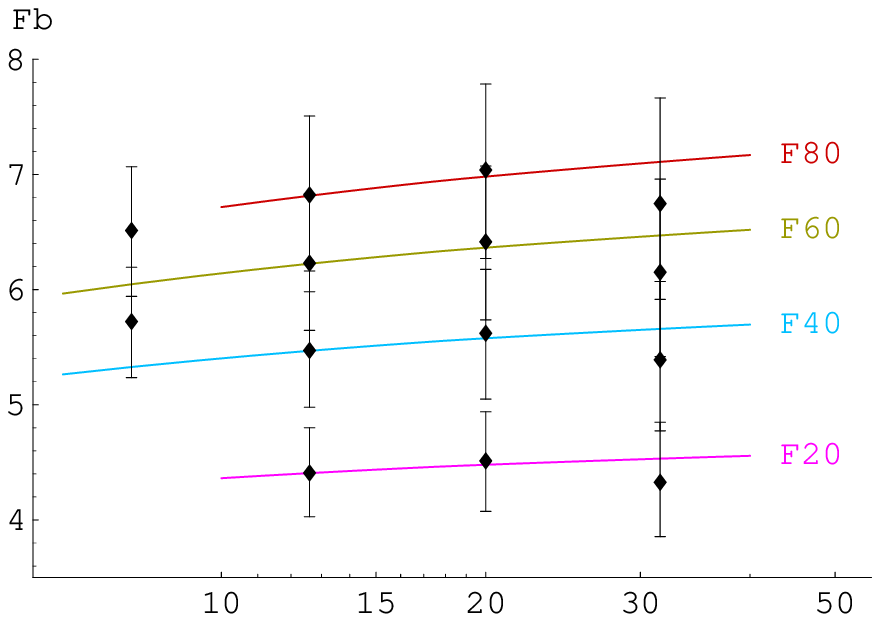}\\
\footnotesize{$Q^{2}/\GeVV$}&\footnotesize{$Q^{2}/\GeVV$}\\\\
\psfrag{Fb}{$F_{nk}$}
\psfrag{F81}{\textcolor{myred}{\small{$F_{81}$}}}
\psfrag{F71}{\textcolor{mygold}{\small{$F_{71}$}}}
\psfrag{F61}{\textcolor{mycyan}{\small{$F_{61}$}}}
\psfrag{F51}{\textcolor{magenta}{\small{$F_{51}$}}}
\psfrag{F41}{\textcolor{blue}{\small{$F_{41}$}}}
\psfrag{F31}{\textcolor{mygreen}{\small{$F_{31}$}}}
\psfrag{F21}{\textcolor{red}{\small{$F_{21}$}}}
\hspace{-.04\textwidth}\includegraphics[width=0.5\textwidth]{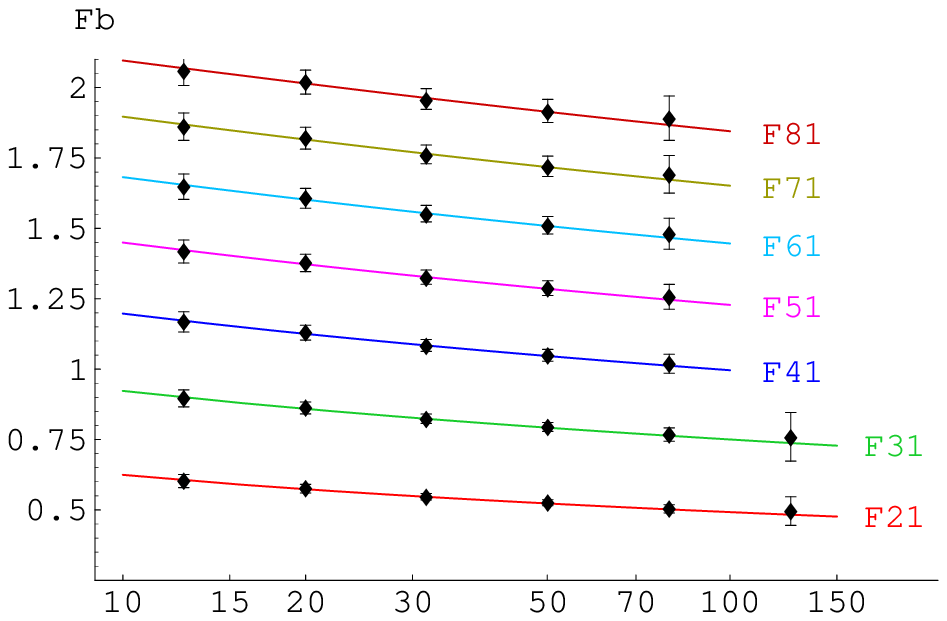}&
\psfrag{Fb}{$F_{nk}$}
\psfrag{F82}{\textcolor{myred}{\small{$F_{82}$}}}
\psfrag{F72}{\textcolor{mygold}{\small{$F_{72}$}}}
\psfrag{F62}{\textcolor{mycyan}{\small{$F_{62}$}}}
\psfrag{F52}{\textcolor{magenta}{\small{$F_{52}$}}}
\psfrag{F42}{\textcolor{blue}{\small{$F_{42}$}}}
\psfrag{F32}{\textcolor{mygreen}{\small{$F_{32}$}}}
\psfrag{F74}{\textcolor{red}{\small{$F_{74}$}}}
\includegraphics[width=0.5\textwidth]{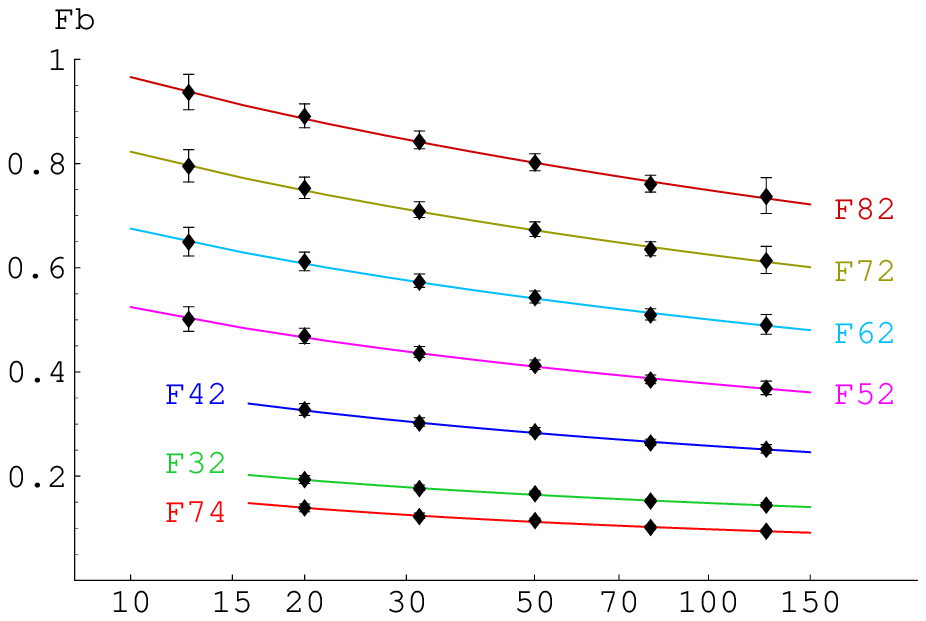}\\
\footnotesize{$Q^{2}/\GeVV$}&\footnotesize{$Q^{2}/\GeVV$}\\\\
\multicolumn{2}{c}{\psfrag{Fb}{$F_{nk}$}
\psfrag{F83}{\textcolor{mycyan}{\small{$F_{83}$}}}
\psfrag{F73}{\textcolor{magenta}{\small{$F_{73}$}}}
\psfrag{F63}{\textcolor{blue}{\small{$F_{63}$}}}
\psfrag{F84}{\textcolor{mygreen}{\small{$F_{84}$}}}
\psfrag{F53}{\textcolor{red}{\small{$F_{53}$}}}
\includegraphics[width=0.5\textwidth]{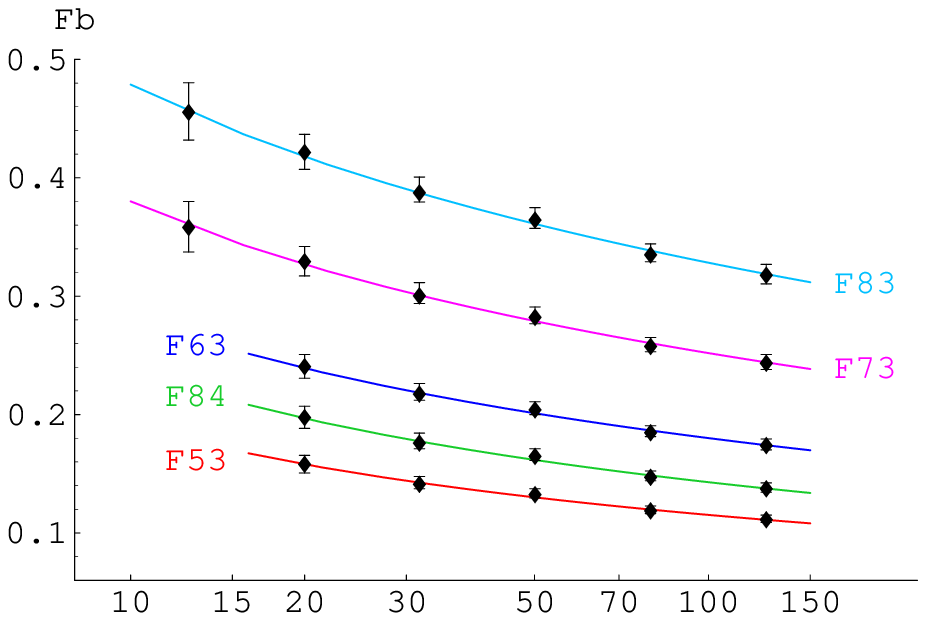}}\\
\multicolumn{2}{c}{\footnotesize{$Q^{2}/\GeVV$}}
\end{tabular}
\end{center}
\caption{CORGI fits for the Bernstein averages, with TMCs included.}
\label{fig:BAfits}
\end{figure}

\begin{figure}[t]
\begin{center}
\begin{tabular}{c c} \psfrag{Fb}{$\tF_{nk}$}
\psfrag{F90}{\textcolor{mygreen}{\small{$\tF_{80}$}}}
\psfrag{F80}{\textcolor{myred}{\small{$\tF_{80}$}}}
\psfrag{F70}{\textcolor{myred}{\small{$\tF_{70}$}}}
\psfrag{F60}{\textcolor{mygold}{\small{$\tF_{60}$}}}
\psfrag{F50}{\textcolor{mycyan}{\small{$\tF_{50}$}}}
\psfrag{F40}{\textcolor{magenta}{\small{$\tF_{40}$}}}
\psfrag{F30}{\textcolor{blue}{\small{$\tF_{30}$}}}
\psfrag{F20}{\textcolor{mygreen}{\small{$\tF_{20}$}}}
\psfrag{F10}{\textcolor{red}{\small{$\tF_{10}$}}}
\hspace{-.04\textwidth}
\includegraphics[width=0.5\textwidth]{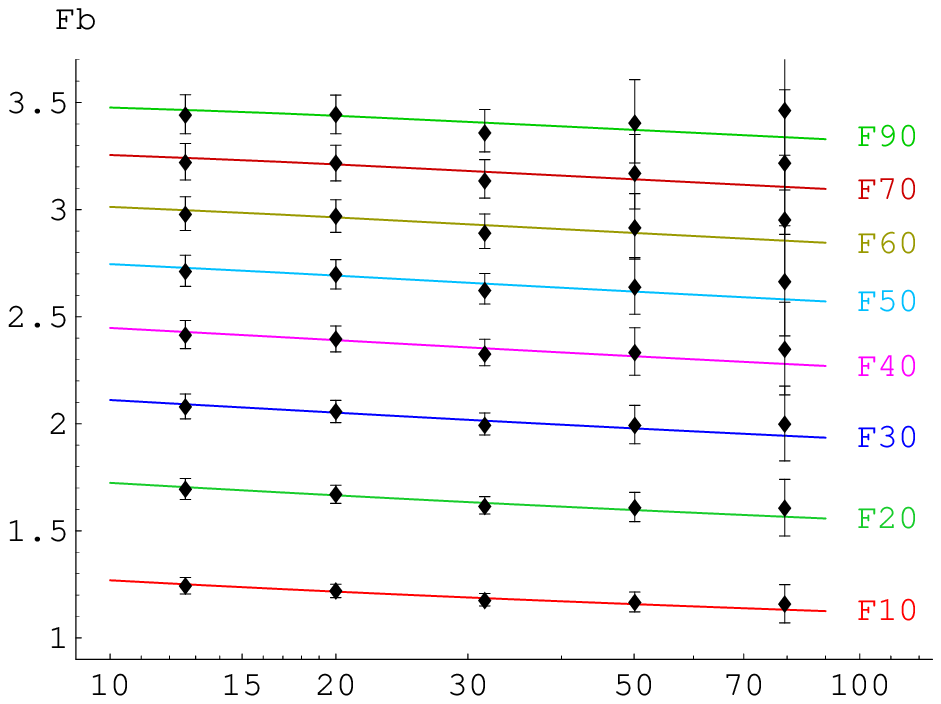}&
\psfrag{Fb}{$\tF_{nk}$}
\psfrag{F81}{\textcolor{myred}{\small{$\tF_{81}$}}}
\psfrag{F71}{\textcolor{mygold}{\small{$\tF_{71}$}}}
\psfrag{F61}{\textcolor{mycyan}{\small{$\tF_{61}$}}}
\psfrag{F51}{\textcolor{magenta}{\small{$\tF_{51}$}}}
\psfrag{F41}{\textcolor{blue}{\small{$\tF_{41}$}}}
\psfrag{F31}{\textcolor{mygreen}{\small{$\tF_{31}$}}}
\psfrag{F21}{\textcolor{red}{\small{$\tF_{21}$}}}
 \includegraphics[width=0.5\textwidth]{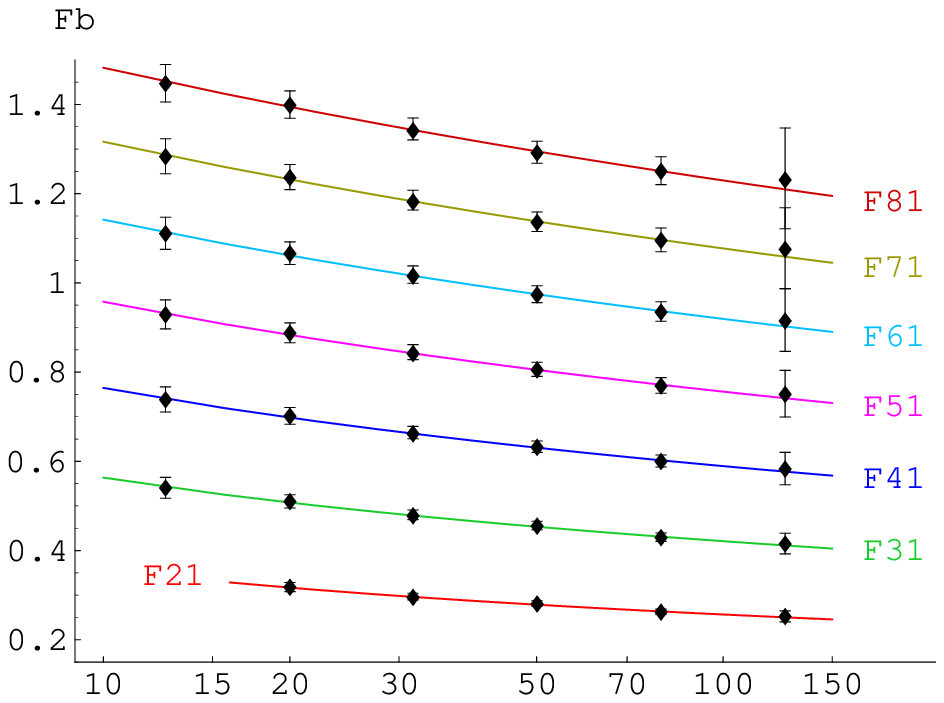}\\
\footnotesize{$Q^{2}/\GeVV$}&\footnotesize{$Q^{2}/\GeVV$}
\\\\
\psfrag{Fb}{$\tF_{nk}$}
\psfrag{F82}{\textcolor{myred}{\small{$\tF_{82}$}}}
\psfrag{F72}{\textcolor{mygold}{\small{$\tF_{72}$}}}
\psfrag{F62}{\textcolor{mycyan}{\small{$\tF_{62}$}}}
\psfrag{F52}{\textcolor{magenta}{\small{$\tF_{52}$}}}
\psfrag{F42}{\textcolor{blue}{\small{$\tF_{42}$}}}
\psfrag{F84}{\textcolor{mygreen}{\small{$\tF_{84}$}}}
\psfrag{F74}{\textcolor{red}{\small{$\tF_{74}$}}}
\hspace{-.04\textwidth}  \includegraphics[width=0.5\textwidth]{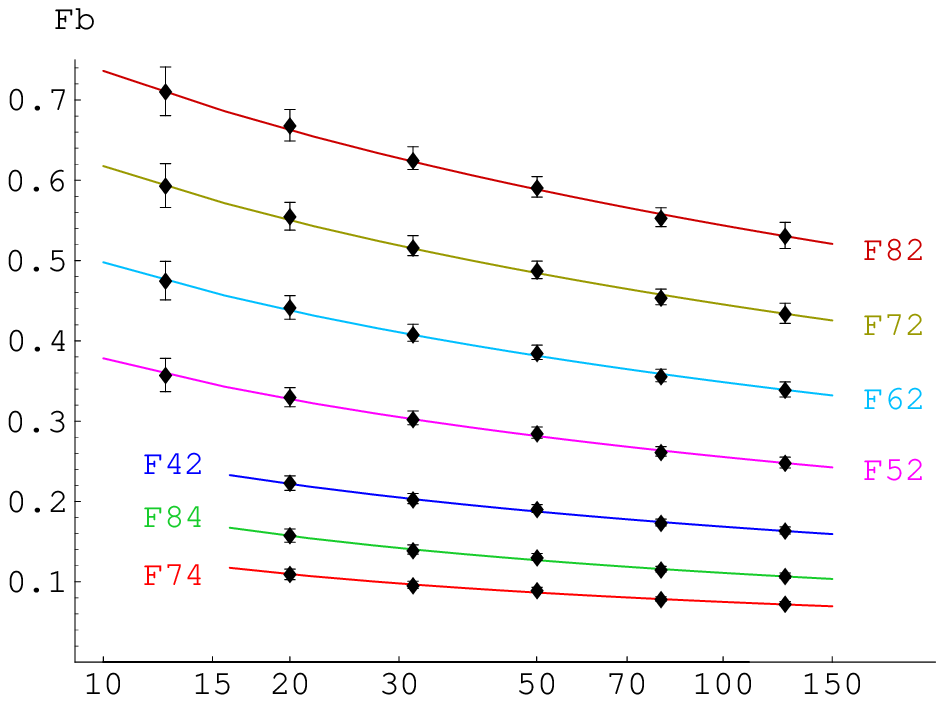}&
\psfrag{Fb}{$\tF_{nk}$}
\psfrag{F83}{\textcolor{magenta}{\small{$\tF_{83}$}}}
\psfrag{F73}{\textcolor{blue}{\small{$\tF_{73}$}}}
\psfrag{F63}{\textcolor{mygreen}{\small{$\tF_{63}$}}}
\psfrag{F53}{\textcolor{red}{\small{$\tF_{53}$}}}
\includegraphics[width=0.5\textwidth]{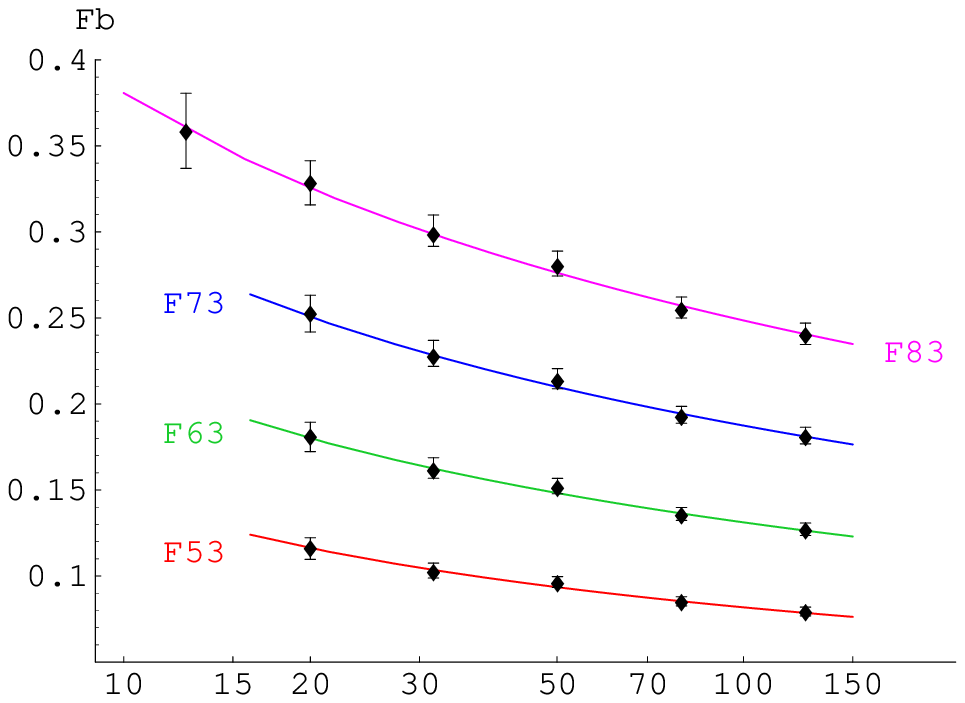}\\
\footnotesize{$Q^{2}/\GeVV$}&\footnotesize{$Q^{2}/\GeVV$}
\end{tabular}
\end{center}
\caption{ CORGI fits for the  modified Bernstein averages, with TMCs
included.} \label{fig:xBA fits}
\end{figure}

Furthermore, in the $\chi^{2}$ function we eventually used, the
errors on the Bernstein averages were rescaled in order to take into
account the `over-counting of degrees of freedom'. As a result, the
number of Bernstein averages is not representative of the true
number of degrees of freedom in this particular $\chi^{2}$ function.
Indeed, the Bernstein averages in the plots in Fig.~\ref{fig:BAfits}
can be constructed from just 58 different moments at different
values of $Q^{2}$ (via Eq.~(\ref{eq:BAsMoms})). Similarly, the
modified Bernstein averages in Fig.~\ref{fig:xBA fits} can be built
from 53 different moments. Hence,
\begin{eqnarray}
\frac{\chi^{2}}{\rm d.o.f.}\;=\;\frac{20.37}{111-21}\;=\;0.226,
\label{eq:6:newchi}
\end{eqnarray}
is more representative of the true value of $\chi^{2}/$d.o.f.~in
this approach. This is a more acceptable value, however we stress
that the {\it true} minimum in $\chi^{2}$ can only be determined by
taking correlation fully into account.

In Fig.~\ref{fig:BAfits} we plot the CORGI predictions for the
Bernstein averages (with TMCs included) fitted to the experimental
values. Figure \ref{fig:xBA fits} shows equivalent plots for the
modified averages.

\begin{table}
\begin{center}
\begin{tabular}{|c||c|c|c|}
\hline&&&\\[-10pt] &$\LMS^{(5)}
(\textrm{MeV})$&$\alpha_{s}(M_{Z})$&$\chi^{2}/{\rm
d.o.f.}$\\[.1cm]\hhline{|=#=|=|=|}

All moments&219.1\pmpm{+23.6}{-22.1}&
0.1189\pmpm{+0.0019}{-0.0019}&$20.37/(271-(20+1))$\\\hhline{-|---|}

Odd moments&210.5\pmpm{+35.0}{-32.7}&
0.1182\pmpm{+0.0029}{-0.0030}&$10.94/(130-(10+1))$\\\hhline{-|---|}

Even moments&229.5\pmpm{+64.7}{-62.1}&
0.1198\pmpm{+0.0048}{-0.0048}&$9.24/(141-(10+1))$\\\hhline{-|---|}

All moments: $Q^{2}>m_{b}^{2}$ only&232.4\pmpm{+34.9}{-33.2}&
0.1200\pmpm{+0.0027}{-0.0027}&$15.59/(228-(20+1))$\\\hline
\end{tabular}
\end{center}
\caption{In this table we present the result of the analysis
performed using
  the {\bf CORGI} approach to perturbation theory with target mass corrections
  included. We compare the results obtained when we include
  all moments (up to $n=20$) with those obtained when we restrict the
  analysis to even or odd moments only. We also show the results from
  performing the analysis with only data points for which
  $Q^{2}>m_{b}^{2}$ ($\nf=5$) included.}
\label{t:1}
\end{table}
In table \ref{t:1} we present the full set of results from the CORGI
analysis. This table shows the results obtained when we include both
odd and even moments (standard and modified Bernstein averages)
together and also when we restrict the analysis to odd and even or
odd moments only. These results allow us to check consistency
between the odd, even and `All' analyses.

We can also use the results of the `odd moments' analysis to check
consistency with previous analyses. In the PS analysis of
Ref.~\cite{r3} (in which only the $n=1,3,5,7,9,11$ and 13 moments
were included) a value of $\LMS^{(4)}=255\pm72$MeV was found,
corresponding to $\LMS^{(5)}=178_{-55}^{+57}$MeV, with a value of
$\chi^{2}/{\rm d.o.f.}=0.007$. Using the same set of moments, the
CORGI analysis of Ref.~\cite{r4} found a value of
$\LMS^{(5)}=228_{-36}^{+35}$MeV, although it must be noted that this
analysis used incorrect values of the coefficients $X_{2}$, and
therefore the result must be regarded as unreliable. Both of those
results are indeed consistent with the `odd moments' analysis
performed here.

We also perform an analysis in which we restrict the CCFR data to
$Q^{2}>m_{b}^{2}$ only, as a check on our method of evolving through
the $b$ quark threshold. These results are also included in table
\ref{t:1}. In table 3 we give the fitted CORGI values for the $A_n$
non-perturbative coefficients for $n=1-20$, together with the values
of the corresponding moments at $Q^2={8.75}\;{\rm{GeV}}^{2}$ and
${12.6}\;{\rm{GeV}}^{2}$.

In table \ref{t:2} we compare the CORGI results with those obtained
using the PS and EC approaches. We also present results obtained
from performing these analyses with and without target mass
corrections. The fact that the number of d.o.f. for the CORGI fits
is $271$ ($272$), as opposed to $273$ for PS and EC, reflects the
fact that for the smallest energy bin $Q^2={7.9}\;{\rm{GeV}}^2$, the
${\Lambda}_{\MM}^{2}$ appearing in the CORGI coupling exceeds $Q^2$
for the highest $n=19,20$ moments, and hence one is below the Landau
pole in the CORGI coupling of Eq.~(\ref{eq:6tHooft}).
Correspondingly $x_{\rmt{CORGI}}$ is significantly less than unity
(see table 1). We simply omit the two affected Bernstein average
points from the CORGI fit.
\begin{table}
\begin{center}
\begin{tabular}{|c|c|c|c|}\hline
$n$&$A_{n}^{(4)}$&$\MM(n;8.75\,\GeVV)$&$\MM(n;12.6\,\GeVV)$\\
\hline 1&2.346&2.494& 2.525$$\\\hline
 2&0.8814& 0.3557&0.3466\\\hline
 3 &0.4133&0.1002&9.545$\times 10^{-2}$\\\hline
 4&0.2217&3.835$\times 10^{-2}$&3.584$\times 10^{-2}$\\\hline
 5&0.1292&1.744$\times 10^{-2}$&1.603$\times 10^{-2}$\\\hline
 6&8.134$\times 10^{-2}$&9.048$\times 10^{-3}$&8.191$\times 10^{-3}$\\\hline
 7&5.241$\times 10^{-2}$&4.988$\times 10^{-3}$&4.452$\times 10^{-3}$\\\hline
 8&3.639$\times 10^{-2}$&3.044$\times 10^{-3}$&2.681$\times 10^{-3}$\\\hline
 9&2.434$\times 10^{-2}$&1.826$\times 10^{-3}$&1.588$\times 10^{-3}$\\\hline
 10&1.822$\times 10^{-2}$&1.246$\times 10^{-3}$&1.07$\times 10^{-3}$\\\hline
 11&1.202$\times 10^{-2}$&7.588$\times 10^{-4}$&6.438$\times 10^{-4}$\\\hline
 12&9.64$\times 10^{-3}$&5.677$\times 10^{-4}$&4.76$\times 10^{-4}$\\\hline
 13&5.935$\times 10^{-3}$&3.289$\times 10^{-4}$&2.726$\times 10^{-4}$\\\hline
 14&5.119$\times 10^{-3}$&2.69$\times 10^{-4}$&2.204$\times 10^{-4}$\\\hline
 15&2.702$\times 10^{-3}$&1.355$\times 10^{-4}$&1.098$\times 10^{-4}$\\\hline
 16&2.535$\times 10^{-3}$&1.22$\times 10^{-4}$&9.768$\times 10^{-5}$\\\hline
 17&9.362$\times 10^{-4}$&4.343$\times 10^{-5}$&3.44$\times 10^{-5}$\\\hline
 18&9.739$\times 10^{-4}$&4.376$\times 10^{-5}$&3.426$\times 10^{-5}$\\\hline
 19&9.807$\times 10^{-10}$&4.284$\times 10^{-11}$&3.317$\times 10^{-11}$\\\hline
 20&7.69$\times 10^{-10}$&3.277$\times 10^{-11}$&2.509$\times 10^{-11}$\\\hline
\end{tabular}
\end{center}
\caption{Fitted values of $A_n$ (in the $\nf=4$ region) together
with the moments evaluated at
  $Q^{2}=8.75$ and $12.6\,\GeVV$, for the CORGI approach.}
\label{t:An}
\end{table}
\begin{table}
\begin{center}
\begin{tabular}{|c|c||c|c|c|}
\hline \multicolumn{2}{|c||}{}
&$\LMS^{(5)}(\textrm{MeV})$&$\alpha_{s}(M_{Z})$&$\chi^{2}/{\rm
d.o.f.}$\\\hhline{|=|=#=|=|=|}

\multirow{2}{2cm}{CORGI} &with
TMC&219.1\pmpm{+23.6}{-22.1}&0.1189\pmpm{+0.0019}{-0.0019}
&$20.37/(271-(20+1))$\\\hhline{~|----|}

&no TMC &280.3\pmpmc{+24.6}{-23.4}&0.1235\pmpm{+0.0017}{-0.0017}
&$24.76/(272-(18+1))$\\\hhline{|=|=#=|=|=|}

\multirow{2}{2cm}{PS} &with TMC&
200.4\pmpmc{+25.8}{-24.8}&0.1173\pmpm{+0.0023}{-0.0022}
&$21.73/(273-(20+1))$\\\hhline{|~|----|}

&no TMC &257.5\pmpmc{+27.8}{-27.6}&0.1219\pmpm{+0.0020}{-0.0020}
&$25.20/(273-(18+1))$\\\hhline{|=|=#=|=|=|}

\multirow{2}{2cm}{EC} &with TMC
&204.5\pmpmc{+19.9}{-18.9}&0.1177\pmpm{+0.0017}{-0.0017}
&$22.71/(273-(20+1))$\\\hhline{|~|----|}

&no TMC &261.5\pmpmc{+19.0}{-18.4}&0.1222\pmpm{+0.0014}{-0.0014}\
&$26.04/(273-(18+1))$\\\hline
\end{tabular}
\end{center}
\caption{In this table we compare the results of the analysis performed with
  the three different approaches to perturbation theory described in section 3,
  CORGI, PS and EC. We also show the results from these analyses  performed  with and without
  target mass corrections.}
\label{t:2}
\end{table}

\section{Discussion and conclusions}\vspace{-\parskip}
\label{s:6Conc} In this paper we have used three different
approaches to perturbation theory to perform a phenomenological
analysis of moments of $F_{3}$ using the method of Bernstein
averages. The three approaches differ in how they deal with the FRS
dependence.  In the CORGI approach, we allow the FRS invariant
quantity $X_{0}(Q)$ to determine the relationship between $M$, $\mu$
and $Q$ for each moment. In so doing, we automatically resum the
subset of terms present in the full perturbative expansion which are
RG-predictable at NNLO. In the physical scale approach we set
$M=\mu=Q$ and adopt the \MSbar~scheme for the subtractions in the
renormalization {\it and} factorization procedures. In the effective
charge approach, we set $M=\mu$ and apply the CORGI approach to the
resulting single-scale effective charge. We described how
predictions are derived in these three approaches and corrected
errors in the CORGI method which were present in Refs.~\cite{r4,r5}.

We described how target mass and higher twist corrections affect
these theoretical predictions and also how we evolve expressions for
the moments through the $b$-quark threshold. We explained how the
Bernstein averages method eliminates any potential dependence of the
analysis on missing data regions in $x$ and $Q^{2}$, and we also
described how this method is generalized to treat both odd and even
moments. We described the fitting procedure used to extract the
optimal values of the QCD scale parameter and how we can implement
various constraints which ensure that the results of this fitting
are consistent with the structure functions being positive definite
functions. We also presented an alternative, and slightly easier
method for deriving the FRS invariant quantities $X_{i}$.

The results of the CORGI analysis presented in table \ref{t:1} show
excellent agreement with the current global average for the strong
coupling evaluated at $Q^{2}=M_{Z}^{2}$ \cite{r16}, and are also in
good agreement with fits based on the Jacobi polynomial approach
\cite{r9C}. From this we conclude that CORGI perturbation theory
performs well when applied to the analysis of moments. The analyses
in which we include only odd or  even moments are consistent with
each other and with the full (all moments) analysis. Furthermore, in
the analysis in which we include all moments, the errors are greatly
reduced. This improvement in the analysis is made possible by the
availability of the full NNLO anomalous dimension calculation and
represents significant improvement on previous analyses.

Excluding data points for which $Q^{2}<m_{b}^{2}$ leads to no
significant change in the results and from this we conclude that the
quark mass threshold method we have applied is suitable to the
moment analysis. The error associated with the exclusion of
higher-twist effects, given in Eq.~(\ref{eq:6:errbrea}), is
relatively small, signifying that these effects are not particularly
important at scales $Q^{2}>7.6\,\GeVV$.

We include in the analysis positivity constraints on the moments
(Eqs.~(\ref{eq:6:con1}) and (\ref{eq:6:con2})), via the
parameter redefinitions defined in section 4. We find that this
implementation has little effect on the prediction of $\LMS$ ($\sim10$ Mev),
but does make a difference to the values of $A_{n}$.

The CORGI predictions for the Bernstein averages (with TMCs
included) are plotted in Figs.~\ref{fig:BAfits} and \ref{fig:xBA
fits} and show excellent agreement with experimental values. This is
reflected by the low value of $\chi^{2}/$d.o.f.~associated with this
fitting, given in Eq.~(\ref{eq:6:chi}). However, as we noted, ideally
the full covariance matrix should be used in constructing $\chi^2$ to
account for correlations between the Bernstein averages used in the fits.
Unfortunately, we found that this matrix is ill-conditioned, having some
eigenvalues close to zero, and so it proved to be numerically intractable
to invert the matrix to construct the true $\chi^2$. We therefore
resorted to the same approximate rescaling of errors employed in
Refs.\cite{r2,r3} to try to compensate for possible correlations.

The results also show consistency between CORGI, PS and EC. The PS
and EC analyses lead to values of $\LMS$ and ${\alpha}_{s}$ slightly
lower than in the CORGI analysis. However, this variation is well
within the error bars on the associated quantities. Inclusion of HT
effects generally results in a small shift in $\LMS$ of about 10
MeV. However, when target mass corrections are included, we see a
shift of approximately $60$ MeV in the predicted value of $\LMS$ and
from this we conclude that these contributions are significant in
the case of $F_{3}$.

An obvious further study would be to apply the same fitting procedure to the recently
released NuTev data \cite{r1a}. In future work we also hope to report on similar fits to
data for the $F_2$ structure function \cite{r18}. This analysis is considerably more complicated due
to the presence of an additional singlet component.

\section*{Note Added in Proof}
At around the same time as our paper was completed a related analysis
of the CCFR data for $xF_3$ also employing Bernstein averages appeared
\cite{r19}.

\section*{Acknowledgements}\vspace{-\parskip}
We would like to thank Andreas Vogt for providing us with computer
code implementing the NNLO anomalous dimension coefficients of
Refs.~\cite{r7,r8}, and also for helpful discussions on quark mass
thresholds. We thank Jeff Forshaw for pointing out the potential importance of taking
correlations between the Bernstein averages into account. 
Andrei Kataev is also thanked for useful discussions. P.M.B.
gratefully acknowledges the receipt of a PPARC UK studentship.

\end{document}